\documentclass[fleqn,usenatbib]{mnras}
\hypersetup{final} 

\usepackage{newtxtext,newtxmath}

\usepackage[T1]{fontenc}
\usepackage{ae,aecompl}

\usepackage{graphicx}	
\usepackage{amsmath}	
\usepackage{amssymb}	

\newcommand{\tess}{{\it TESS}}

\newcommand{\gaia}{{\it Gaia}}

\newcommand{\NGTS}{NGTS}

\newcommand{\gpe}{{GP-EBOP}}
\newcommand{\LSO}{La Silla Observatory}

\newcommand{\Euler}{Eulercam}

\newcommand{\SHOCa}{SHOCa}
\newcommand{\SHOCh}{SHOCh}

\newcommand{\kms}{\mbox{km\,s$^{-1}$}}
\newcommand{\ms}{\mbox{m\,s$^{-1}$}}
\newcommand{\mas}{\mbox{mas}}
\newcommand{\masy}{\mbox{mas\,y$^{-1}$}}
\newcommand{\mpl}{\mbox{M$_{\rm p}$}}
\newcommand{\rpl}{\mbox{R$_{\rm p}$}}
\newcommand{\rhopl}{\mbox{$\rho_{\rm p}$}}
\newcommand{\mstar}{\mbox{M$_{\rm s}$}}
\newcommand{\rstar}{\mbox{R$_{\rm s}$}}
\newcommand{\mjup}{\mbox{M$_{\rm J}$}}
\newcommand{\rjup}{\mbox{R$_{\rm J}$}}

\newcommand{\msun}{\mbox{$M_{\odot}$}}
\newcommand{\rsun}{\mbox{$R_{\odot}$}}

\newcommand{\gccc}{g\,cm$^{-3}$}

\newcommand{\teff}{$T_{\rm eff}$}
\newcommand{\teqpl}{$T_{\rm eq}$}
\newcommand{\tdur}{$T_{\rm 14}$}
\newcommand{\logg}{$\log g$}
\newcommand{\feh}{[Fe/H]}
\newcommand{\av}{$A_{V}$}
\newcommand{\Qs}{$Q'_{\rm s}$}

\newcommand{\logQs}{$\log Q'_{\rm s}$}
\newcommand{\logQpl}{$\log Q'_{\rm p}$}

\newcommand{\vsini}{$v \sin i_\star$}
\newcommand{\vmacro}{$v_{\mathrm{macro}}$}
\newcommand{\NStar}{NGTS-10}
\newcommand{\NStarB}{G-6880}
\newcommand{\NGaiaId}{2911987212510959232}
\newcommand{\NRa}{\mbox{$06^{\rmn{h}} 07^{\rmn{m}} 29\fs3472$}} 
\newcommand{\NDec}{\mbox{$-25\degr 35\arcmin 41\farcs 6268$}} 
\newcommand{\NPropRa}{\mbox{$-2.323\pm0.343$}} 
\newcommand{\NPropDec}{\mbox{$10.527\pm0.395$}} 
\newcommand{\NRaB}{\mbox{$06^{\rmn{h}} 07^{\rmn{m}} 29\fs3118$}} 
\newcommand{\NDecB}{\mbox{$-25\degr 35\arcmin 40\farcs 6118$}} 
\newcommand{\NPropRaB}{\mbox{$-1.120\pm0.219$}} %
\newcommand{\NPropDecB}{\mbox{$9.671\pm0.161$}} %
\newcommand{\NStarMassBoy}{\mbox{$0.696\pm0.040$}} 
\newcommand{\NStarRadiusBoy}{\mbox{$0.697\pm0.036$}} 
\newcommand{\NLoggBoy}{\mbox{$4.595\pm0.019$}} 
\newcommand{\NTeffSed}{\mbox{$4400\pm100$}} 
\newcommand{\NScaleSed}{\mbox{$0.287\pm0.02 \times10^{-20}$}} 
\newcommand{\NAvSed}{\mbox{$0.0067\,^{+0.0174}_{-0.0098}$}} 
\newcommand{\NErrorInflSed}{\mbox{$0.0498\,^{+0.048}_{-0.030}$}} 
\newcommand{\NTeffBSed}{\mbox{$6263\,^{+206}_{-212}$}} 
\newcommand{\NScaleBSed}{\mbox{$0.178\,^{+0.0070}_{-0.0044}\times10^{-20}$}} 
\newcommand{\NLoggBSed}{\mbox{$4.0\pm0.26$}} 
\newcommand{\NAvBSed}{\mbox{$0.0366\,^{+0.074}_{-0.040}$}} 
\newcommand{\NMetalBarry}{\mbox{$-0.02\pm0.12$}} 
\newcommand{\NActivityIndexBarry}{\mbox{$-4.70\pm0.19$}} 
\newcommand{\NTeffBarry}{\mbox{$4600\pm150$}} 
\newcommand{\NLoggBarry}{\mbox{$4.5\pm0.2$}} 
\newcommand{\NAstrometricNoiseExcess}{\mbox{$2.1518$}} 
\newcommand{\NDist}{\mbox{$325\pm29$}} 
\newcommand{\NProt}{\mbox{$17.290\pm0.008$}}
\newcommand{\Nage}{\mbox{$10.4\pm2.5$}}

\newcommand{\NBmag}{\mbox{$15.274\pm0.044$}}

\newcommand{\Ngmag}{\mbox{$14.780\pm0.040$}}
\newcommand{\Nrmag}{\mbox{$13.980\pm0.050$}}
\newcommand{\Nimag}{\mbox{$13.637\pm0.015$}}
\newcommand{\NGaiaMagB}{\mbox{$15.593\pm0.005$}}
\newcommand{\NGaiaMag}{\mbox{$14.260\pm0.005$}}
\newcommand{\NNmag}{\mbox{$13.626\pm0.010$}}
\newcommand{\NJmag}{\mbox{$12.392\pm0.026$}}
\newcommand{\NHmag}{\mbox{$11.878\pm0.028$}}
\newcommand{\NKmag}{\mbox{$11.728\pm0.025$}}
\newcommand{\NWmag}{\mbox{$11.644\pm0.024$}}
\newcommand{\NWWmag}{\mbox{$11.672\pm0.022$}}
\newcommand{\NVmag}{\mbox{$14.340\pm0.015$}}

\newcommand{\NDNgts}{\mbox{$0.22\,^{+0.03}_{-0.02}$}}
\newcommand{\NDi}{\mbox{$0.19\,^{+0.03}_{-0.02}$}}
\newcommand{\NDV}{\mbox{$0.30\,^{+0.03}_{-0.02}$}}
\newcommand{\NDB}{\mbox{$0.42\,^{+0.04}_{-0.03}$}}
\newcommand{\NDz}{\mbox{$0.18\,^{+0.03}_{-0.02}$}}

\newcommand{\CompanionGaiaID}{2911987212508106880}
\newcommand{\CompanionSeperation}{$1.2$}
\newcommand{\CompanionGaiaMag}{$15.59$}
\newcommand{\CompanionPositionAngle}{$334.74$}
\newcommand{\CompanionAstromertricNoiseExcess}{$0.0644$}

\newcommand{\FArasum}{\mbox{$0.2637\,^{+0.0105}_{-0.0081}$}}%
\newcommand{\FArr}{\mbox{$0.1765\,^{+0.0110}_{-0.0070}$}}%
\newcommand{\FAcosi}{\mbox{$0.1890\,^{+0.0142}_{-0.0071}$}}%
\newcommand{\FAb}{\mbox{$0.8523\,^{+0.03158}_{-0.0197}$}}%
\newcommand{\FAperiod}{\mbox{$0.7668944\pm0.0000003$}}%
\newcommand{\FAPeriodShort}{\mbox{$0.76689$}}%
\newcommand{\FAepoch}{\mbox{$2457518.84377\pm0.00017$}}%
\newcommand{\FADilNgts}{\mbox{$0.210\,^{+0.021}_{-0.027}$}}
\newcommand{\FADilI}{\mbox{$0.200\pm0.026$}}
\newcommand{\FADilV}{\mbox{$0.297\pm0.026$}}
\newcommand{\FADilB}{\mbox{$0.398\,^{+0.051}_{-0.045}$}}
\newcommand{\FADilz}{\mbox{$0.193\pm0.034$}}
\newcommand{\FAvsys}{\mbox{$39.0931\pm0.0057$}}%
\newcommand{\FAkp}{\mbox{$0.5949\,^{+0.0077}_{-0.0063}$}}%

\newcommand{\NPlanet}{NGTS-10b}
\newcommand{\NTDur}{\mbox{$1.091\pm0.019$}}%
\newcommand{\NMass}{\mbox{$2.162\,^{+0.092}_{-0.107}$}}%
\newcommand{\NRadius}{\mbox{$1.205\,^{+0.117}_{-0.083}$}}%
\newcommand{\NDensity}{\mbox{$1.430\,^{+0.354}_{-0.404}$}}%
\newcommand{\NTeq}{\mbox{$1332\,^{+49}_{-54}$}}%
\newcommand{\NAAu}{\mbox{$0.0143\pm0.0010$}}%
\newcommand{\NARs}{\mbox{$4.447\,^{+0.123}_{-0.141}$}}%
\newcommand{\NARoche}{\mbox{$1.462\pm0.179$}}
\newcommand{\NARocheAbs}{\mbox{$1.46\pm0.18$}}
\newcommand{\NInsol}{\mbox{$1.091\pm0.214$}} 

\newcommand{\NPeriodNGTSVI}{\mbox{$0.882058\pm0.000001$}}
\newcommand{\NAAuNGTSVI}{\mbox{$0.01662\pm0.00050$}}
\newcommand{\NARsNGTSVI}{\mbox{$3.1268\pm0.3789$}}
\newcommand{\NARocheNGTSVI}{\mbox{$1.1044\pm0.2149$}}
\newcommand{\NInsolNGTSVI}{\mbox{$0.723\pm0.189$}}

\newcommand{\NPeriodWXIIX}{\mbox{$0.941450\pm0.000001$}}
\newcommand{\NAAuWXIIX}{\mbox{$0.02030\pm0.00700$}}
\newcommand{\NARsWXIIX}{\mbox{$3.3838\pm1.1742$}}
\newcommand{\NARocheWXIIX}{\mbox{$2.8370\pm1.0109$}}
\newcommand{\NInsolWXIIX}{\mbox{$8.470\pm5.884$}}

\newcommand{\NPeriodWXIX}{\mbox{$0.788838989\pm0.000000040$}}
\newcommand{\NAAuWXIX}{\mbox{$0.01634\pm0.00024$}}
\newcommand{\NARsWXIX}{\mbox{$3.4996\pm0.0758$}}
\newcommand{\NARocheWXIX}{\mbox{$1.0481\pm0.0392$}}
\newcommand{\NInsolWXIX}{\mbox{$4.450\pm0.298$}}

\newcommand{\NPeriodWXXXXIII}{\mbox{$0.813475\pm0.000001$}}
\newcommand{\NAAuWXXXXIII}{\mbox{$0.01420\pm0.00040$}}
\newcommand{\NARsWXXXXIII}{\mbox{$5.0891\pm0.3683$}}
\newcommand{\NARocheWXXXXIII}{\mbox{$1.8735\pm0.1637$}}
\newcommand{\NInsolWXXXXIII}{\mbox{$0.821\pm0.191$}}

\newcommand{\NPeriodWCIII}{\mbox{$0.925542\pm0.000019$}}
\newcommand{\NAAuWCIII}{\mbox{$0.01985\pm0.00021$}}
\newcommand{\NARsWCIII}{\mbox{$2.9724\pm0.1121$}}
\newcommand{\NARocheWCIII}{\mbox{$1.1725\pm0.0631$}}
\newcommand{\NInsolWCIII}{\mbox{$8.944\pm1.155$}}

\newcommand{\NPeriodKXVI}{\mbox{$0.9689951\pm0.0000024$}}
\newcommand{\NAAuKXVI}{\mbox{$0.02044\pm0.00026$}}
\newcommand{\NARsKXVI}{\mbox{$3.2318\pm0.1575$}}
\newcommand{\NARocheKXVI}{\mbox{$1.6032\pm0.1042$}}
\newcommand{\NInsolKXVI}{\mbox{$8.209\pm0.849$}}

\newcommand{\NPeriodHXIIX}{\mbox{$0.83784340\pm0.00000047$}}
\newcommand{\NAAuHXIIX}{\mbox{$0.01761\pm0.00027$}}
\newcommand{\NARsHXIIX}{\mbox{$3.7125\pm0.2151$}}
\newcommand{\NARocheHXIIX}{\mbox{$1.3797\pm0.1108$}}
\newcommand{\NInsolHXIIX}{\mbox{$4.046\pm0.583$}}

\newcommand{\Nhdsep}{\mbox{$40.73$}}
\newcommand{\Nhdmagv}{\mbox{$9.32$}}
\newcommand{\Nhdmagg}{\mbox{$9.00$}}

\newcommand{\RvMask}{K5}
\newcommand{\RvBisSlope}{$-0.003\pm0.042$}

\newcommand{\NVsini}{$4.0\pm0.6$} 
\newcommand{\NKAmp}{\mbox{$595\,^{+8}_{-6} $}}
\newcommand{\NGamma}{\mbox{$39.0931\,^{+0.0054}_{-0.0057}$}}
\newcommand{\NVsys}{$39.0931$}

\newcommand{\NMedInspiralTimeQVII}{$38$}
\newcommand{\NPeriodDecayFiveYear}{$2$}
\newcommand{\NPeriodDecayDecade}{$7$}
\newcommand{\NInertia}{$0.2756\pm0.006$}

\title[\NPlanet]{\NPlanet: The shortest period hot Jupiter yet discovered}

\author[J. McCormac et al.]{
\parbox{\textwidth}{
James~McCormac,$^{1, 2, *}$
Edward~Gillen,$^{3,\dagger}$
James~A.~G.~Jackman,$^{1, 2}$
David~J.~A.~Brown,$^{1, 2}$
Daniel~Bayliss,$^{1, 2}$
Peter~J.~Wheatley,$^{1, 2}$
David~R.~Anderson,$^{1, 2}$ 
David~J.~Armstrong,$^{1, 2}$
Fran\c{c}ois~Bouchy,$^{4}$
Joshua~T.~Briegal,$^{3}$
Matthew~R.~Burleigh,$^{5}$
Juan~Cabrera,$^{6}$
Sarah~L.~Casewell,$^{5}$
Alexander~Chaushev,$^{5, 1, 2}$
Bruno~Chazelas,$^{4}$
Paul~Chote,$^{1, 2}$
Benjamin~F.~Cooke,$^{1, 2}$
Jean C. Costes,$^{11}$
Szil\'ard~Csizmadia,$^{6}$
Philipp~Eigm\"uller,$^{6}$
Anders~Erikson,$^{6}$
Emma~Foxell,$^{1, 2}$
Boris~T.~G\"ansicke,$^{1, 2}$
Michael~R.~Goad,$^{5}$
Maximilian~N.~G{\"u}nther,$^{8, 9}$
Simon~T.~Hodgkin,$^{10}$
Matthew~J.~Hooton,$^{11}$
James~S.~Jenkins,$^{12, 13}$
Gregory~Lambert,$^{3}$
Monika~Lendl,$^{4, 17}$
Emma~Longstaff,$^{5}$ 
Tom~Louden,$^{1, 2}$
Maximiliano Moyano,$^{14}$
Louise~D.~Nielsen,$^{4}$
Don~Pollacco,$^{1, 2}$
Didier~Queloz,$^{3}$
Heike~Rauer,$^{6, 7, 15}$
Liam~Raynard,$^{5}$
Alexis~M.~S.~Smith,$^{6}$
Barry~Smalley,$^{16}$
Maritza~Soto,$^{12}$
Oliver~Turner,$^{4}$ 
St\'{e}phane~Udry,$^{4}$
Jose~I.~Vines,$^{12}$
Simon~R.~Walker,$^{1, 2}$
Christopher~A.~Watson,$^{11}$
Richard~G.~West,$^{1, 2}$
}
\\
$^{1}$Centre for Exoplanets and Habitability, University of Warwick, Gibbet Hill Road, Coventry CV4 7AL, UK\\
$^{2}$Dept.\ of Physics, University of Warwick, Gibbet Hill Road, Coventry CV4 7AL, UK\\
$^{3}$Astrophysics Group, Cavendish Laboratory, J.J. Thomson Avenue, Cambridge CB3 0HE, UK\\
$^{4}$Observatoire de Gen{\`e}ve, Universit{\'e} de Gen{\`e}ve, 51 Ch. des Maillettes, 1290 Sauverny, Switzerland\\
$^{5}$Department of Physics and Astronomy, University of Leicester, University Road, Leicester, LE1 7RH, UK\\
$^{6}$Institute of Planetary Research, German Aerospace Center, Rutherfordstrasse 2, 12489 Berlin, Germany\\
$^{7}$Center for Astronomy and Astrophysics, TU Berlin, Hardenbergstr. 36, D-10623 Berlin, Germany\\
$^{8}$Department of Physics, and Kavli Institute for Astrophysics and Space Research, Massachusetts Institute of Technology,\\
Cambridge, MA 02139, USA $^{9}$Juan Carlos Torres Fellow\\
$^{10}$Institute of Astronomy, University of Cambridge, Madingley Road, Cambridge CB3 0HA, UK\\
$^{11}$Astrophysics Research Centre, School of Mathematics and Physics, Queen's University Belfast, BT7 1NN Belfast, UK\\
$^{12}$Departamento de Astronomia, Universidad de Chile, Casilla 36-D, Santiago, Chile\\
$^{13}$ Centro de Astrof\'isica y Tecnolog\'ias Afines (CATA), Casilla 36-D, Santiago, Chile.\\
$^{14}$Instituto de Astronom\'ia, Universidad Cat\'olica del Norte, Angamos 0610, 1270709 Antofagasta, Chile\\
$^{15}$Institute of Geological Sciences, FU Berlin, Malteserstr. 74-100, D-12249 Berlin, Germany\\
$^{16}$ Astrophysics Group, Keele University, Staffordshire ST5 5BG, UK\\
$^{17}$Austrian Academy of Sciences, Space Research Institute, Schmiedlstr. 6, 8042 Graz, Austria\\
$^{\dagger}$ Winton Fellow\\
$^{*}$ j.j.mccormac@warwick.ac.uk
}
\date{Published: 20$^{\rm th}$ Feb 2020 - Accepted: 10$^{\rm th}$ Jan 2020 - Received: 29$^{\rm th}$ Sept 2019}
\pubyear{2020}

\begin{document}
\label{firstpage}
\pagerange{\pageref{firstpage}--\pageref{lastpage}}
\maketitle

\begin{abstract}
We report the discovery of a new ultra-short period transiting hot Jupiter from the Next Generation Transit Survey (\NGTS). \NPlanet{} has a mass and radius of \NMass\,\mjup~and \NRadius\,\rjup~and orbits its host star with a period of \FAperiod\,days, making it the shortest period hot Jupiter yet discovered. The host is a \Nage\,Gyr old K5V star ($T_\mathrm{eff}$=\NTeffSed\,K) of Solar metallicity ([Fe/H] = \NMetalBarry\,dex) showing moderate signs of stellar activity. \NPlanet{} joins a short list of ultra-short period Jupiters that are prime candidates for the study of star-planet tidal interactions. \NPlanet{} orbits its host at just \NARocheAbs{} Roche radii, and we calculate a median remaining inspiral time of \NMedInspiralTimeQVII\,Myr and a potentially measurable orbital period decay of \NPeriodDecayDecade{} seconds over the coming decade, assuming a stellar tidal quality factor \Qs$=2\times10^{7}$.
\end{abstract}

\begin{keywords}
techniques: photometric, stars: individual: \NStar, planetary systems
\end{keywords}

\section{Introduction}
\label{sec:intro}

To date over 4000 transiting exoplanets have been discovered\footnotemark\footnotetext{https://exoplanetarchive.ipac.caltech.edu (2019 Sept 24)}, 389 of which have been detected by ground-based surveys such as WASP \citep{waspproject}, HATNet \citep{hatproject}, HAT-South \citep{hatsproject} and KELT \citep{keltproject1, keltproject2}. The majority (84\%) of the ground-based discoveries are hot Jupiters, planets with masses in the range \mbox{$0.1<M_{\mathrm{p}}<13$ M$_{\mathrm{Jup}}$} and periods $\lesssim 10$ days. Given their relatively large transit depth and geometrically increased transit probability, hot Jupiters are amongst the easiest transiting planets to detect, especially from the ground. Ultra-short period (USP) hot Jupiters, those with periods $<1$ day, are theoretically the easiest to detect but have proven to be extremely rare; only 6 of 389 hot Jupiters detected by ground-based surveys have periods $<1$ day. Such short period hot Jupiters are ideal targets for studying star-planet interactions and atmospheric characterisation through phase curve and secondary eclipse measurements, as well as transmission spectroscopy. This explains why these six planets (namely WASP-18b \citealt{wasp18}; WASP-19b \citealt{wasp19}; WASP-43b \citealt{wasp43}; WASP-103b \citealt{wasp103}; HATS-18b \citealt{hats18} and KELT-16b \citealt{kelt16}) are some of the most studied systems.

WASP-18b was proposed to undergo rapid orbital period decay through tidal interactions with its host star \citep{wasp18}. \citet{2017ApJ...836L..24W} searched for the period decay. Their joint analysis of published transit and secondary eclipse times, along with new data spanning a 9 year baseline, found no evidence of departure from a linear ephemeris, indicating that the tidal quality factor for WASP-18 is \Qs$\geq1\times10^{6}$ at 95\% confidence. \citet{2019arXiv191011930P} report a null detection of orbital period decay for the WASP-19b system. They analysed 62 archival and 12 new transit observations spanning a decade, establishing upper limits on the rate of orbital period decay of $\dot{P}=-2.294$ ms/yr for WASP-19b, and on stellar tidal quality factor \Qs$=(1.23\pm0.23)\times10^{6}$ for WASP-19. WASP-43b has been the subject of several studies that calculated the rate of orbital period decay \citep{2014ApJ...781..116B, 2014A&A...563A..41M, 2014A&A...563A..40C, 2015PASP..127..143R, 2016AJ....151...17J, 2016AJ....151..137H}. \citet{2016AJ....151..137H} analysed all available transit light curves (52 from the literature and 15 new) and ruled out the existence of any decay, placing limits of $\dot{P}=-0.02\pm6.6$ ms/yr and \Qs$>10^{5}$ on the system. \citet{2018AcA....68..371M} found no evidence of orbital period decay for WASP-103b and KELT-16b, placing lower limits on \Qs{} of $>10^{6}$ and $>1.1\times10^{5}$ with $>95\%$ confidence, respectively. WASP-12b \citep{wasp12} is another well studied short period (P=$1.09142$\,d; \citealt{2019AJ....158...39C}) hot Jupiter and is currently the only giant planet demonstrating significant orbital period decay. \citet{2018AcA....68..371M} measured a period shift of approximately 8\,minutes over the past decade and they derived a highly efficient tidal quality factor of \Qs$=(1.82\pm0.32)\times10^{5}$ for the host star. 

Hot Jupiters are also prime targets for atmospheric characterisation. The planet's close proximity to their host stars leads to increased equilibrium temperatures which aid in the detection of phase curves and secondary eclipses, and which may also drive a large atmospheric scale heights, increasing the strength of transmission spectroscopy signals. Several USP hot Jupiters have recently been the target of extensive atmospheric studies, e.g. WASP-18b (\citealt{2019A&A...626A.133H, 2019AJ....157..178S, 2019A&A...625A.136A, 2018ApJ...855L..30A, 2017ApJ...850L..32S, 2016ApJ...821...16K}, etc); WASP-19b (\citealt{2019MNRAS.482.2065E, 2019MNRAS.482.1485P, 2017Natur.549..238S, 2016Natur.529...59S}, etc); WASP-43 (\citealt{2018MNRAS.474..271G, 2018AJ....155..150M, 2018ApJ...869..107M, 2017ApJ...849L...5K}, etc), and WASP-103b (\citealt{2017AJ....153...34C, 2017A&A...606A..18L}) to list but a few.

\NGTS{} \citep{project2018,McCormac2017,2013EPJWC..4713002W,Chazelas2012} has been in routine operation on Paranal since April 2016. To date we have published 8 new transiting exoplanets: a rare hot Jupiter orbiting an M star, NGTS-1b \citep{2018MNRAS.475.4467B}; the sub-Neptune sized planet NGTS-4b \citep{2019MNRAS.486.5094W}; the highly inflated Saturn NGTS-5b \citep{2019A&A...625A.142E}; several other hot Jupiters (NGTS-2b, \citealt{2018MNRAS.481.4960R}; NGTS-3Ab, \citealt{2018MNRAS.478.4720G}; NGTS-8b/NGTS-9b, Costes et al. MNRAS in press), and the discovery of an USP tidally locked brown dwarf orbiting an M star, NGTS-7Ab \citep{2019arXiv190608219J}. We recently published the discovery of another USP hot Jupiter, NGTS-6b \citep[P=$0.88$\,days;][]{2019arXiv190407997V}, bringing the total known USP hot Jupiter population to seven.

Here we present the 10$^{\mathrm{th}}$ discovery (9$^{\mathrm{th}}$ planet) from \NGTS. \NPlanet{} is the shortest period  hot Jupiter yet found ($P=0.766891$ days), and is thus both a good candidate for studying star-planet interactions. With $H=11.9$ it is also a good candidate for atmospheric characterisation with the James Webb Space Telescope (JWST). In \S\ref{sec:obs} we describe the NGTS discovery photometry and the subsequent follow-up photometry/spectroscopy, after which we discuss the analysis of our data and the determination of the host star's parameters in \S\ref{sec:analysis} . Our global modelling process is described in \S\ref{sec:global}, while in \S\ref{sec:tides} we model the tidal evolution of the system. In \S\ref{sec:discussion} we discuss our results, and we close in \S\ref{sec:conclusions} with our conclusions.

\section{Observations}
\label{sec:obs}

\begin{table*}
	\centering
	\caption{A summary of the follow-up photometry of \NPlanet{} transits. The FWHM and the aperture photometry radius (R$_{\mathrm{aper}}$) values are given in units of binned pixels, if binning was applied. Both R$_{\mathrm{aper}}$ and the number of comparison stars (N$_{\mathrm{comp}}$) were chosen to minimise the RMS in the scatter out of transit (RMS$_{\rm OOT}$).}
	\label{tab:followup_phot}
	\begin{tabular}{ccccccccccc} 
		Night & Instrument & N$_{\mathrm{images}}$ & Exptime & Binning & Filter & FWHM & R$_{\mathrm{aper}}$ & N$_{\mathrm{comp}}$ & RMS$_{\rm OOT}$ & Comment \\
        &  &  & (seconds) & (X$\times$Y) & & (pixels) & (pixels) &  & (\%) & \\
		\hline
        2016-11-27 & \SHOCh   & 630  & 30  & 4$\times$4 & z' & 1.22 & 4.0  & 5 & 0.57 & full transit    \\
        2017-09-29 & \SHOCa   & 778  & 8   & 4$\times$4 & V  & 2.04 & 4.5  & 2 & 1.24 & partial transit \\
        2017-10-11 & Eulercam & 89   & 120 & 1$\times$1 & I  & 7.32 & 15.0 & 4 & 0.11 & full transit    \\
        2017-11-27 & \SHOCa   & 242  & 30  & 4$\times$4 & V  & 2.1  & 5.0  & 5 & 0.63 & partial transit \\
        2017-12-16 & Eulercam & 103  & 90  & 1$\times$1 & V  & 8.15 & 17.0 & 7 & 0.20 & full transit    \\
    	2018-01-29 & \SHOCa   & 3278 & 4   & 4$\times$4 & B  & 1.06 & 2.7  & 1 & 0.79 & full transit    \\
		\hline
	\end{tabular}
\end{table*}

\begin{figure}
	\includegraphics[width=\columnwidth]{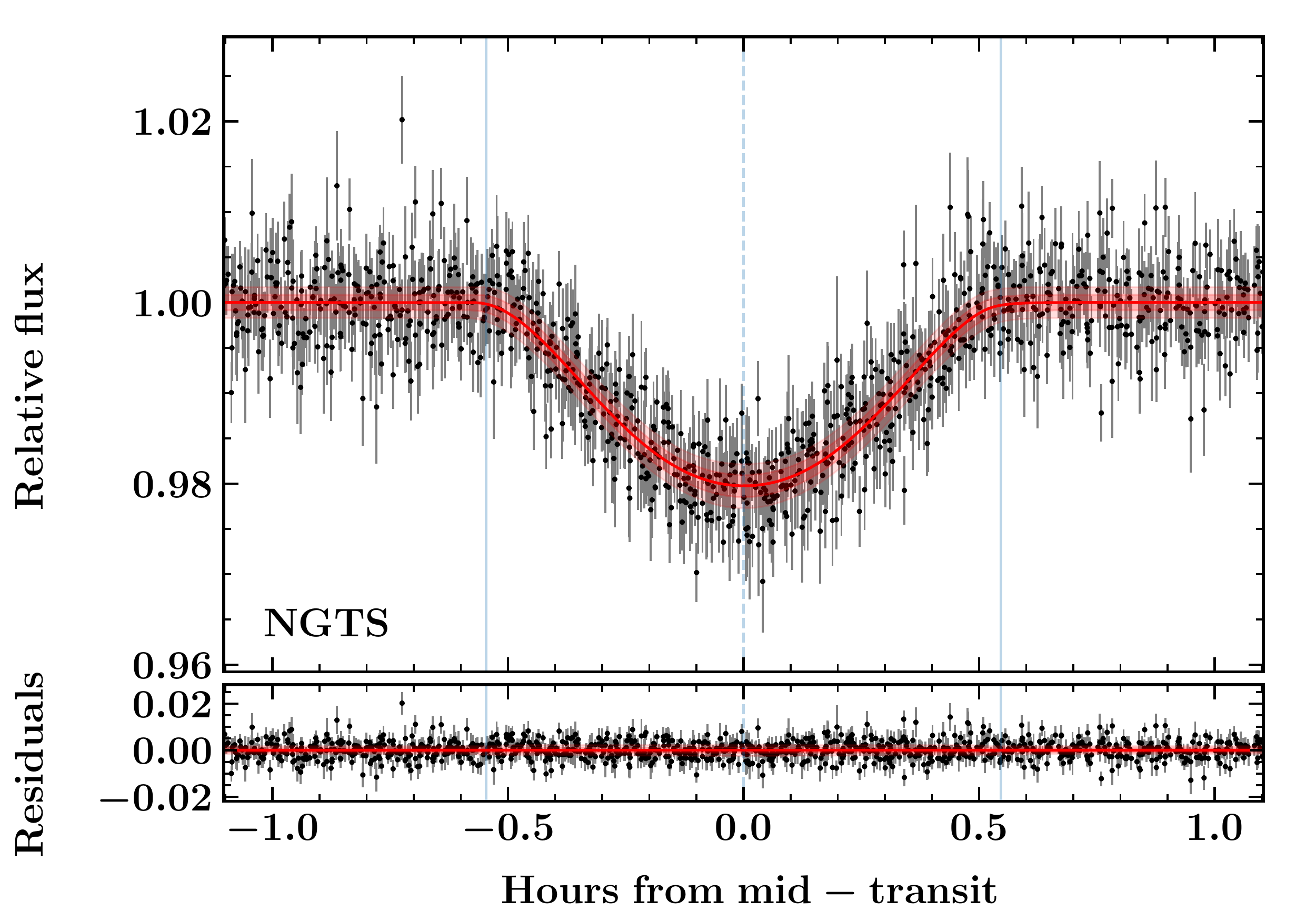}
    \caption{Top: Forty-six phase folded and detrended transits of \NPlanet{} as observed by \NGTS. The data have been binned in time to five minutes for clarity and then phase folded. The best fitting transit model is over plotted in red. Bottom: Residuals after removing the best fitting model from the top panel. The RMS of the scatter out of transit is $0.41\%$.}
    \label{fig:ngts_phot}
\end{figure}

\subsection{\NGTS{} photometry}
\label{sub:ngtsphot}

\NGTS{} consists of an array of twelve 20 cm telescopes and the system is optimised for detecting small planets around K and early M stars. \NStar{} was observed using a single \NGTS{} camera over a 237 night baseline between 2015 September 21 and 2016 May 14. The observations were acquired as part of the commissioning of the facility. Routine science operations began at the ESO Paranal observatory in April 2016. A total of $220\,918$ images were obtained, each with an exposure time of $10$ s. The data were taken using the custom \NGTS{} filter (550 -- 880\,nm) and the telescope was autoguided using an improved version of the DONUTS autoguiding algorithm \citep{donuts}. The root mean square (RMS) of the field tracking errors was $0.057$ pixels over the 237 night baseline. The data were reduced and aperture photometry was extracted using the CASUTools\footnote{\url{http://casu.ast.cam.ac.uk/surveys-projects/software-release}} photometry package. The data were then detrended for nightly trends, such as atmospheric extinction, using our implementation of the SysRem algorithm \citep{Tamuz2005}. We refer the reader to \citet{project2018} for more details on the NGTS facility, the data acquisition and reduction processes. The data were searched for transit-like signals using ORION; our implementation of the box-fitting least squares (BLS) algorithm \citep{Kovacs2002}. A strong signal was found at a period of \FAPeriodShort\,d. The \NGTS{} data have been phased on this period in Figure~\ref{fig:ngts_phot}. The RMS in the scatter out of transit in the \NGTS{} data is $0.41\%$, which masks any possible detection of a secondary eclipse ($\sim0.1\%$). We conducted an additional BLS search on the \NGTS{} data after masking the transits of \NPlanet{}. No other significant detections were found. The \NGTS{} data set along with all photometry and Radial Velocities (RVs) presented below are available in a machine-readable format from the online journal. 

To help eliminate the possibility of the transit signal originating from another object we conduct several checks for all \NGTS{} candidates, including multi-colour follow up photometry of the transits. This enables us to measure transit depth variations which may be indicative of a false positive detection (e.g. blended eclipsing binary). We search all archival catalogues surrounding the candidate for additional stellar contamination, which may lead to transit depth dilution. We also use the centroid vetting procedure of \citet{Guenther17b} to look for contamination from additional unresolved objects. This technique is able to detect sub-milli-pixel shifts in the photometric centre-of-flux during transit and can identify blended eclipsing binaries at separations $<1$\arcsec, well below the size of individual \NGTS\ pixels ($5$\arcsec). We find no centroid variation during the transits of \NPlanet, indicating that transit signal originates from \NStar.

For \NStar{} we found one spurious detection of a neighbour in the Guide Star Catalogue v2.3 \citep{gsc23_2008} and one real blended neighbour in \gaia{} DR2 \citep{gaiadr2}. We discuss these objects further in \S \ref{sub:third_light} and outline the treatment of the photometric contamination in \S \ref{subsub:third_light_double_ccf} \& \S\ref{sec:global}. We note that the background contaminating star falls within the photometric aperture in both the discovery photometry (this section) and the all the follow-up photometry presented in this paper (\S \ref{sub:eulerphot}-\ref{sub:saaophot}).

\subsection{Eulercam photometry}
\label{sub:eulerphot}
Two follow-up light curves of the transit of \NPlanet{} were obtained on 2017 October 11 and 2017 December 16 with \Euler~on the 1.2\,m Euler Telescope \citep{Lendl2012} at \LSO. In October a total of $89$ images with $120$ s exposure time were obtained using the Cousins I-band filter. In December a total of $103$ images with $90$ s exposure time were obtained in the Gunn V-band. Both observations were made in focus. The data were reduced using the standard procedure of bias subtraction and flat field correction. Aperture photometry was performed with the \emph{phot} routine from IRAF. The comparison stars and the photometry aperture radius were chosen to minimise the RMS in the scatter out of transit. A summary of the \Euler~observations is given in Table {\ref{tab:followup_phot}}. The \Euler~light curve and best fitting transiting exoplanet model from \S\ref{sec:global} are shown in Fig. \ref{fig:followup_phot_all_detrended}. The undetrended follow up Eulercam data is presented in Figure \ref{fig:followup_phot_all_undetrended}.

\subsection{SHOC photometry}
\label{sub:saaophot}

Four additional transit light curves of \NPlanet{} were obtained using 2 of the 3 Sutherland High-speed Optical Cameras \citep[SHOC]{Coppejans:2013gx} - SHOC'n'awe (hereafter \SHOCa) and SHOC'n'horror (hereafter \SHOCh). All observations were obtained with the cameras mounted on the 1\,m telescope at SAAO. Observations were taken in focus with $4 \times 4$ binning. 

\begin{figure*}
	\includegraphics[width=5.8cm]{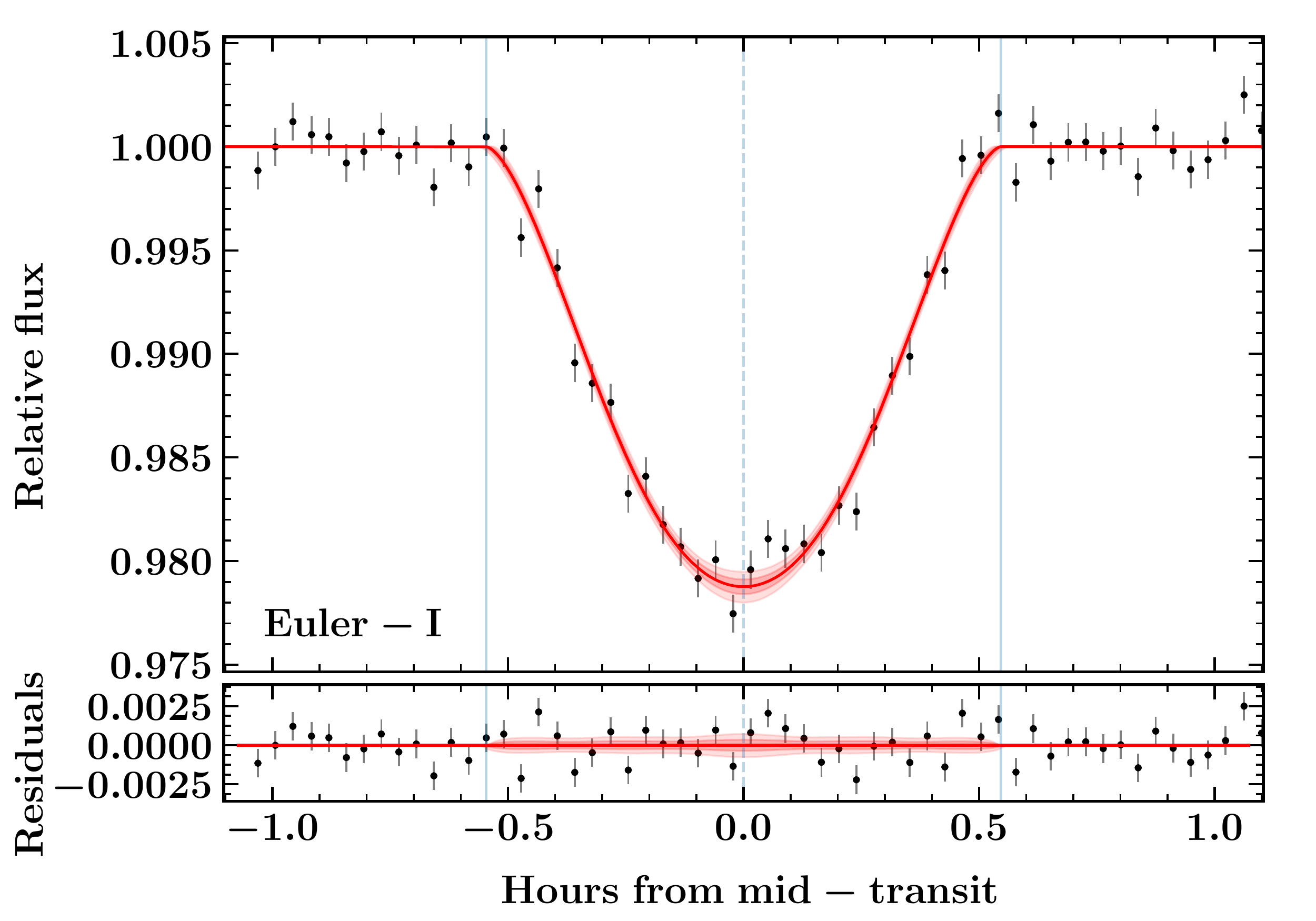}
	\includegraphics[width=5.8cm]{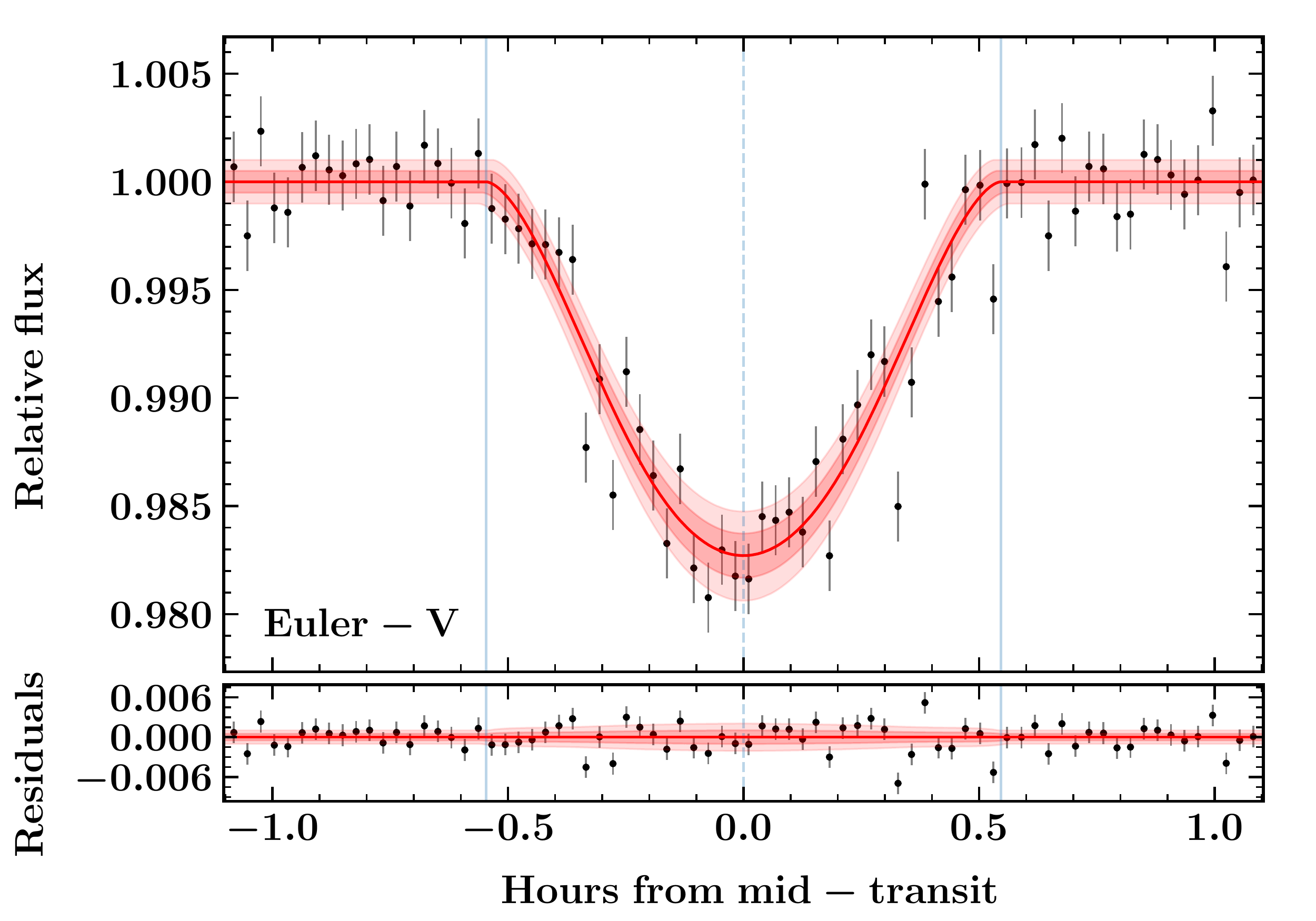}
	\includegraphics[width=5.8cm]{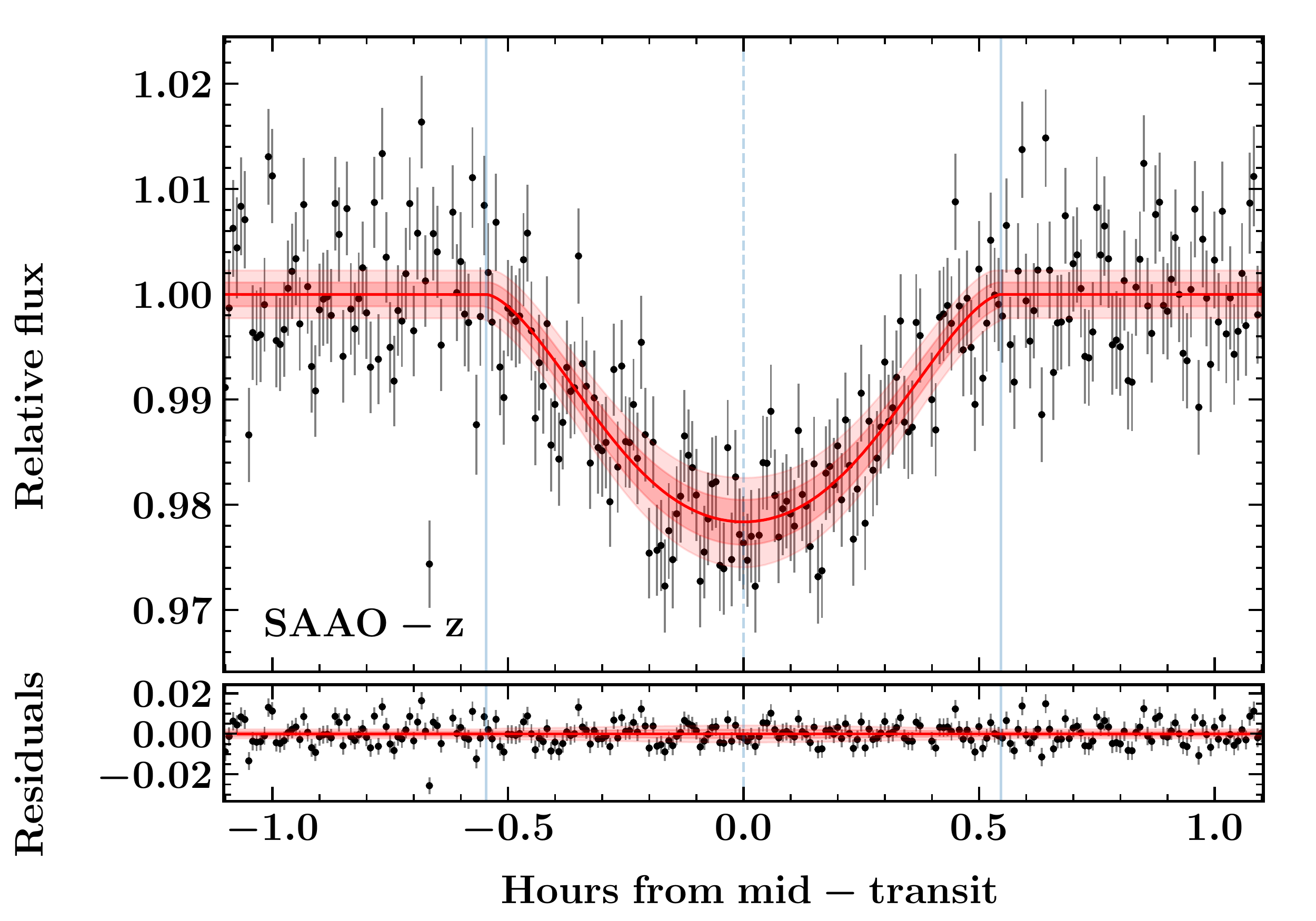}
	\includegraphics[width=5.8cm]{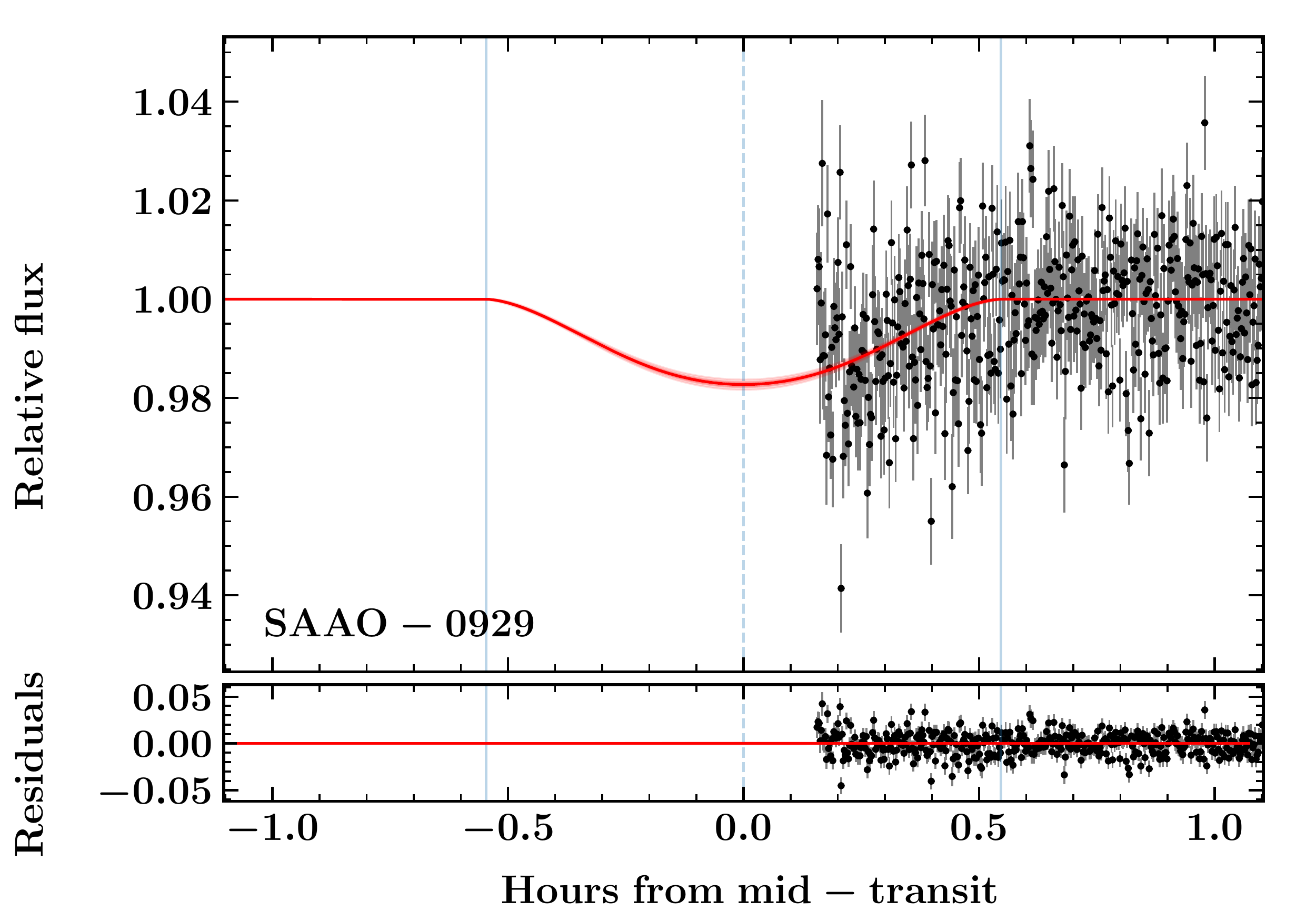}
    \includegraphics[width=5.8cm]{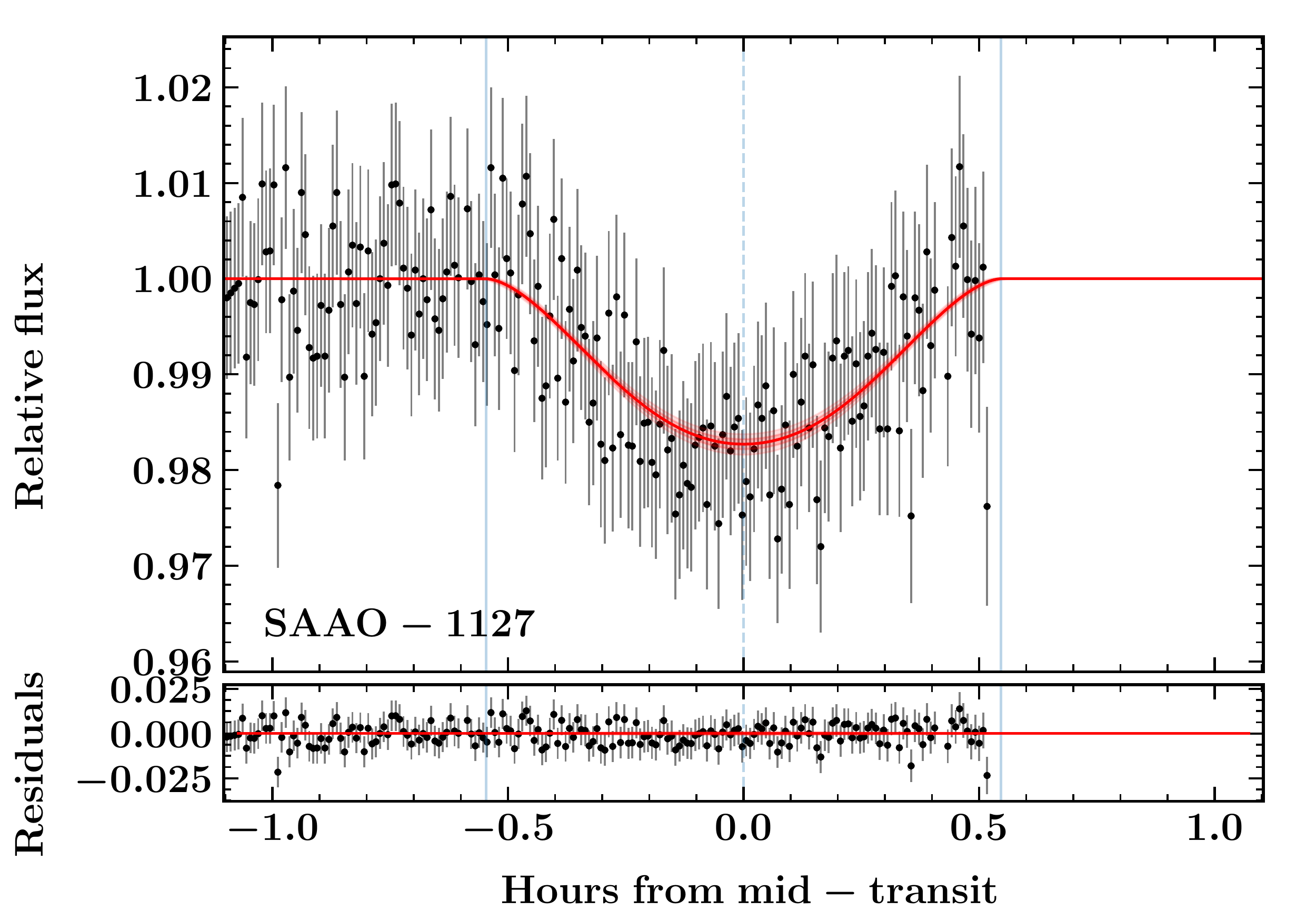}
    \includegraphics[width=5.8cm]{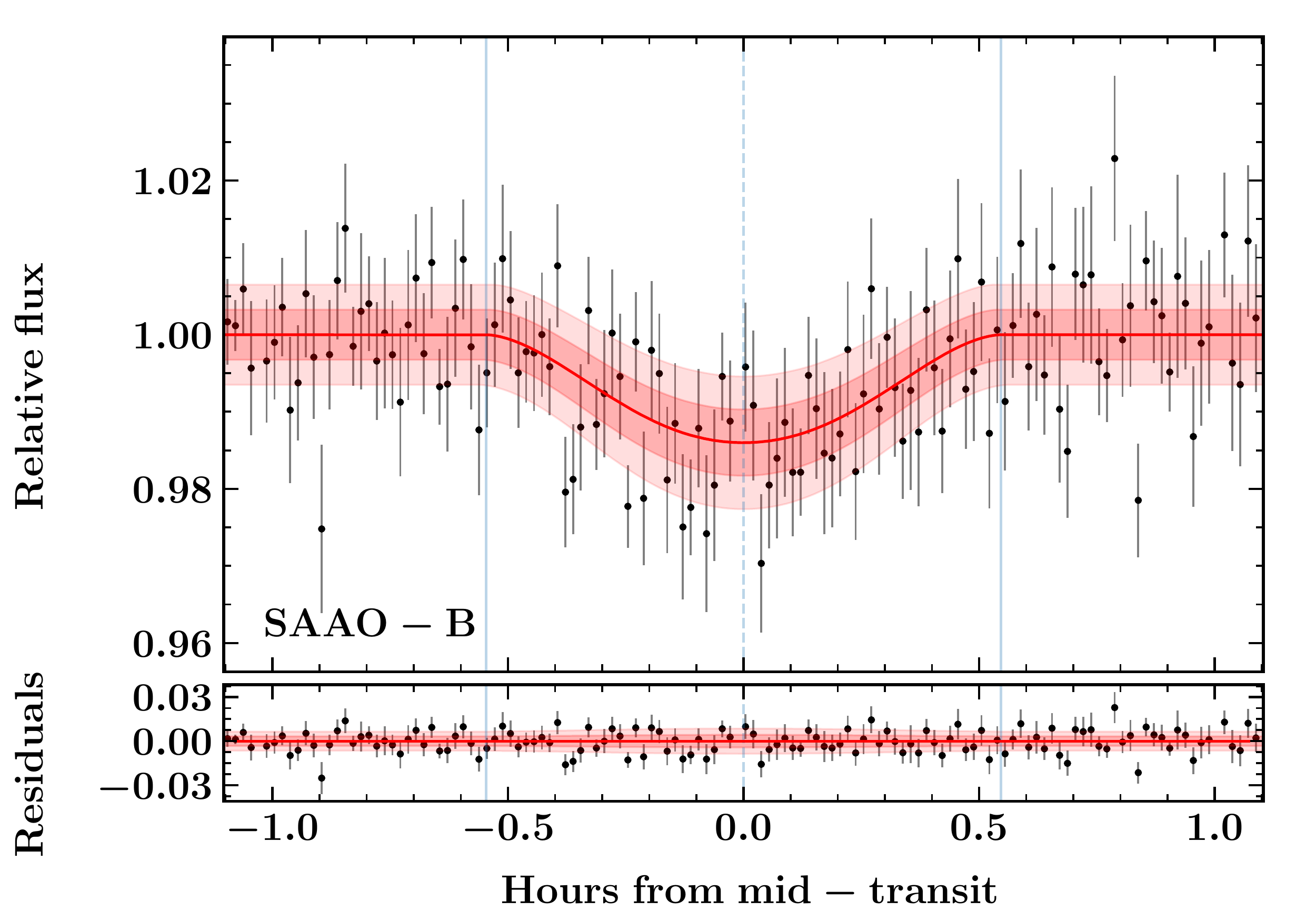}
    \caption{Top: From left to right we plot the detrended follow up light curves from Eulercam in the I and V bands on 2017 Oct 11 and 2017 Dec 16, respectively, followed by a z-band transit in from \SHOCh{} on 2016 Nov 27. Bottom: From left to right we plot the detrended follow up light curves from \SHOCa{} in the V-band from nights 2017 Sept 29 and 2017 Nov 27, followed by a B-band transit on the night of 2018 Jan 29. Each light curve is over plotted with the best fitting model from \S\,\ref{sec:global}, where the red line and pink shaded regions represent the median and 1 \& 2\,$\sigma$ confidence intervals of the \gpe\ posterior transit model. The lower panel below each plot shows the residuals to the fit. The vertical blue lines highlight the start, middle and end of each transit. The undetrended follow up data is presented in Figure \ref{fig:followup_phot_all_undetrended}}
    \label{fig:followup_phot_all_detrended}
\end{figure*}

The data were bias and flat field corrected via the standard procedure using the CCDPROC package \citep{2015ascl.soft10007C} in Python. Aperture photometry was extracted using the SEP package \citep{Barbary16, 1996A&AS..117..393B} and the sky background was measured and subtracted using the SEP background map. The number of comparison stars, aperture radius and sky background interpolation parameters were chosen to minimise the RMS in the scatter out of transit. A summary of the four follow-up light curves, two of which were complete and two of which were partial transits, is given in Table \ref{tab:followup_phot}. The \SHOCa~and \SHOCh~light curves are shown in Fig. \ref{fig:followup_phot_all_detrended}. The undetrended follow up SHOC data is presented in Figure \ref{fig:followup_phot_all_undetrended}.

\subsection{\emph{TESS} photometry}
\label{sub:tessphot}

We inspected the \tess{} full frame images (FFIs) in the area surrounding \NStar{} and find no evidence of the target nor the bright neighbour HD 42043. The TESScut\footnotemark\footnotetext{https://mast.stsci.edu/tesscut/} tool returns blank FFIs for this region of sky. We checked with the \tess{} team, who confirmed that \NStar{} fell in the overscan region of that particular camera. We therefore ignore \tess{} in the remainder of this paper.

\subsection{Spectroscopy}
\label{sub:spect}

We obtained multi-epoch spectroscopy for \NStar\ with the HARPS spectrograph \citep{2003Msngr.114...20M} on the ESO 3.6\,m telescope at \LSO, Chile, between 2016 November 3 and 2017 December 22. \NStar{} was observed under programme IDs 098.C-0820(A) and 0100.C-0474(A) using the HARPS target ID of NG0612-2518-44284. Due to the relatively faint optical magnitude of \NStar{} (V=\NVmag), we used the HARPS in the high efficiency (EGGS) mode. EGGS mode employs a fibre with a larger, 1.4\arcsec~aperture, compared to standard 1.0\arcsec~fibre. This allows for higher signal-to-noise spectra at slightly lower resolution of R=85000, compared to R=110000 in standard (HAM) mode. 

We used the standard HARPS data reduction software (DRS) to the measure the radial velocity of \NStar{} at each epoch. This was done via cross-correlation with the \RvMask{} binary mask. The exposure times for each spectrum ranged between 1800 and 3600\,s. The radial velocities are listed, along with their associated error, FWHM, bisector span and exposure time in Table\,\ref{tab:rvs}.

\begin{figure}
	\includegraphics[width=\columnwidth]{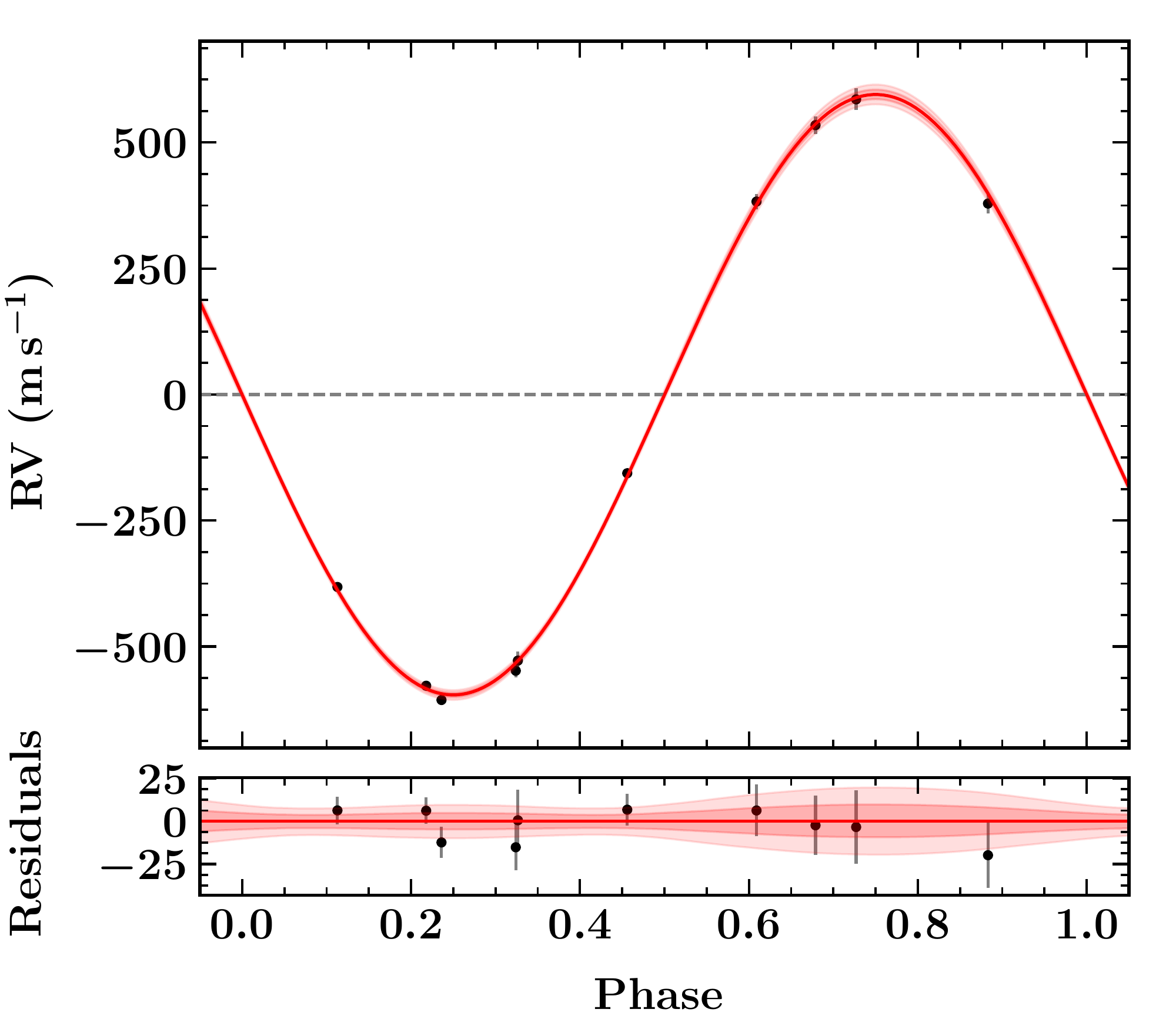}
    \caption{Top: RV measurements of \NStar{} over plotted with the best fitting model from \S\ref{sec:global}. The systemic velocity \mbox{$V_{\mathrm{sys}}=$\NVsys\,\kms} has been subtracted from the RVs. Bottom: Residuals after the removal of the model in the upper panel. The RMS of the residuals is $9.79$\,\ms\ and is a combination of both jitter from stellar activity and instrumental noise.}
    \label{fig:harps_rvs}
\end{figure}

The radial velocities show a variation in-phase with the photometric period detected by ORION with semi-amplitude of \mbox{$K=$\NKAmp\,\ms}. Figure~\ref{fig:harps_rvs} shows the phase folded radial velocities over plotted with the best fitting exoplanet model from \S\ref{sec:global}. 

\begin{figure}
	\includegraphics[width=\columnwidth]{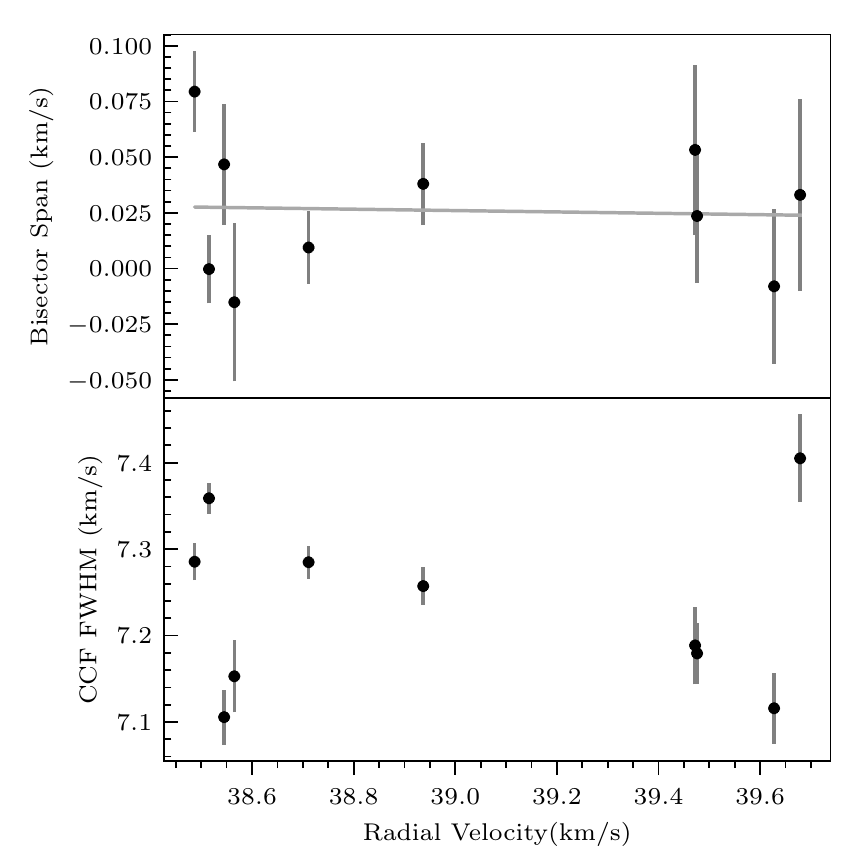}
    \caption{CCF bisector spans (top panel) and CCF full widths at half maximum (bottom panel) measured from HARPS spectra plotted against the radial velocity of \NStar{}. No correlations are found between either pair of measurements.}
    \label{fig:bisectors}
\end{figure}

To ensure that the radial velocity signal is not caused by stellar activity we analyse the HARPS cross correlation functions (CCFs) using the line bisector technique of \citet{queloz2001}. We find no evidence for a correlation between the radial velocity and the bisector spans. Fitting a line to the bisector spans in Figure \ref{fig:bisectors}, we find a gradient of \RvBisSlope, indicating that the radial velocity signal is coming from orbital motion of \NStar{} around the system barycentre rather than from stellar activity. The error on the gradient in Figure \ref{fig:bisectors} is estimated via a bootstrapping technique. We resample the bisector spans in Table \ref{tab:rvs}, with replacement, a total of $1000$ times. We fit a straight line to each resampled set and estimate the error on the slope as the standard deviation of the slopes from the $1000$ samples. 

\begin{table}
	\centering
	\caption{HARPS Radial Velocities for \NStar{}}
	\label{tab:rvs}
	\begin{tabular}{cccccc} 
	BJD	&	RV	& RV$_{\rm{err}}$ &	FWHM & BIS    & T$_{\mathrm{exp}}$ \\
	-2450000	& \kms & \kms &\kms & \kms & s \\
	\hline
    7696.84979 & 38.7113 & 0.0082 & 7.458 & 0.011 & 3600\\
    7753.69446 & 38.4870 & 0.0091 & 7.446 & 0.078 & 3600\\
    7756.82948 & 38.5451 & 0.0136 & 7.347 & 0.056 & 3600\\
    7877.50407 & 39.6274 & 0.0174 & 7.338 & -0.019 & 3600\\
    8052.85264 & 38.5652 & 0.0178 & 7.336 & -0.004 & 2400\\
    8053.83621 & 39.4759 & 0.0151 & 7.372 & 0.026 & 2400\\
    8054.81329 & 39.4719 & 0.0192 & 7.375 & 0.061 & 2400\\
    8055.83695 & 38.5153 & 0.0077 & 7.513 & 0.005 & 2700\\
    8056.78642 & 38.9369 & 0.0093 & 7.428 & 0.038 & 2700\\
    8110.67686 & 39.6787 & 0.0216 & 7.578 & 0.043 & 1800\\
	\hline
	\end{tabular}
\end{table}

\section{Analysis}
\label{sec:analysis}

We begin this section by addressing the photometric contamination caused by a star nearby to \NStar{} (third-light). We describe the treatment of the third-light in our spectroscopy in \S \ref{subsub:third_light_double_ccf}, photometry in \S \ref{subsub:third_light_photometry} and then continue with the derivation of  stellar properties in \S \ref{sub:stellar}-\ref{sub:age}. Finally, in \S \ref{sub:gaia_scans} we hypothesise on the source of excess astrometric noise in the \gaia{} DR2 \citep{gaiadr2} measurements of \NStar.

\begin{figure}
	\includegraphics[width=\columnwidth]{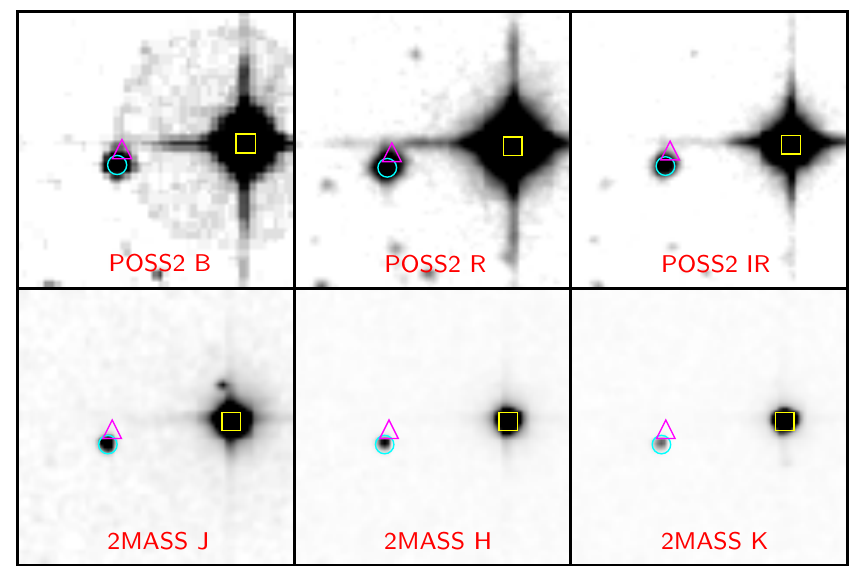}
    \caption{Archival images from the Digital Sky Survey of the area surrounding \NStar. HD42043 is highlighted by the yellow square, \NStar{} is circled in cyan and the spurious companion \mbox{GSC23-S3GL019224} is marked with a magenta triangle. As can be seen, the latter appears to come from the intersection of a diffraction spike from HD42043 and \NStar. \NStarB{} is not highlighted as the source is fully encapsulated within the profile of \NStar. Each thumbnail is 0.75\arcmin~square. North is up and East is left.} 
    \label{fig:false_companion}
\end{figure}

\subsection{Treatment of third-light}
\label{sub:third_light}

An additional object (\mbox{GSC23-S3GL019224}, J=17.63) is reported in the Guide Star Catalogue v2.3 (GSC2.3) with a separation of 5.13\arcsec~from \NStar. It is flagged as class 3 (non-stellar). As this object would be enclosed inside our photometry aperture we consulted archival imaging of this area. Closer inspection of the photographic plates, overlaid with GSC2.3, reveals \mbox{GSC23-S3GL019224} to be a spurious detection caused by the intersection of \NStar{} and a diffraction spike of HD42043 (located \Nhdsep\arcsec~from \NStar{}, V=\Nhdmagv~ \citealt{tycho2}, G=\Nhdmagg~\citealt{gaiadr1}), see Fig.\ \ref{fig:false_companion}. \mbox{GSC23-S3GL01922} is not reported in 2MASS and more recently \gaia{} DR2 \citep{gaiadr2} does not report the existence of this source. Hence we ignore \mbox{GSC23-S3GL019224} in our treatment of third-light below. 

However, \gaia{} DR2 does report a G=\CompanionGaiaMag~mag object (object ID: \CompanionGaiaID; hereafter \NStarB{}) located \CompanionSeperation$\arcsec$ from \NStar{} with a position angle of \CompanionPositionAngle$^{\circ}$. The \gaia{} DR2 parallax of the companion \mbox{($0.297\pm0.081$ mas)} places it at a much greater distance than \NStar{} \mbox{($3.080\pm0.261$ mas)}. We also note a significant level of astrometric noise for \NStar{} in \gaia{} DR2, which we discuss in \S\,\ref{sec:discussion}. \NStarB{} was also seen on the HARPS guide camera and we see evidence for it in the cross correlation functions from our HARPS radial velocities. Treatment of the RV CCF and photometric contamination is discussed in the following subsections.

\subsubsection{CCF analysis}
\label{subsub:third_light_double_ccf}

\begin{figure}
	\includegraphics[width=\columnwidth]{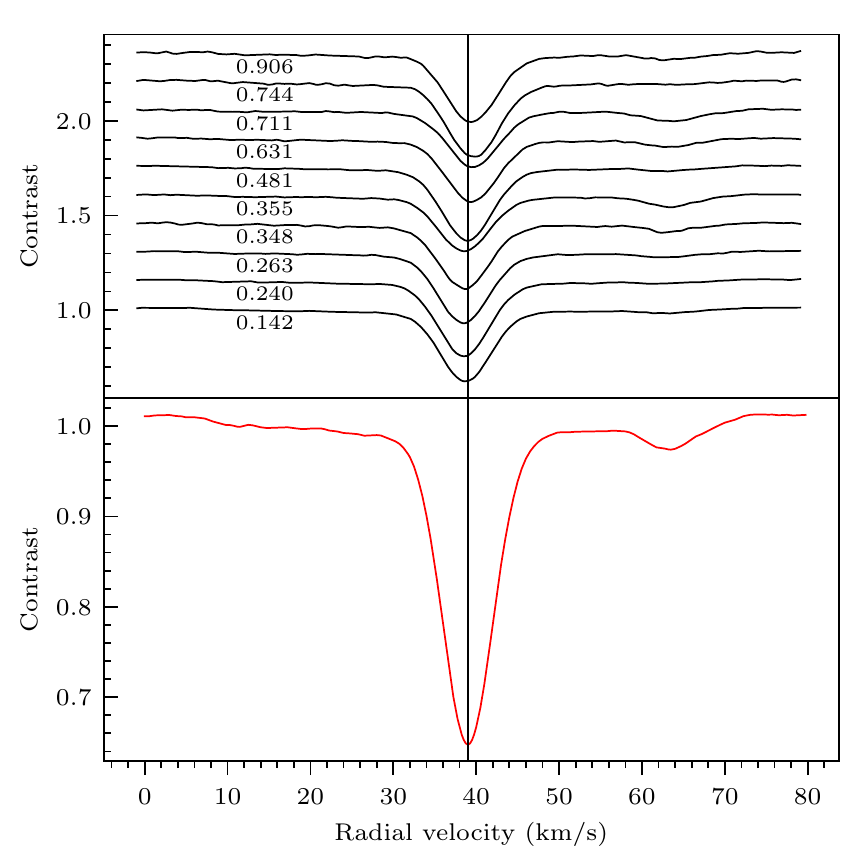}
    \caption{Top: Ten HARPS cross correlation functions obtained using the K5 mask. A shallow peak can be seen near $62$\,\kms~in some of the CCFs. The orbital phase of each CCF is listed below the trace. Each CCF has been offset vertically by $0.15$ for clarity. Bottom: The combined CCFs from the top panel. The vertical line in each plot shows the average radial velocity of \NStar}
    \label{fig:harps_ccfs}
\end{figure}

A second shallow peak is occasionally visible in the CCF of our HARPS spectra (see Fig. \ref{fig:harps_ccfs}). This comes from third light entering the fibre from the nearby object \NStarB. The strength of the peak correlates with the instantaneous seeing at La Silla. To demonstrate that the companion is not the host of the eclipsing body we fit a double Voigt profile to the two peaks in the CCF and measure the radial velocity of each component. 

\begin{figure}
	\includegraphics[width=\columnwidth]{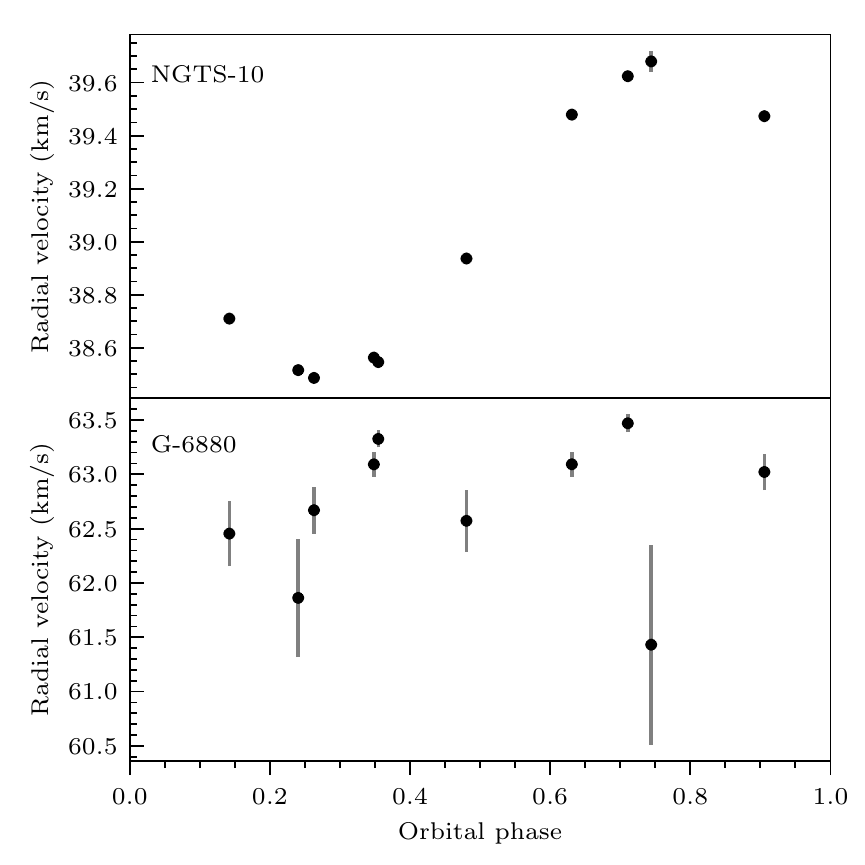}
    \caption{Radial velocities of both peaks in the \NStar~CCFs. The top panel shows the radial velocity signal of the main peak from \NStar~phased on the orbital period detected by ORION. The bottom panel shows the radial velocity signal from \NStarB{} phased on the same period.}
    \label{fig:double_ccf_fit}
\end{figure}

Figure \ref{fig:double_ccf_fit} shows the radial velocities of both peaks phased to the orbital period from the \NGTS~photometry. It is clear that the main CCF peak (coming from \NStar) is moving in phase with the photometry as expected and the second shallow peak is noisy and incoherent with the \NGTS~photometry. Given the above and the lack of a centroid shift measured during transit, we are certain that the photometric and radial velocity signals are not caused by \NStarB.

\subsubsection{Dilution of transit depth by third-light}
\label{subsub:third_light_photometry}

\begin{figure}
	\includegraphics[width=\columnwidth,angle=0]{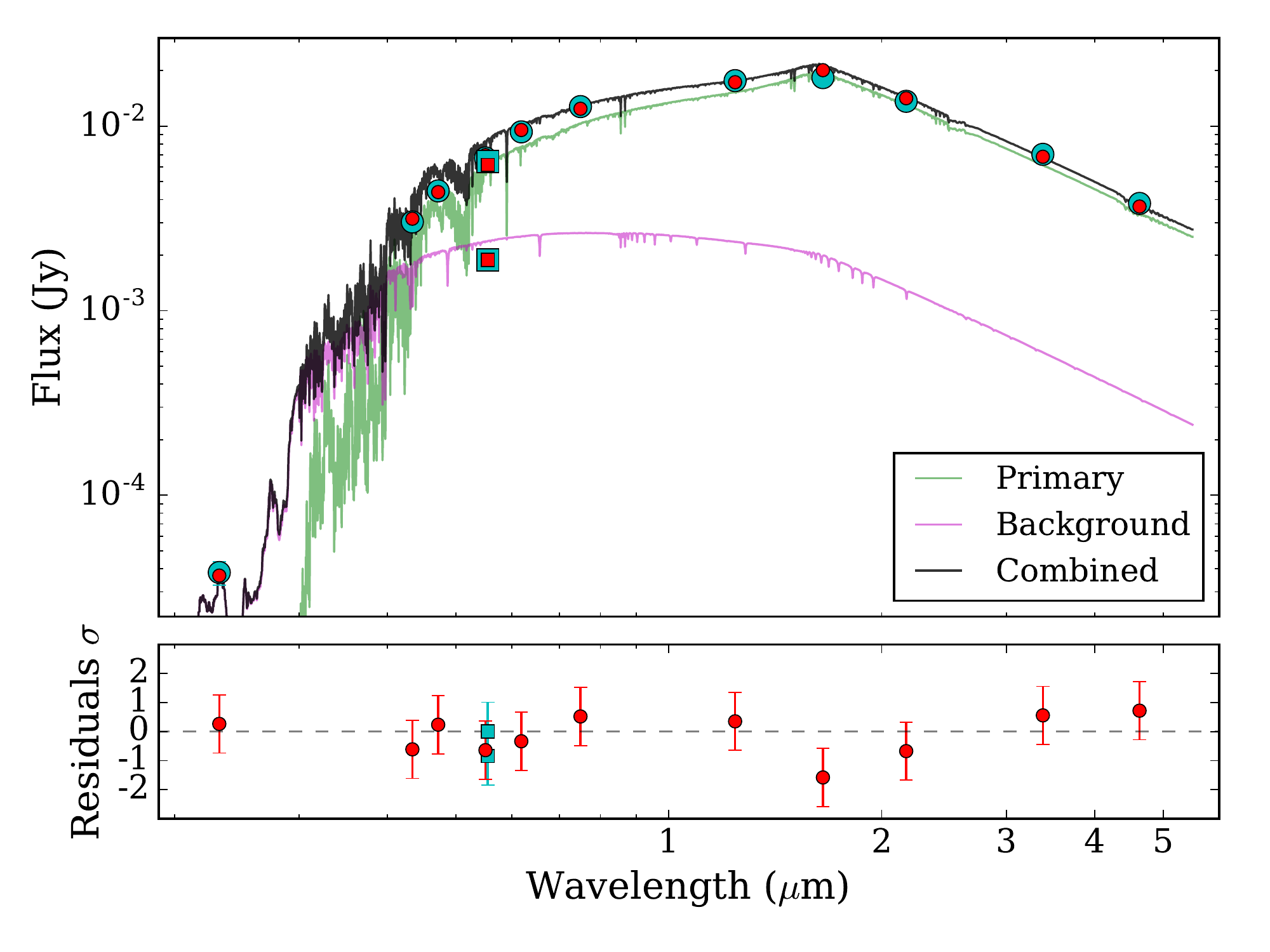}
	\vspace{-0.65cm}
    \caption{Top: Double SED fit to the available catalogue photometry of \NStar{} and \NStarB. Note, that the two stars are only resolved in the \gaia{} G band. The SED of \NStar{} is plotted in green, \NStarB{} is plotted in magenta and the combined SEDs of both stars is plotted in black. Cyan markers represent the observed catalogue photometry, while the red markers are our combined synthetic photometry. The \gaia{} points are plotted with square markers, all others are circles. We have inflated the cyan marker size as they were often hidden behind the red markers. 
    Bottom: The residuals to the double SED fit versus wavelength. The red circles represent the difference between our catalogue and synthetic photometry. The \gaia{} photometric points are given as cyan squares.}
    \label{fig:double_sed}
\end{figure}

In order to estimate the photometric dilution caused by \NStarB{} we simultaneously fit the Spectral Energy Distribution (SED) of both stars. The method is based on the analysis of NGTS-7Ab by \citet{2019arXiv190608219J}. For completeness we outline the process here. We used the PHOENIX v2 set of stellar models \citep[][]{Husser13} for both stars. We initially convolved these models with the bandpasses given in Table\,\ref{tab:stellar} in order to generate a grid of fluxes in \teff\ and \logg\ space, assuming a Solar metallicity. This grid was then used for fitting. As the two stars are blended in all catalogue photometry except \gaia{} G (which is obtained through fitting the line spread function (LSF)), we used the combined synthetic flux from the two stars for comparison with the observed values in all other bands.

\gaia{} DR2 reports an astrometric noise excess of \mbox{\NAstrometricNoiseExcess\,\mas} for \NStar. The \gaia{} DR2 release notes warn that the parallax measurements are compromised when the astrometric noise excess exceeds $2$ mas, hence we chose to ignore the \gaia{} parallax and instead fit for \teff, the extinction \av, and the scale factor $S$ ($S = R^{2}/D^{2}$, where $R$ and $D$ are the radius and distance of the star, respectively) for each star. 

When fitting for \av{} we used the extinction law of \citet{Fitzpatrick99} with the improvements of \citet{Indebetouw05}. We also fit for \logg{} of \NStarB{} but fixed the \logg{} of \NStar{} at the value determined from our stellar spectra described in \S\,\ref{sub:stellar}. Finally we fit an uncertainty inflation term $\sigma_{f}$, which is used to inflate the uncertainties on observed fluxes and account for potentially underestimated errors. 

When fitting we required the synthetic \gaia{} G band flux for each source to match the observed values. This was achieved via a Gaussian prior for each source. We required the extinction of \NStarB{} to be greater than that of \NStar, to match our expectations from their relative distances. In order to fully explore the posterior parameter space we used EMCEE \citep[][]{foremanmackey13} to generate an MCMC process using 100 walkers for 10,000 steps, conservatively taking the final 1000 steps to sample the distribution. As the parameters derived from the SED fit are model dependent we inflate the formal error bars by a factor of 2 to account for potential differences in stellar models. 

To calculate the dilution of the background source in each of the filters given in Table \ref{tab:followup_phot} (and in the NGTS filter) we used the posterior distribution directly from our SED fitting. Each SED model was used to generate synthetic fluxes in each filter, which were then used to calculate the dilution. The resulting dilution values and stellar parameters are given in Table \ref{tab:stellar} and the SEDs are plotted in Fig \ref{fig:double_sed}. We note that the dilution factors calculated here are lower limits based on the completeness of \gaia{} DR2.

\subsection{Stellar Properties}
\label{sub:stellar}

 The HARPS spectra were ordered by increasing seeing. We combined the 6 with the sharpest seeing and the least evidence of contamination by \NStarB{} into a higher SNR spectrum. Using methods similar to those described by \citet{2013MNRAS.428.3164D} we determined values for the stellar effective temperature $T_\mathrm{eff}$, surface gravity $\log g$, stellar metallicity [Fe/H], and the projected stellar rotational velocity \vsini.
 In determining \vsini{} we assumed a zero macroturbulent velocity, as it is below that of thermal broadening \citep{2008oasp.book.....G}. Hence \mbox{\vsini{}=\NVsini\,\kms} is an upper limit. 
 Lithium is not seen in the spectra. We find \teff=\NTeffBarry{} K and \logg=\NLoggBarry{} which are consistent with our double SED fit from \S \ref{subsub:third_light_photometry}. We note that the \teff{} obtained from the combined spectroscopy has a relatively large uncertainty due to the low SNR of the combined spectrum (SNR$\sim30$) and the weak Balmer line. If we adopt the 
 effective temperature for \NStar{} from the \S\,\ref{subsub:third_light_photometry}, the \logg{} value decreases to $4.3\pm0.2$, but is still consistent with a main sequence star. 
 
 In order to obtain a consistent set of stellar parameters for the third light calculation and the global analysis in \S\,\ref{sec:global} we adopt the effective temperature of \NStar{} from the double SED fit in \S\,\ref{subsub:third_light_photometry}. Given the excess astrometric noise for \NStar{} we also ignore the \gaia{} distance. Without the distance we are forced to assume that the star is on the main sequence. If instead the \gaia{} parallax and the distance (\NDist\,pc) were correct, our fitted scaling factor S from \S\,\ref{subsub:third_light_photometry} would imply a stellar radius of \mbox{$0.77\pm0.07$ \rsun}, which is still consistent with a main sequence star within the uncertainties. Additionally, as seen below in \S\,\ref{sub:kinematics} the host star's kinematics show that is consistent with a thin disc object regardless of whether we trust the \gaia{} DR2 parallax or not, which strengthens our assumption of a main sequence host. Finally, we obtain main sequence mass and radius for a \NTeffSed{} K star using the relations from \citet{Boyajian12} \& \citet{Boyajian17}. We list the stellar parameters in Table \ref{tab:stellar}.  
 
\begin{table}
	\centering
	\scriptsize
	\caption{Stellar Properties of \NStar{} and \NStarB.  The dilution factors in each bandpass are calculated as \mbox{$f_{\mathrm{B}} / (f_{\mathrm{A}} + f_{\mathrm{B}})$}, where star A is \NStar{} and star B is \NStarB. The stellar mass, radius and density assumed for the host are obtained from BJ1217 for a main sequence host star with $T_{\rm eff}=$\NTeffSed\,K. Parameters listed as Double SED and HARPS Spectra as described in \S \ref{subsub:third_light_photometry} \& \ref{sub:stellar}, respectively. $^{\dagger}$ Note that all catalogue photometry of \NStar{} is blended with \NStarB, except for \gaia{} G. $^{\ddagger}$ \vmacro=0\,\kms. }
	\begin{tabular}{lcc} 
	Property	&	Value		&Source\\
	\hline
    \multicolumn{3}{l}{{\bf \NStar{} Astrometric \& photometric properties:}}\\
    I.D.	& 06072933-2535417 & 2MASS \\
    I.D.	& \NGaiaId         & \gaia{} DR2 \\
    R.A.	& \NRa	           & \gaia{} DR2	\\
	Dec		& \NDec	           & \gaia{} DR2	\\
    $\mu_{{\rm R.A.}}$ (\masy) & \NPropRa & \gaia{} DR2 \\
	$\mu_{{\rm Dec.}}$ (\masy) & \NPropDec & \gaia{} DR2 \\
	V (mag)		& \NVmag$^{\dagger}$ 	& APASS\\
	B (mag)		& \NBmag$^{\dagger}$	& APASS\\
	g (mag)		& \Ngmag$^{\dagger}$	& APASS\\
	r (mag)		& \Nrmag$^{\dagger}$	& APASS\\
	i (mag)		& \Nimag$^{\dagger}$	& APASS\\
    G (mag)		& \NGaiaMag             & \gaia{} DR2\\
    NGTS (mag)	& \NNmag$^{\dagger}$	& NGTS photometry\\
    J (mag)		& \NJmag$^{\dagger}$	& 2MASS	\\
   	H (mag)		& \NHmag$^{\dagger}$	& 2MASS	\\
	K (mag)		& \NKmag$^{\dagger}$	& 2MASS	\\
    W1 (mag)	& \NWmag$^{\dagger}$	& WISE	\\
    W2 (mag)	& \NWWmag$^{\dagger}$	& WISE	\\
    \\
    \multicolumn{3}{l}{{\bf \NStarB{} Astrometric \& photometric properties:}}\\
    I.D.                       & \CompanionGaiaID & \gaia{} DR2 \\
    R.A.		               & \NRaB		& \gaia{} DR2	\\
	Dec			               & \NDecB		& \gaia{} DR2	\\
    $\mu_{{\rm R.A.}}$ (\masy) & \NPropRaB  & \gaia{} DR2 \\
	$\mu_{{\rm Dec.}}$ (\masy) & \NPropDecB & \gaia{} DR2 \\
    G (mag)		               & \NGaiaMagB	& \gaia{} DR2\\
    \\
    \multicolumn{3}{l}{{\bf Dilution parameters:}} \\
    $\delta$NGTS     & \NDNgts        & Double SED \\
    $\delta$i        & \NDi           & Double SED \\
    $\delta$V        & \NDV           & Double SED \\
    $\delta$B        & \NDB           & Double SED \\
    $\delta$z        & \NDz           & Double SED \\
    $\sigma_{\rm f}$ & \NErrorInflSed & Double SED \\
    \\
    \multicolumn{3}{l}{{\bf \NStar{} Derived properties:}}\\
    T$_{\rm eff}$ (K)     & \NTeffSed        & Double SED \\
    S                     & \NScaleSed       & Double SED \\
    Av (mag)              & \NAvSed          & Double SED \\
    Spectral Type         & K5V              & Double SED \\
    Age (Gyr)             & \Nage            & Double SED  \\
    T$_{\rm eff}$ (K)     & \NTeffBarry      & HARPS Spectra \\
    $\left[M/H\right]$    & \NMetalBarry     & HARPS Spectra \\
    $\log({\rm g})$       & \NLoggBarry      & HARPS Spectra \\
    ${\rm v}\sin{\rm i}_\star$ (\kms)$^{\ddagger}$ & $\leq$\NVsini & HARPS Spectra \\
    $\gamma_{RV}$ (\kms)  & \NGamma          & HARPS Spectra \\
    log R'$_{\mathrm{HK}}$ & \NActivityIndexBarry & HARPS Spectra \\
    \mstar (\msun)        & \NStarMassBoy    & B1217 \\
    \rstar (\rsun)        & \NStarRadiusBoy  & B1217 \\
    log g                 & \NLoggBoy        & B1217 \\
    P$_{\mathrm{rot}}$ (days)    & \NProt    & NGTS photometry \\
    \\
    \multicolumn{3}{l}{{\bf \NStarB{} Derived properties:}}\\
    T$_{\rm eff}$ (K)   & \NTeffBSed    & Double SED \\
    S                   & \NScaleBSed   & Double SED \\
    log g               & \NLoggBSed    & Double SED \\
    Av (mag)            & \NAvBSed      & Double SED \\

	\hline
    \multicolumn{3}{l}{2MASS \citep{2MASS}; APASS \citep{APASS};}\\
    \multicolumn{3}{l}{WISE \citep{WISE}; \gaia{} \citep{GAIA}}\\
    \multicolumn{3}{l}{B1217 = \citet{Boyajian12, Boyajian17}}\\
	\end{tabular}
    \label{tab:stellar}
\end{table}

\subsection{Kinematics}
\label{sub:kinematics}

\begin{figure}
	\includegraphics[width=\columnwidth,angle=0]{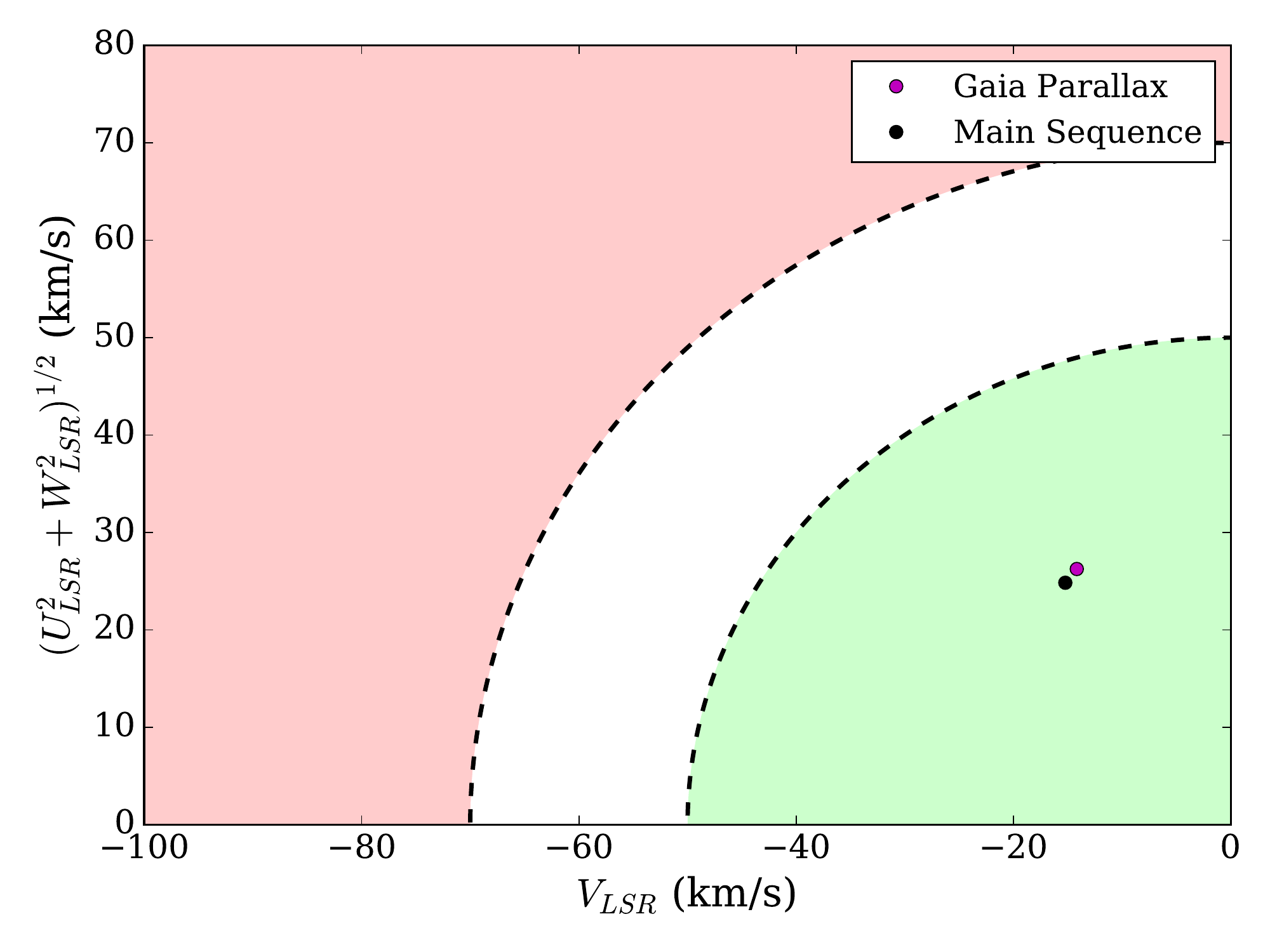}
	\vspace{-0.65cm}
    \caption{Toomre diagram for \NStar\ for our two scenarios. The purple marker corresponds to the solution when we use the \gaia{} parallax and the black marker is when we assume \NStar\ is on the main sequence. The green and red regions represent the expected total velocity distributions for the thin and thick disc respectively, using the values from \citep[][]{Bensby14}. The white area is a region of intermediate probability. Note that in both scenarios \NStar\ is well within the expected velocity range for a thin disc source.} 
    \label{fig:toomre}
\end{figure}

To check whether \NStar\ is consistent with belonging to the galactic thin disc we calculated its kinematics using the stellar parameters from Table \ref{tab:stellar}. We compared the solution to the selection criteria of \citet{Bensby03} and calculated the thick-disc to thin-disc ($P_{\mathrm{thick}}/P_{\mathrm{thin}}$) relative probability. As a 
check, this was repeated using the slightly larger stellar radius ($0.77\pm0.07$\rsun) implied by the \gaia{} parallax. We found that for both scenarios \NStar\ was more likely to belong to the thin disc ($P_{\mathrm{thick}} < 0.1 \times P_{\mathrm{thin}}$) than the thick disc. This is also shown in the Toomre diagram in Fig.\,\ref{fig:toomre}, supporting our assumption that the host star is on the Main Sequence.

\subsection{Stellar Activity and Rotation}
\label{sub:activity_rotation}

We verified the stellar rotation period by calculating the Generalised Autocorrelation Function (G-ACF) of the NGTS photometric time series (Kreutzer et al. [in prep.]). The autocorrelation function is a proven method for extracting stellar variability from photometric light curves \citep[as in][]{mcquillan2014}, and this generalisation allows analysis of irregularly sampled data. This method has been used on NGTS data to successfully extract rotation periods from a large numbers of stars within the Blanco 1 open cluster \citep{2019arXiv191109705G}.

We first binned the time series to $20$ minutes, giving $2432$ data points. As the G-ACF does not return an error on the rotation period directly, we employ a bootstrapping technique. We randomly select 2000 data points from the binned NGTS time series and run the G-ACF analysis. This was repeated 1024 times, giving a rotation period and error of \NProt{} days, where the error is the standard deviation in the periods from the $1024$ runs, divided by $\sqrt{1024}$.

\begin{figure}
	\includegraphics[width=\columnwidth,angle=0]{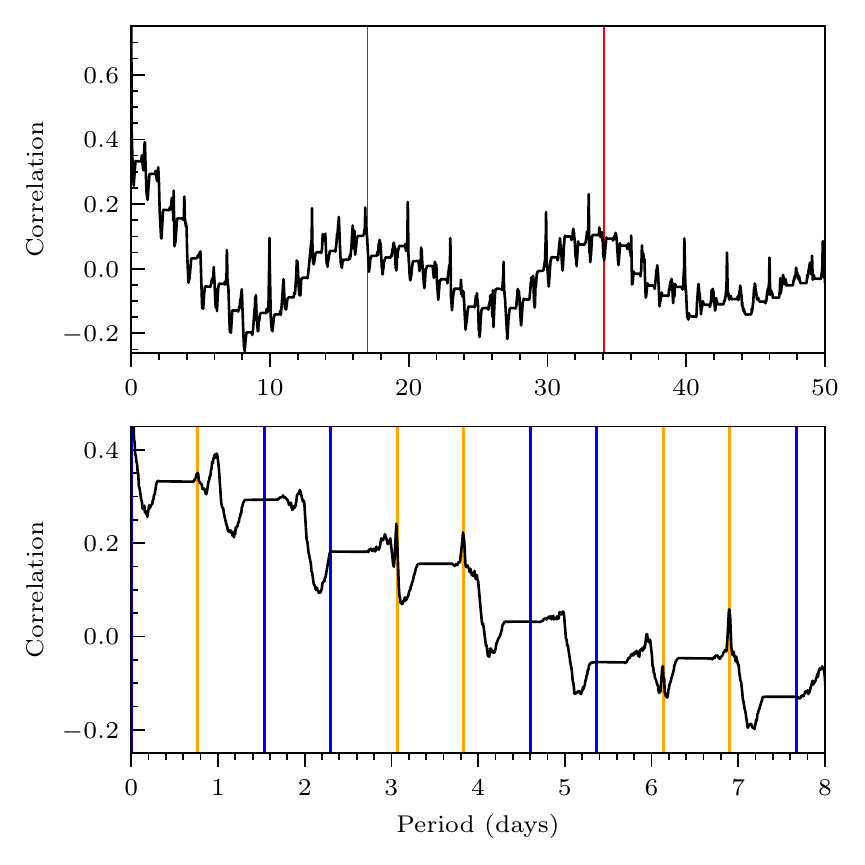}
    \vspace{-0.65cm}
    \caption{Top: The Generalised Autocorrelation Function (G-ACF) of the \NGTS{} light curve. The red lines highlight the \NProt{}\,day periodicity detected. Bottom: Zooming further in on the G-ACF reveals the planet transit signal at $P=\FAPeriodShort$\,d as predicted. The vertical lines show clear periodic increases in the autocorrelation. Orange lines indicate we observe a transit, and blue lines indicate a gap in the data. }
    \label{fig:rotation}
\end{figure}

Figure \ref{fig:rotation} shows this clear periodic signal. This period was verified to be unique for objects within the vicinity of \NStar{} on the \NGTS{} CCD, providing strong evidence that this is not a systematic feature. We note that the rotation period of \NProt{} implies a stellar rotation velocity of $2.04\pm0.11$\,\kms which is smaller than the \NVsini{} upper limit quoted in \S \ref{sub:stellar}. Extrapolation of the \citep{2014MNRAS.444.3592D} calibration suggests a value for macroturbulence could be \vmacro$\sim3.5$\,\kms, which would give a \vsini$\sim2.0$\,\kms, which is in line with
the photometric rotation period above. This high value of macroturbulence, is supported by the relation used by the Gaia-ESO Survey which gives \vmacro$=3.8$\,\kms. Additionally, we find evidence for line core emission in the Ca II H and K lines with an activity index of \mbox{log R'$_{\mathrm{HK}}$ = \NActivityIndexBarry}. This evidence for stellar activity may additionally support the case for a non-zero macroturbulent velocity.

\subsection{Stellar Age}
\label{sub:age}
We place constraints on the age of the host star using the Bayesian fitting process described in \citet{2015AA...575A..36M}, and which is available as the open source \textsc{BAGEMASS}\footnote{\url{https://sourceforge.net/projects/bagemass/}} code. \textsc{BAGEMASS} uses the GARSTEC models of \citet{2008ApSS.316...99W}, as computed by \citet{2013MNRAS.429.3645S}, and works in $[\log(L_{\rm star}), T_{\rm eff}]$. 

\begin{figure}
	\includegraphics[width=\columnwidth,angle=0]{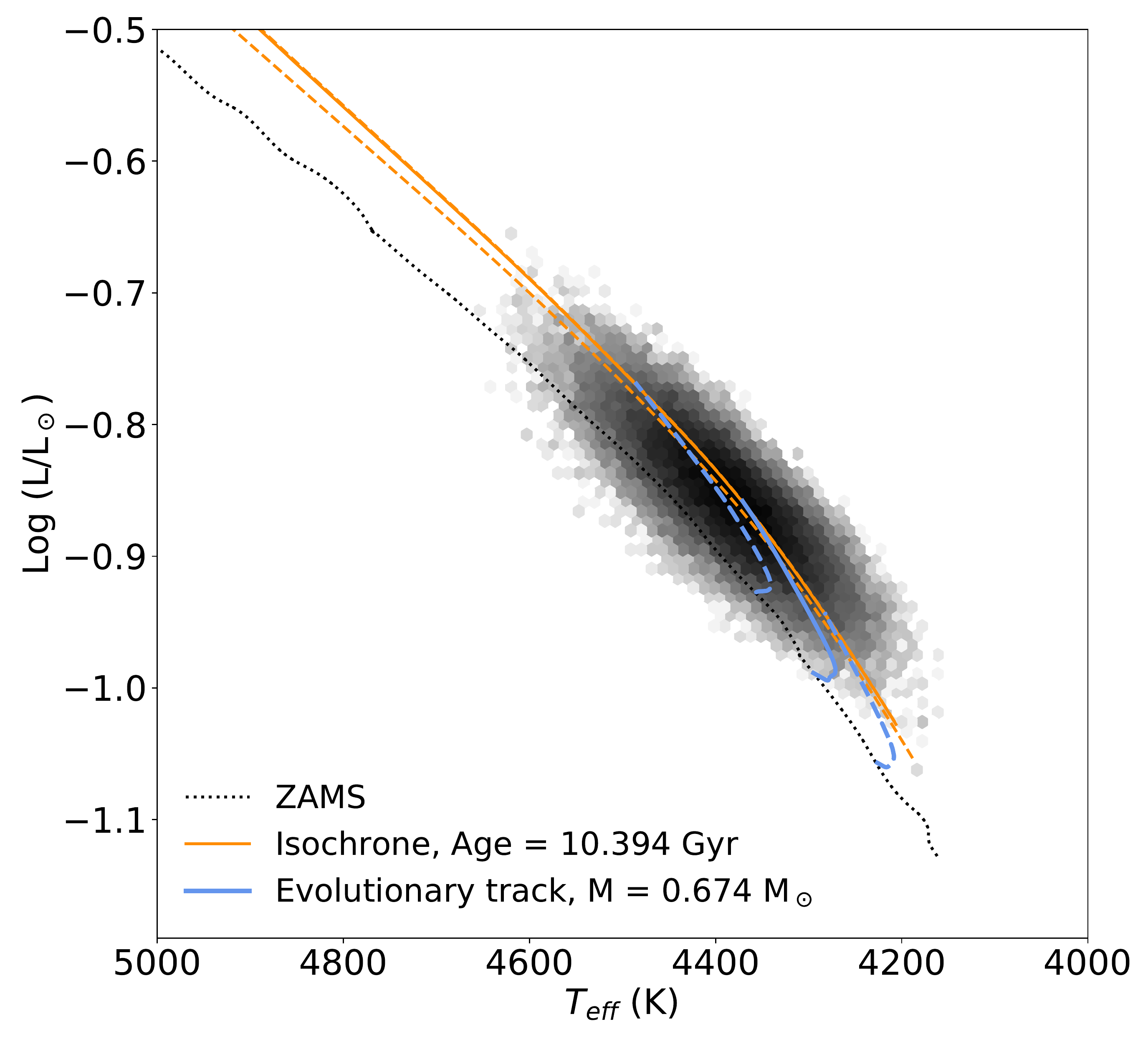}
    \vspace{-0.65cm}
	\caption{The results from stellar model fitting using \textsc{BAGEMASS}, showing the best-fitting stellar models and the posterior probability distribution of the MCMC fitting process, the colour scale of which represents the density of points. The ZAMS is shown as a dotted black line. The solid blue line is the best-fitting stellar evolutionary tracks, with the blue dashed lines representing evolutionary tracks for the $1\sigma$ limits on stellar mass. The solid orange line is the stellar isochrone, with the orange dashed lines representing isochrones for the $1\sigma$ limits on stellar age.}
	\label{fig:bagemass}
\end{figure}

We perform a stellar model fit using all three sets of model grids available within \textsc{BAGEMASS}, deriving age estimates that agree within $1\sigma$. In Figure\,\ref{fig:bagemass} we show the posterior probability distribution for the fit to one of those grids. We adopt a final age of \Nage\,Gyr, calculating as the weighted average of the three fits, which is consistent with the age of the thin disc ($8.8\pm1.7$\,Gyr \citealt{2005A&A...440.1153D}) within the uncertainties. 

\subsection{Analysis of \gaia{} scan angles}
\label{sub:gaia_scans}

\begin{figure}
	\includegraphics[width=\columnwidth]{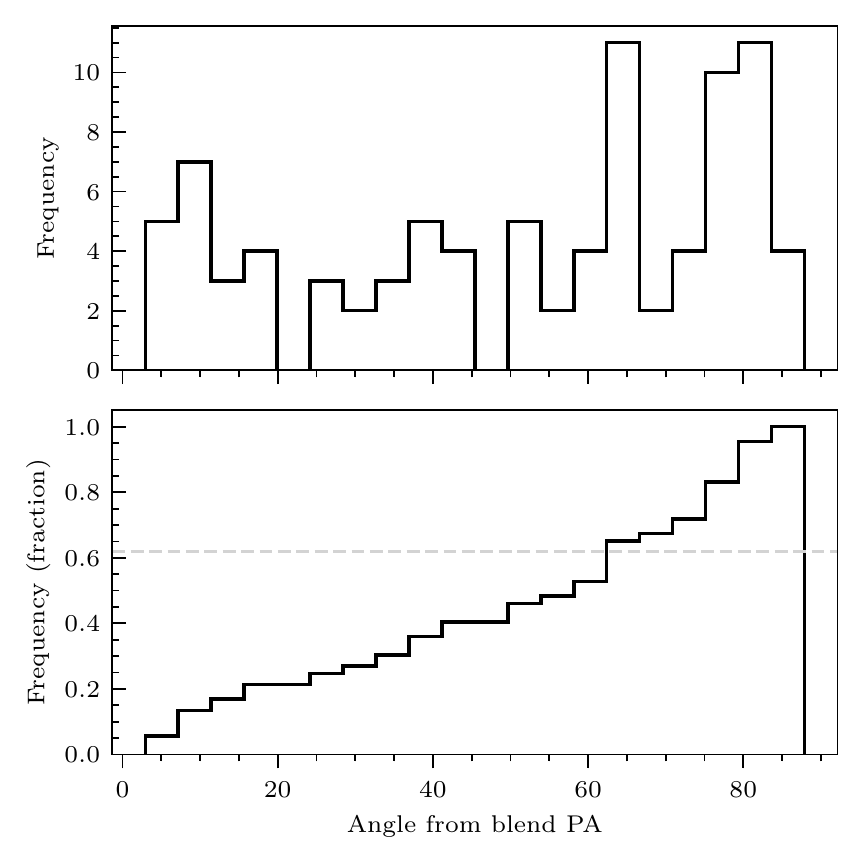}
	\vspace{-0.65cm}
    \caption{Top: Histogram showing the distribution of \gaia{} scan angles with respect to the position angle of the \NStar{} \& \NStarB{} blend. \gaia{} scanned across the blend over a range of angles from co-planar to perpendicular. Bottom: A cumulative frequency distribution of the data in the top panel. The dashed grey line shows the fraction of measurements where \gaia{} measures astrometry for \NStar{} only. To explain the discrepancy between the numbers of measurements of the two sources in the blend, we hypothesise that once the scan angle of \gaia{} is within approximately $60^{\circ}$ of the blend PA, the two sources appear to become confused.}
    \label{fig:gaia_scan_angles}
\end{figure}

\gaia{} DR2 quotes an astrometric noise of \mbox{\NAstrometricNoiseExcess\,\mas} for \NStar{} but only \CompanionAstromertricNoiseExcess\,\mas{} for \NStarB, which was initially puzzling. With the current data release we are unable to analyse the astrometric data from individual scans separately to draw more detailed conclusions. Below we hypothesise on what may be happening. \gaia{} DR2 contains $332$ and $122$ astrometric measurements of \NStar{} and \NStarB, respectively. Of these, $320$ and $122$ are flagged as being \emph{good}, respectively. This discrepancy between the numbers of astrometric measurements for the two sources may point to \NStarB{} being unresolved in $62$\% of scans and could help explain the source of the astrometric noise excess. We inspected $89$ scan angles available via Gost and measured the angular offsets between the position angle of the \NStar/\NStarB{} blended pair (\CompanionPositionAngle$^{\circ}$) and each individual scan angle. The distribution of angular offsets is shown in Figure \ref{fig:gaia_scan_angles}. Propagating our assumption above we draw a line at $62$\% on the histogram in the lower panel of Fig. \ref{fig:gaia_scan_angles}, indicating that once the scan angle is within $60^{\circ}$ of the blend PA, the two sources appear to become confused. We plan to revisit this issue when the individual data astrometric measurements become available in \gaia{} DR4.

\section{Global Modelling}
\label{sec:global}

We modelled the NGTS and follow-up data with \gpe\ \citep{Gillen17} to determine fundamental and orbital parameters of \NPlanet. The full data set comprises the NGTS discovery light curve (containing 46 transits), six follow-up transit light curves (in four photometric bands), and 10 HARPS RVs. 
\gpe\ comprises a central transiting planet and eclipsing binary model, which is coupled with a Gaussian process (GP) model to simultaneously account for correlated noise in the data, and uses Markov Chain Monte Carlo (MCMC) to explore the posterior parameter space. Limb darkening is treated using the analytic prescription of \citet{Mandel02} for the quadratic law. 

Each light curve bandpass possesses its own stellar variability and each transit observation posses its own atmospheric/instrumental noise properties, which will affect the apparent transit shape and hence the inferred planet parameters. To account for this, \gpe\ simultaneously models the variability and systematics using GPs, at the same time as fitting the planet transits, which gives a principled framework for propagating uncertainties in the noise modelling through into the planet posterior parameters. 
We chose a Matern-32 kernel for all light curves given the reasonably low level of stellar variability but clear instrument systematics and/or atmospheric variability.
Limb darkening profile priors were generated with LDtk \citep{Parviainen15} assuming the \teff, \logg\ and \feh\ values from the double SED fit given in Table \ref{tab:stellar}. We placed Gaussian priors on the dilution factors in each photometry band using the values in Table \ref{tab:stellar}. 

The NGTS data was binned to 5 min and the \gpe\ model binned accordingly. The SAAO B band light curve was binned to 1 min cadence but we opted not to integrate the \gpe\ model given the short resulting cadence. All other light curves were modelled at the cadences reported in Table \ref{tab:followup_phot}.

Given the sparse RV coverage (10 data points over 400 days), we opted not to include a GP noise model in the \gpe\ RV model, and instead incorporated a white noise jitter term, under penalty, that was added in quadrature to the observational uncertainties. Given the limited information contained within the RV data and the extremely short period, we assumed a circular Keplerian orbit for the planet. We tried fitting the RVs with a linear drift in time but found a zero slope ($8.0\pm8.6$\,\ms) and a reduced $\chi^{2}<0$ (overfitting), so we opted to exclude the linear drift from the RV fit in the final MCMC run. Using the method of \citet{2017MNRAS.468.4772S} for a star of log R'$_{\mathrm{HK}}$= \NActivityIndexBarry\ we calculate a stellar jitter of $2.9$\,\ms. The RMS of the RV residuals (see lower panel in Figure \ref{fig:harps_rvs}) is $9.79$\,\ms and the average error on each RV measurement is $14$\,\ms. Hence, the precision of the RV measurements is limited by the instrumental performance and not by stellar jitter.

\begin{figure*}
	\includegraphics[width=17cm,angle=0]{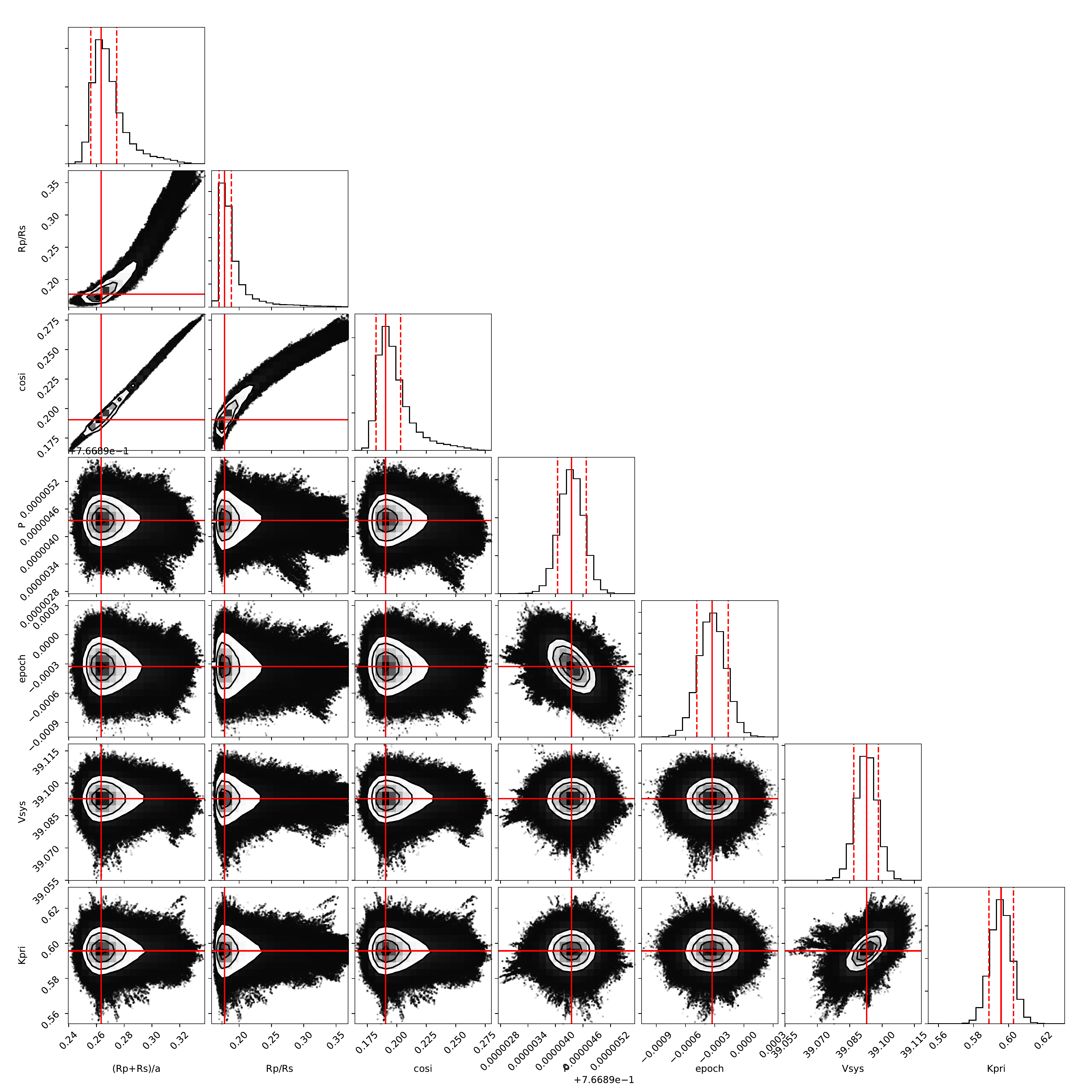}
    \caption{MCMC posterior distributions. A total of 300 walkers and 200,000 steps were used in this fit. We plot the MCMC chain for the final 100,000 steps showing every 500$^{\mathrm{th}}$ element only. The solid red lines mark the mode of each parameter's distribution while the dashed red lines mark the full width at half maximum. We use the mode as the grazing nature of the system causes the distributions of $(R_{\rm p} + R_{\rm s}) / a$, $R_{\rm p}/R_{\rm s}$ and $\cos\,i$ to be skewed. Both the median and mean of the distributions overestimate the parameters and do not represent the peak of each distribution.} 
    \label{fig:mcmc_posteriors}
\end{figure*}

We performed 200,000 MCMC steps with 300 walkers, discarding the first 100,000 steps as a conservative burn in. The resulting chains yielded posterior values for the fundamental and orbital planet parameters, which are reported in Table \ref{tab:globalfit}. The posteriors can be seen in Fig. \ref{fig:mcmc_posteriors}. The grazing nature of the system resulted in skewed distributions of $(R_{\rm p} + R_{\rm s}) / a$, $R_{\rm p}/R_{\rm s}$ and $\cos\,i$, hence we extracted the best fitting values and their uncertainties using the mode and full width at half maximum of each distribution. We find that \NPlanet\ has a mass and radius of \mpl\ = \NMass\ \mjup\ and \rpl\ = \NRadius\ \rjup, and orbits its host star in \FAperiod\ days at a distance of \NAAu\ AU. 

\begin{table}  
 \scriptsize
 \caption[Model parameters]{Best fitting and derived parameters from the global modelling of \NPlanet.} 
 \label{tab:globalfit}  
 \begin{tabular}{l c c c}  
 Parameter  &   Symbol  &  Unit  & Value \\  
\hline 
 \multicolumn{4}{c}{\emph{Transit parameters}} \\ [1.0ex]
 Sum of radii            & $(\rm{R}_{\rm{s}}+\rm{R}_{\rm{p}})/\rm{a}$ & ---  & \FArasum   \\ [1.0ex]
 Radius radio            & $\rm{R}_{\rm{p}}/\rm{R}_{\rm{s}}$          & ---  & \FArr      \\ [1.0ex]
 Cosine inclination      & $\cos i$                                   & ---  & \FAcosi    \\ [1.0ex]
 Impact parameter        & b                                          & ---  & \FAb       \\ [1.0ex]
 Epoch                   & T$_{0}$                                    & HJD  & \FAepoch   \\ [1.0ex]
 Period                  & P                                          & days & \FAperiod  \\ [1.0ex]
 Eccentricity            & e                                          & ---  & 0 (fixed)  \\ [1.0ex] 
 Dilution NGTS    & $\delta$NGTS                               & ---  & \FADilNgts \\ [1.0ex]
 Dilution I-band  & $\delta$I                                  & ---  & \FADilI    \\ [1.0ex]
 Dilution V-band  & $\delta$V                                  & ---  & \FADilV    \\ [1.0ex]
 Dilution z-band  & $\delta$z                                  & ---  & \FADilz    \\ [1.0ex]
 Dilution B-band  & $\delta$B                                  & ---  & \FADilB    \\ [1.0ex]
\multicolumn{4}{c}{\emph{Radial velocity parameters}} \\  [1.0ex]
Systemic velocity  & V$_{\rm{sys}}$  & \kms & \FAvsys       \\ [1.0ex]
RV semi-amplitude  & K               & \kms & \FAkp         \\ [1.0ex]
\multicolumn{4}{c}{\emph{Planet parameters:}} \\  [1.0ex]
Planet mass              & \mpl      & \mjup   & \NMass    \\ [1.0ex]
Planet radius            & \rpl      & \rjup   & \NRadius  \\ [1.0ex]
Planet density           & \rhopl    & \gccc   & \NDensity \\ [1.0ex]
Semi-major axis          & a         & AU      & \NAAu     \\ [1.0ex]
Semi-major axis          & a/\rstar  & ---     & \NARs     \\ [1.0ex]
Transit duration         & \tdur     & hours   & \NTDur    \\ [1.0ex]
Equilibrium Temp.  & \teqpl    & K       & \NTeq     \\ [1.0ex]
 \noalign{\smallskip} \noalign{\smallskip}  
 \hline  
 \end{tabular}  
 \end{table} 

\section{Tidal Modelling}
\label{sec:tides}

Owing to the ultra-short orbital period of \NPlanet, it is likely to undergo strong tidal interactions with its host star, and ultimately undergo orbital decay. We therefore model the tidal interactions in the system to constrain the remaining lifetime of the planet. We adopt the tidal evolution model of \citet{brown2011}, which implements the equilibrium tide theory of \citet{eggleton1998} as parametrised by \citet{dobbsdixon2004}.

We define a set of values for the stellar tidal quality factor $Q'_{s}$, $\{$\logQs $| 5,6,7,8,9,10\}$ to explore for the system; these are assumed to be constant with time. It is very likely that the value of $Q'_{s}$ evolves over time as the structure of the star changes, particularly the radial extent of its radiative and convective regions, which are known to affect the efficiency of tidal dissipation. For example, \citet{BolmontMathis16} found that young stars are much more dissipative than main sequence stars owing to their extended convective envelopes. However, detailed modelling of the dynamical tide (e.g. following \citealt{OgilvieLin07}), which would give a more accurate picture of the future evolution of this system, is beyond the scope of this paper.

We set the planetary tidal quality factor as constant, \logQpl$=8$, as \citeauthor{brown2011} found this parameter to have a negligible effect on semi-major axis evolution. Without knowledge of the planet's interior structure, estimating the moment of inertia constant is difficult. As the density of \NPlanet{} (\NDensity\,\gccc) is similar to that of Jupiter ($1.33$\,\gccc), we assume the Jovian moment of inertia \NInertia{} from \citet{2018A&A...613A..32N}. \citet{Alvarado19} showed that such an assumption can lead to mis-representation of the true tidal evolution of a system, and that the physical evolution of giant planets plays a key role in determining the strength of tidal interactions in hot Jupiter systems. However, they also note that the tidal dissipative properties of exoplanets and stars are poorly defined, that constant properties can be a reasonable approximation given the lack of information on interior properties.

For each value of \logQs, we draw $1000$ random samples from Gaussian distributions in system age, semi-major axis, eccentricity (to avoid divide-by-zero errors, we assume a negligible but non-zero eccentricity of $10^{-6}$), and stellar rotation frequency. Each Gaussian is centered on the model value from Table\,\ref{tab:globalfit} and has a width equal to the listed $1\sigma$ uncertainty. Where the uncertainty is asymmetric, a skewed Gaussian was used.

For each set of random samples, we use a fourth-order Runge-Kutta integrator \citep[adapted from algorithms in][]{press1992} to integrate the set of equations (1), (6), (7), and (12) from \citet{brown2011} forwards from the sampled age until twice the host star's estimated main sequence lifetime has elapsed, or until the planet reaches the Roche limit, which we define as

\begin{equation}
a_{\rm Roche} = 2.46\rpl\left(\frac{\mstar}{\mpl}\right)^{1/3}.
\end{equation}

\begin{figure}
	\includegraphics[width=\columnwidth,angle=0,trim=1cm 2cm 3.5cm 1cm]{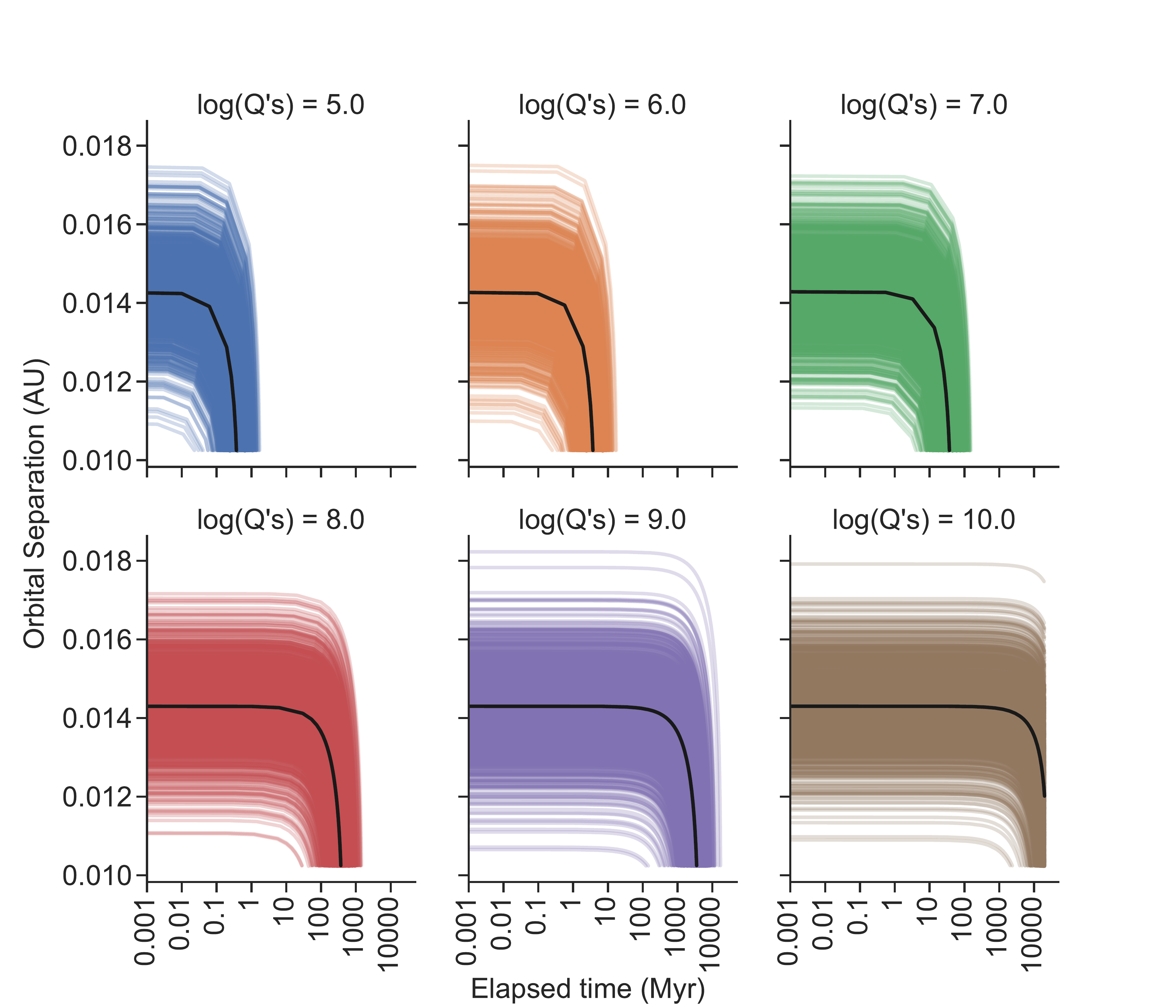}
    \caption{Semi-major axis as a function of elapsed time for different values of \logQs. In each panel, each coloured line represents a possible history based on a set of parameters drawn randomly from distributions based on the observed parameters. The black line in each panel denotes the evolution taking the observed parameters at face value. The tracks cut-off when the planet reaches the Roche limit, or twice the estimated main sequence lifetime of the host star.}
    \label{fig:atraces}
\end{figure}

\begin{figure}
	\includegraphics[width=\columnwidth,angle=0,trim=1.7cm 2cm 3.5cm 1cm]{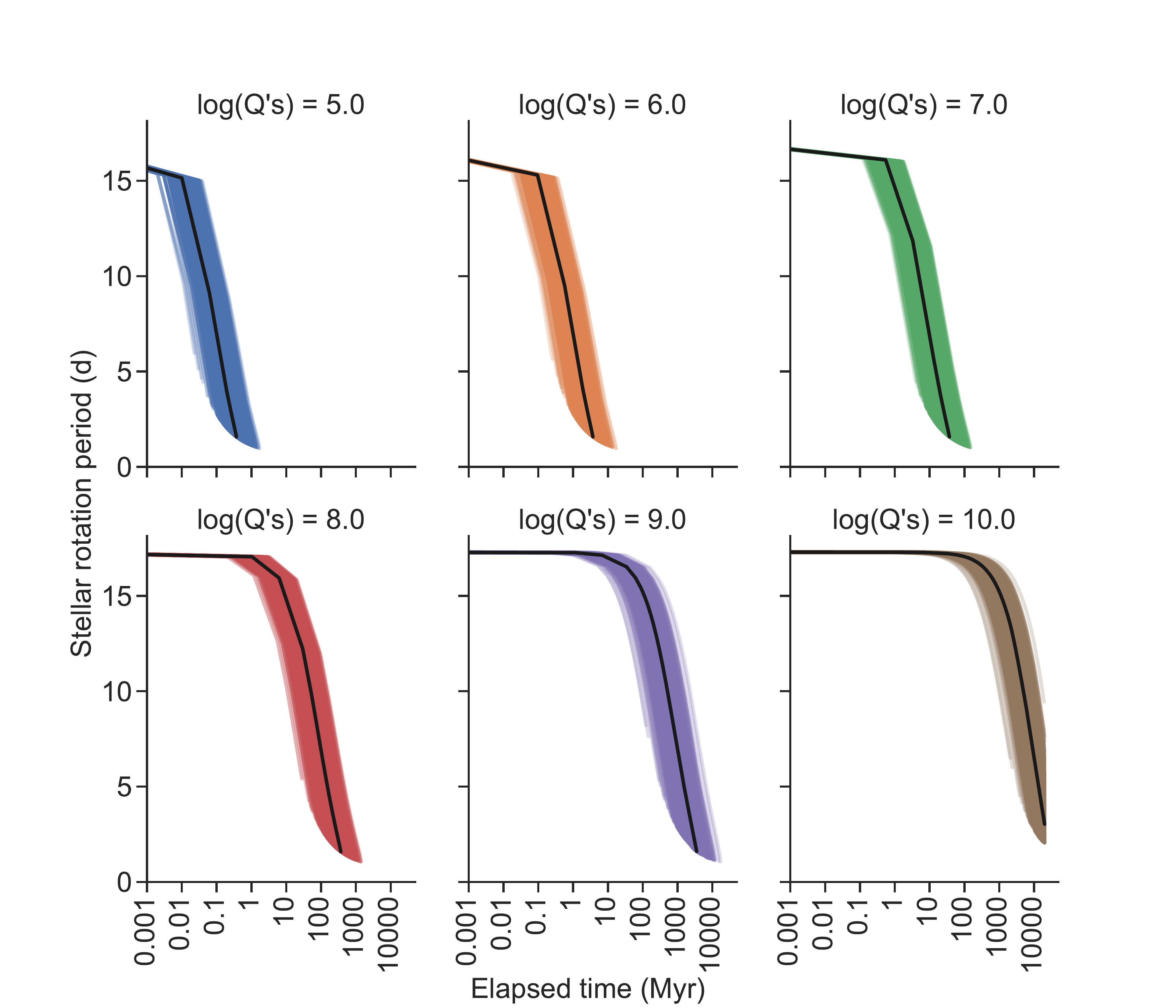}
    \caption{Stellar rotation period as a function of elapsed time for different values of \logQs. Format is as for Figure\,\ref{fig:atraces}.}
    \label{fig:ptraces}
\end{figure}

\begin{figure}
	\includegraphics[width=\columnwidth,angle=0,trim=1cm 2cm 1cm 2cm]{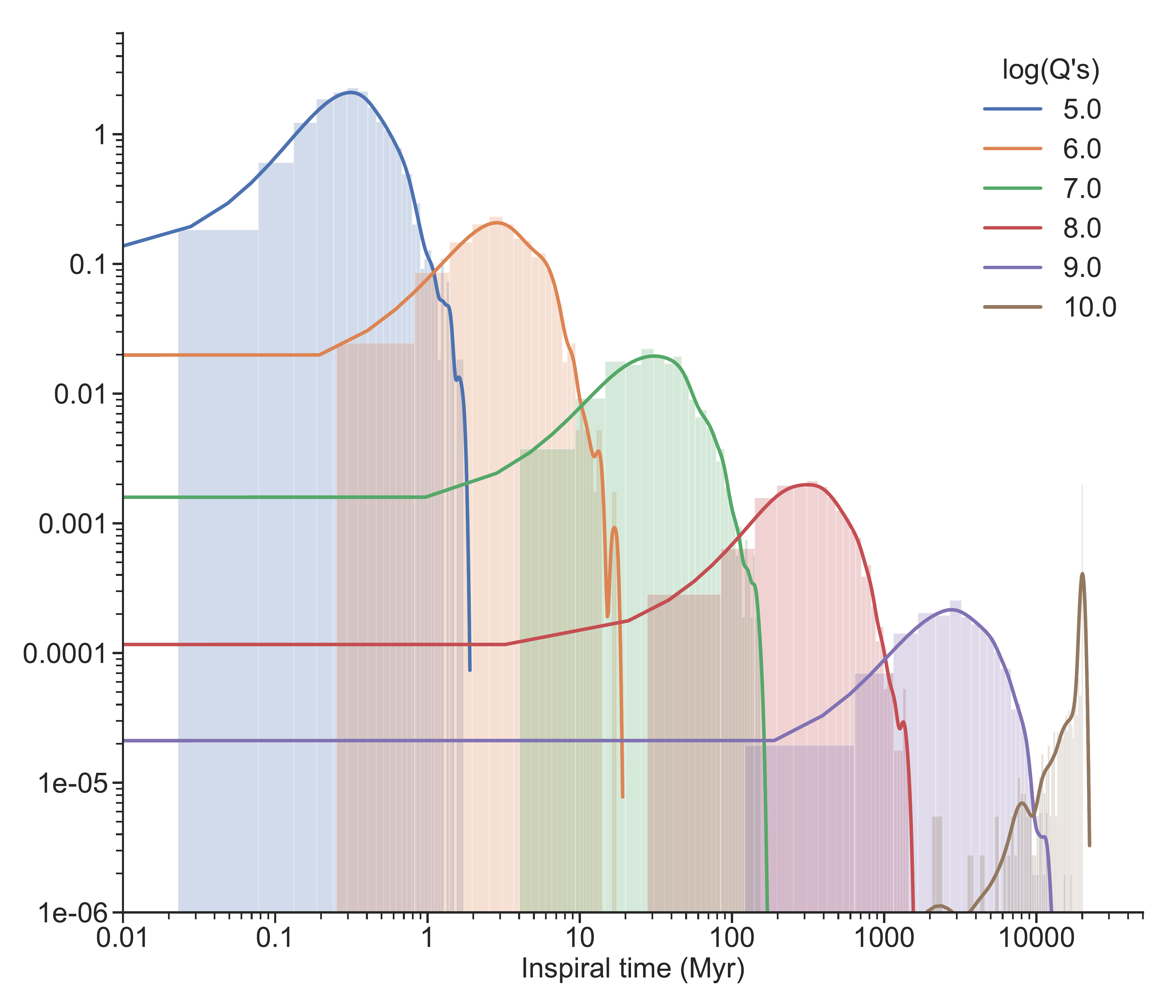}
    \caption{Histograms of inspiral time for different values of \logQs. As the efficiency of tidal dissipation decreases, the spread in modelled inspiral time increases. Note that the y-axis is log-formatted for display purposes}
    \label{fig:tspiralHist}
\end{figure}

Figure\,\ref{fig:atraces} and \ref{fig:ptraces} show the resulting ensemble evolutionary traces for semi-major axis and stellar rotation period, respectively. Also plotted are `baseline' traces which use the values from Table\,\ref{tab:globalfit} as their starting point. For \logQs$=10$, i.e. very inefficient dissipation and thus very weak tidal forces, we find that many of the traces reach maximum runtime rather than ending at the Roche limit. Figure\,\ref{fig:tspiralHist} plots histograms of the remaining time to reach the Roche limit for the different values of \logQs. For values of \logQs{} more consistent with those found for other hot Jupiters, we find that the planet reaches the Roche limit on timescales of the order Myr (\logQs$=6$) to hundreds of Myr (\logQs$=8$).

In all cases, the host star is spun up during the tide-driven evolution of the planet's orbit. For the `baseline' traces, the stellar rotation period when the planet reaches the Roche limit is between $1.4$\,d and $1.9$\,d. The full range of final stellar rotation periods is $1.0$\,d$\leq P_{\rm rot}\leq9.4$\,d, though we note that this includes traces that halt at the maximum runtime rather than the Roche limit, and have thus experienced less evolution of the system parameters.

\section{Discussion}
\label{sec:discussion}

\begin{figure}
	\includegraphics[width=\columnwidth, trim=0cm 0.8cm 0cm 0.5cm]{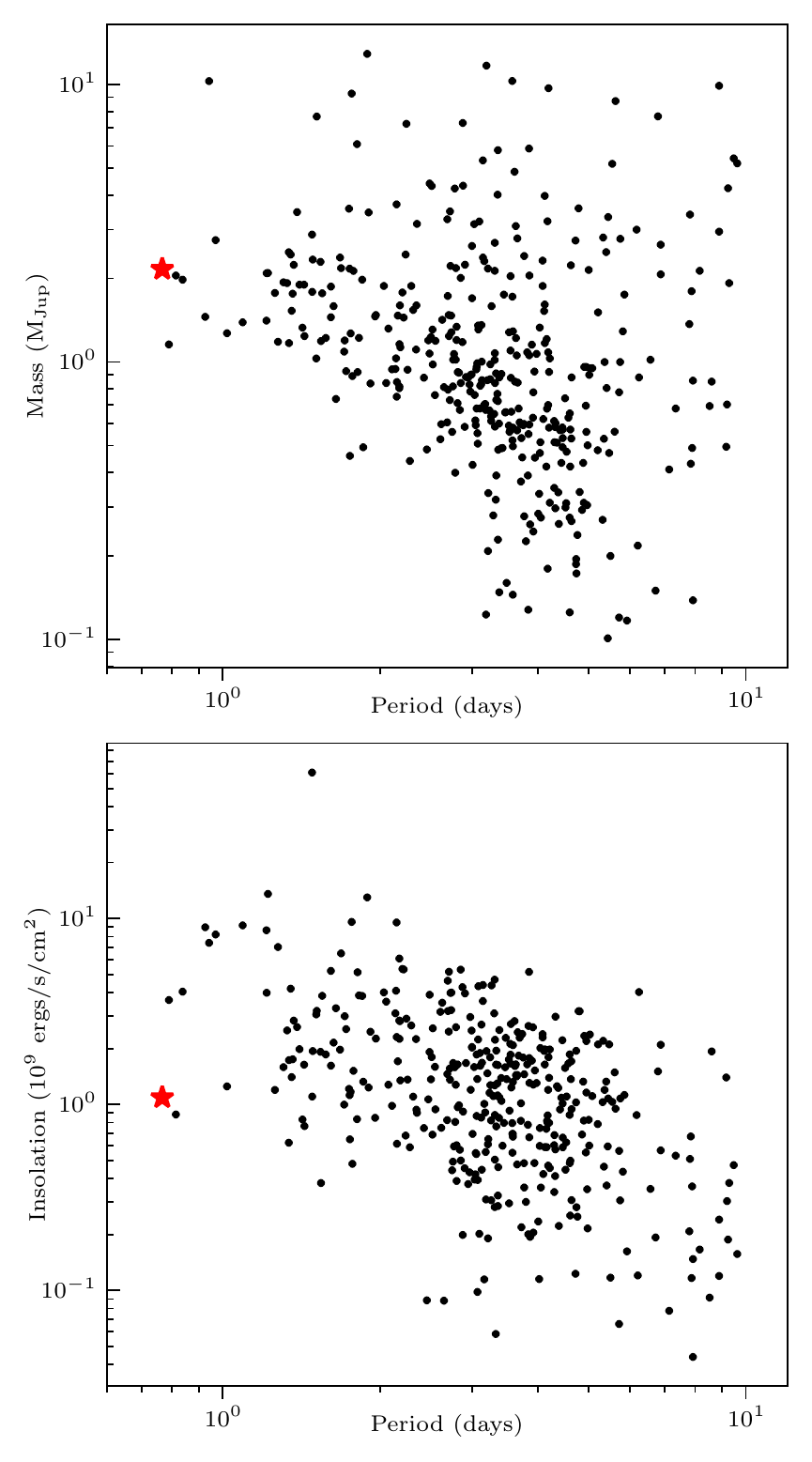}
    \caption{Top: Planet mass vs orbital period. Bottom: Received insolation vs orbital period. In each plot we show planets with masses and radii determined to 20\% precision or better, periods $<10$ days and masses in the range $0.1$--$13$M$_{\mathrm{Jup}}$. Error bars are excluded for clarity and \NPlanet{} is highlighted with a red star.}
    \label{fig:planet_context}
\end{figure}

Our observations and modelling show that NGTS-10 hosts a hot Jupiter with the shortest orbital period yet found. It orbits its K5 host star with a period of only $18.4$ hours (\FAperiod\,days). We have determined the mass (\NMass\,\mjup) and radius (\NRadius\,\rjup) of the planet to better than 5\% and 10\% precision, respectively. Figure \ref{fig:planet_context} shows \NPlanet{} in the context of the current population of transiting hot Jupiters with precisely determined masses and radii ($\leq20$\% precision). The top panel highlights \NPlanet{} as an extreme object, pushing the bounds of planetary mass/period parameter space. \NPlanet{} orbits its host star at only \NARs{} stellar radii (or \NARoche{} Roche radii).

Hot Jupiters are typically prime candidates for atmospheric characterisation. The close proximity to their host star increases the level of reflected light and raises their equilibrium temperatures leading to increased thermal emission. This in turn permits atmospheric characterisation via secondary eclipse and phase curve measurements. \citet{2019AJ....157..178S} recently used this technique to measure the low albedo and the inefficient redistribution of energy in the atmosphere of WASP-18b. Atmospheric characterisation is also possible via transmission spectroscopy. The transmission spectroscopy signal strength is increased for planets with larger atmospheric scale heights ($H$) and smaller host star radii. A planet's atmospheric scale height is driven by its equilibrium temperature and surface gravity. Although \NPlanet{} is the shortest period Jupiter yet discovered, the host star is relatively cool (\teff$=4400$\,K) which leads to a lower insolation (see bottom panel of Figure \ref{fig:planet_context} and Table \ref{tab:uspgp}) and a lower equilibrium temperature when compared to other USP hot Jupiters. We calculate a fairly typical scale height of $H=135$\,km for \NPlanet{} but given it orbits a relatively small K5V star, the transmission spectroscopy signal strength is comparable to that of the well studied USP hot Jupiters WASP-18b and WASP-103b. Atmospheric characterisation of \NPlanet{} in the optical will be challenging given the faintness of the host star ($V=14.3$). However, in the infrared the host star is much brighter ($H=11.9$), this combined with the expected strength of its transmission spectroscopy signal makes \NPlanet{} a potential candidate for atmospheric characterisation using JWST and NIRSpec.

A major problem with investigations of tidal evolution is placing constraints on \Qs. Previous studies have attempted to do this on a system by system basis \citep[e.g.][]{brown2011, 2016AJ....152..127P}, but a recent paper by \citet{penev2018} described a simple relationship between \Qs{} and the tidal forcing period, which itself is a simple function of the orbital and stellar rotation periods. We use equations (1) and (2) of \citet{penev2018} to calculate a value of \Qs$=2\times10^7$ for \NStar. Based on this, we estimate a median inspiral time of \NMedInspiralTimeQVII\,Myr for \NPlanet{} from the distribution for \logQs$=7$ in Figure\,\ref{fig:tspiralHist}. 

However, recent work by \citet{Heller19} predicted that we are unlikely to observe tidally-driven orbital decay in systems such as \NPlanet{} owing to extremely inefficient tidal dissipation in the convective envelope of main-sequence, Sun-like stars. \citet{Heller19} also demonstrated that the pile-up of hot Jupiters around $0.05$\,AU could be reproduced by a combination of the dynamical tide and type-II planet migration. \NPlanet{}'s location at much shorter semi-major axis thus suggests either an old age, supporting the BAGEMASS prediction, or stronger tides than are typical for a star of this type.

\begin{figure}
	\includegraphics[width=\columnwidth]{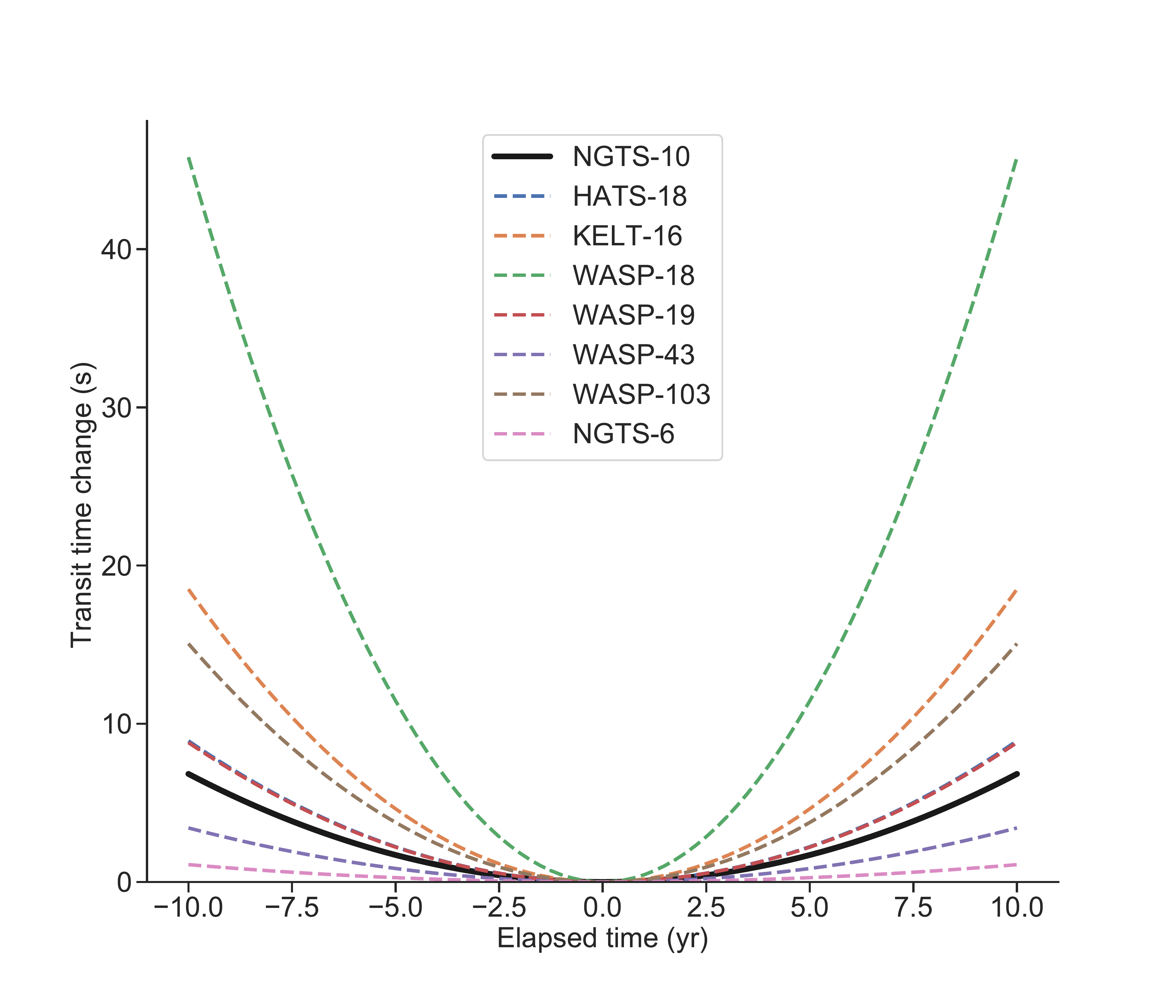}
    \caption{Change in transit time as a function of elapsed time over a 20 year baseline for \NPlanet, assuming a value of \Qs$=2\times10^{7}$. Also plotted are the equivalent curves for the other currently known USP hot Jupiters listed in Table\,\ref{tab:uspgp}.}
    \label{fig:deltat}
\end{figure}

If we assume that \NPlanet{} is likely to have a decaying orbit, it is interesting to consider on what timescale might we be able to detect changes in orbital period or transit times. We use Eqn.\,27 of \citet{acc2018} to determine the quadratic change in transit time on human-observable timescales of order one decade. We predict a change of \NPeriodDecayFiveYear\,s over 5\,years, and \NPeriodDecayDecade\,s over 10\,years, assuming \Qs$=2\times10^7$ (see Fig.\,\ref{fig:deltat}). \NPlanet{} therefore joins WASP-19 and HATS-18 as reasonable candidates for the direct measurement of orbital period decay over the coming years. We note that our current precision on T$_{0}$ is approximately 15\,s. Therefore, detecting the comparably small variation in transit timing (\NPeriodDecayDecade\,s) will require higher cadence and higher quality transit observations if it is to be measured accurately. We also note that an ongoing RV survey would also be needed to rule out the influence of additional longer period planets in the system.

\begin{table*}
    \scriptsize
	\centering
	\caption{Orbital parameters of the currently known USP hot Jupiters.}
	\label{tab:uspgp}
	\begin{tabular}{ccccccc}
	Name  & Period & Semi-major & $a/R_{\mathrm{star}}$ & $a/R_{\mathrm{Roche}}$ & Insolation           & Reference \\
          & (days) & axis (AU)  &                       &                        & ($\times10^{9}$ ergs\,$s^{-1}$\,cm$^{-2}$) &           \\ 
	\hline
    \NPlanet{} & \FAperiod        & \NAAu         & \NARs         & \NARoche         & \NInsol         & This work           \\ [1.0ex]
    NGTS-6b    & \NPeriodNGTSVI   & \NAAuNGTSVI   & \NARsNGTSVI   & \NARocheNGTSVI   & \NInsolNGTSVI   & \citet{2019arXiv190407997V} \\ [1.0ex]
    WASP-18b   & \NPeriodWXIIX    & \NAAuWXIIX    & \NARsWXIIX    & \NARocheWXIIX    & \NInsolWXIIX    & \citet{stassun2017} \\ [1.0ex]
    WASP-19b   & \NPeriodWXIX     & \NAAuWXIX     & \NARsWXIX     & \NARocheWXIX     & \NInsolWXIX     & \citet{wong2016}    \\ [1.0ex]
    WASP-43b   & \NPeriodWXXXXIII & \NAAuWXXXXIII & \NARsWXXXXIII & \NARocheWXXXXIII & \NInsolWXXXXIII & \citet{wasp43}      \\ [1.0ex]
    WASP-103b  & \NPeriodWCIII    & \NAAuWCIII    & \NARsWCIII    & \NARocheWCIII    & \NInsolWCIII    & \citet{wasp103}     \\ [1.0ex]
	KELT-16b   & \NPeriodKXVI     & \NAAuKXVI     & \NARsKXVI     & \NARocheKXVI     & \NInsolKXVI     & \citet{kelt16}      \\ [1.0ex]
	HATS-18b   & \NPeriodHXIIX    & \NAAuHXIIX    & \NARsHXIIX    & \NARocheHXIIX    & \NInsolHXIIX    & \citet{hats18}      \\ [1.0ex]
	\hline
	\end{tabular}
\end{table*}

\section{Conclusions}
\label{sec:conclusions}

We report the discovery of the shortest period hot Jupiter to date, \NPlanet. We have determined the planetary mass and radius to better than $5$\% and $10$\% precision, respectively. \NStar{} is determined to be a K5 main sequence star with an effective temperature of \NTeffSed\,K. Our tidal analysis determined a quality factor \Qs$=2\times10^{7}$ for \NStar{} and a median inspiral time of \NMedInspiralTimeQVII\,Myr for \NPlanet. We calculate a potentially-measurable transit time change of \NPeriodDecayDecade\,s over the coming decade. We aim to obtain high precision transit observations over the coming years to directly measure the efficiency of the tidal dissipation. Due to the blended nature of the system, with a contaminating source, the precise distance to \NStar\ remains uncertain. Using data from future \gaia{} releases we plan to revisit this outstanding issue. \\

\section*{Acknowledgements}
Based on data collected under the NGTS project at the ESO La Silla Paranal Observatory.  The construction of the NGTS facility was funded by 
the University of Warwick,
the University of Leicester,
Queen's University Belfast,
the University of Geneva,
the Deutsches Zentrum f\" ur Luft- und Raumfahrt e.V. (DLR; under the `Gro\ss investition GI-NGTS'),
the University of Cambridge
and the UK Science and Technology Facilities Council (STFC; project reference ST/M001962/1).
The NGTS facility is operated by the consortium institutes with 
support from STFC under projects ST/M001962/1 and ST/S002642/1. 
This paper uses observations made at the South African Astronomical Observatory (SAAO). 
The contributions at the University of Warwick by PJW, RGW, DLP, DJA, BTG and TL have been supported by STFC through consolidated grants ST/L000733/1 and ST/P000495/1.
Contributions at the University of Geneva by FB, BC, LDN, ML, OT and SU were carried out within the framework of the National Centre for Competence in Research "PlanetS" supported by the Swiss National Science Foundation (SNSF).
The contributions at the University of Leicester by MRG and MRB have been supported by STFC through consolidated grant ST/N000757/1.
EG gratefully acknowledges support from the David and Claudia Harding Foundation in the form of a Winton Exoplanet Fellowship.
CAW acknowledges support from the STFC grant ST/P000312/1.
TL was also supported by STFC studentship 1226157.
MNG acknowledges support from MIT's Kavli Institute as a Torres postdoctoral fellow.
DJA acknowledges support from the STFC via an Ernest Rutherford Fellowship (ST/R00384X/1).
SLC acknowledges support from the STFC via an Ernest Rutherford Fellowship (ST/R003726/1).
JSJ acknowledges support by Fondecyt grant 1161218 and partial support by CATA-Basal (PB06, CONICYT).
JIV acknowledges support of CONICYT-PFCHA/Doctorado Nacional-21191829, Chile.
DJAB, JMcC acknowledge support by the UK Space Agency.
PE, ACh, and HR acknowledge the support of the DFG priority program SPP 1992 "Exploring the Diversity of Extrasolar Planets" (RA 714/13-1).
This project has received funding from the European Research Council (ERC) under the European Union's Horizon 2020 research and innovation programme (grant agreement No 681601).
The research leading to these results has received funding from the European Research Council under the European Union's Seventh Framework Programme (FP/2007-2013) / ERC Grant Agreement n. 320964 (WDTracer).
This work has made use of data from the European Space Agency (ESA) mission
\gaia{} (\url{https://www.cosmos.esa.int/gaia}), processed by the \gaia{}
Data Processing and Analysis Consortium (DPAC,
\url{https://www.cosmos.esa.int/web/gaia/dpac/consortium}). Funding for the DPAC
has been provided by national institutions, in particular the institutions
participating in the \gaia{} Multilateral Agreement.
The research here makes use of the following software packages: Astropy \citep{astropy:2018}, NumPy \citep{oliphant2006guide}, SciPy \citep{scipy2001}, Matplotlib \citep{hunter2007matplotlib}, Jupyter Notebook \citep{Kluyver:2016aa}, Astroquery \citep{2019AJ....157...98G}.


\bibliographystyle{mnras}
\bibliography{paper} 

\begin{thebibliography}{}
\makeatletter
\relax
\def\mn@urlcharsother{\let\do\@makeother \do\$\do\&\do\#\do\^\do\_\do\%\do\~}
\def\mn@doi{\begingroup\mn@urlcharsother \@ifnextchar [ {\mn@doi@}
  {\mn@doi@[]}}
\def\mn@doi@[#1]#2{\def\@tempa{#1}\ifx\@tempa\@empty \href
  {http://dx.doi.org/#2} {doi:#2}\else \href {http://dx.doi.org/#2} {#1}\fi
  \endgroup}
\def\mn@eprint#1#2{\mn@eprint@#1:#2::\@nil}
\def\mn@eprint@arXiv#1{\href {http://arxiv.org/abs/#1} {{\tt arXiv:#1}}}
\def\mn@eprint@dblp#1{\href {http://dblp.uni-trier.de/rec/bibtex/#1.xml}
  {dblp:#1}}
\def\mn@eprint@#1:#2:#3:#4\@nil{\def\@tempa {#1}\def\@tempb {#2}\def\@tempc
  {#3}\ifx \@tempc \@empty \let \@tempc \@tempb \let \@tempb \@tempa \fi \ifx
  \@tempb \@empty \def\@tempb {arXiv}\fi \@ifundefined
  {mn@eprint@\@tempb}{\@tempb:\@tempc}{\expandafter \expandafter \csname
  mn@eprint@\@tempb\endcsname \expandafter{\@tempc}}}

\bibitem[\protect\citeauthoryear{{Alvarado-Montes} \&
  {Garc{\'\i}a-Carmona}}{{Alvarado-Montes} \&
  {Garc{\'\i}a-Carmona}}{2019}]{Alvarado19}
{Alvarado-Montes} J.~A.,  {Garc{\'\i}a-Carmona} C.,  2019, \mn@doi [\mnras]
  {10.1093/mnras/stz1081}, \href
  {https://ui.adsabs.harvard.edu/abs/2019MNRAS.486.3963A} {486, 3963}

\bibitem[\protect\citeauthoryear{{Arcangeli} et~al.,}{{Arcangeli}
  et~al.}{2018}]{2018ApJ...855L..30A}
{Arcangeli} J.,  et~al., 2018, \mn@doi [\apjl] {10.3847/2041-8213/aab272},
  \href {https://ui.adsabs.harvard.edu/abs/2018ApJ...855L..30A} {855, L30}

\bibitem[\protect\citeauthoryear{{Arcangeli} et~al.,}{{Arcangeli}
  et~al.}{2019}]{2019A&A...625A.136A}
{Arcangeli} J.,  et~al., 2019, \mn@doi [\aap] {10.1051/0004-6361/201834891},
  \href {https://ui.adsabs.harvard.edu/abs/2019A&A...625A.136A} {625, A136}

\bibitem[\protect\citeauthoryear{{Bakos}, {Noyes}, {Kov{\'a}cs}, {Stanek},
  {Sasselov}  \& {Domsa}}{{Bakos} et~al.}{2004}]{hatproject}
{Bakos} G.,  {Noyes} R.~W.,  {Kov{\'a}cs} G.,  {Stanek} K.~Z.,  {Sasselov}
  D.~D.,   {Domsa} I.,  2004, \mn@doi [\pasp] {10.1086/382735}, \href
  {http://adsabs.harvard.edu/abs/2004PASP..116..266B} {116, 266}

\bibitem[\protect\citeauthoryear{{Bakos} et~al.,}{{Bakos}
  et~al.}{2013}]{hatsproject}
{Bakos} G.~{\'A}.,  et~al., 2013, \mn@doi [\pasp] {10.1086/669529}, \href
  {http://adsabs.harvard.edu/abs/2013PASP..125..154B} {125, 154}

\bibitem[\protect\citeauthoryear{Barbary}{Barbary}{2016}]{Barbary16}
Barbary K.,  2016, {SEP: Source Extractor as a library}, The Journal of Open
  Source Software, \mn@doi{10.21105/joss.00058}

\bibitem[\protect\citeauthoryear{{Bayliss} et~al.,}{{Bayliss}
  et~al.}{2018}]{2018MNRAS.475.4467B}
{Bayliss} D.,  et~al., 2018, \mn@doi [\mnras] {10.1093/mnras/stx2778}, \href
  {https://ui.adsabs.harvard.edu/abs/2018MNRAS.475.4467B} {475, 4467}

\bibitem[\protect\citeauthoryear{{Bensby}, {Feltzing}  \&
  {Lundstr{\"o}m}}{{Bensby} et~al.}{2003}]{Bensby03}
{Bensby} T.,  {Feltzing} S.,   {Lundstr{\"o}m} I.,  2003, \mn@doi [\aap]
  {10.1051/0004-6361:20031213}, \href
  {https://ui.adsabs.harvard.edu/#abs/2003A&A...410..527B} {410, 527}

\bibitem[\protect\citeauthoryear{{Bensby}, {Feltzing}  \& {Oey}}{{Bensby}
  et~al.}{2014}]{Bensby14}
{Bensby} T.,  {Feltzing} S.,   {Oey} M.~S.,  2014, \mn@doi [\aap]
  {10.1051/0004-6361/201322631}, \href
  {https://ui.adsabs.harvard.edu/#abs/2014A&A...562A..71B} {562, A71}

\bibitem[\protect\citeauthoryear{{Bertin} \& {Arnouts}}{{Bertin} \&
  {Arnouts}}{1996}]{1996A&AS..117..393B}
{Bertin} E.,  {Arnouts} S.,  1996, \mn@doi [\aaps] {10.1051/aas:1996164}, \href
  {http://adsabs.harvard.edu/abs/1996A%26AS..117..393B} {117, 393}

\bibitem[\protect\citeauthoryear{{Blecic} et~al.,}{{Blecic}
  et~al.}{2014}]{2014ApJ...781..116B}
{Blecic} J.,  et~al., 2014, \mn@doi [\apj] {10.1088/0004-637X/781/2/116}, \href
  {https://ui.adsabs.harvard.edu/abs/2014ApJ...781..116B} {781, 116}

\bibitem[\protect\citeauthoryear{{Bolmont} \& {Mathis}}{{Bolmont} \&
  {Mathis}}{2016}]{BolmontMathis16}
{Bolmont} E.,  {Mathis} S.,  2016, \mn@doi [Celestial Mechanics and Dynamical
  Astronomy] {10.1007/s10569-016-9690-3}, \href
  {https://ui.adsabs.harvard.edu/abs/2016CeMDA.126..275B} {126, 275}

\bibitem[\protect\citeauthoryear{{Boyajian} et~al.,}{{Boyajian}
  et~al.}{2012}]{Boyajian12}
{Boyajian} T.~S.,  et~al., 2012, \mn@doi [\apj] {10.1088/0004-637X/757/2/112},
  \href {https://ui.adsabs.harvard.edu/abs/2012ApJ...757..112B} {757, 112}

\bibitem[\protect\citeauthoryear{Boyajian et~al.,}{Boyajian
  et~al.}{2017}]{Boyajian17}
Boyajian T.~S.,  et~al., 2017, \mn@doi [The Astrophysical Journal]
  {10.3847/1538-4357/aa8362}, 845, 178

\bibitem[\protect\citeauthoryear{{Brown}, {Collier Cameron}, {Hall}, {Hebb}  \&
  {Smalley}}{{Brown} et~al.}{2011}]{brown2011}
{Brown} D.~J.~A.,  {Collier Cameron} A.,  {Hall} C.,  {Hebb} L.,   {Smalley}
  B.,  2011, \mn@doi [\mnras] {10.1111/j.1365-2966.2011.18729.x}, \href
  {http://adsabs.harvard.edu/abs/2011MNRAS.415..605B} {415, 605}

\bibitem[\protect\citeauthoryear{Bucciarelli et~al.,}{Bucciarelli
  et~al.}{2008}]{gsc23_2008}
Bucciarelli B.,  et~al., 2008, \mn@doi [Proceedings of The International
  Astronomical Union] {10.1017/S1743921308019443}, 248, 316

\bibitem[\protect\citeauthoryear{{Cartier} et~al.,}{{Cartier}
  et~al.}{2017}]{2017AJ....153...34C}
{Cartier} K. M.~S.,  et~al., 2017, \mn@doi [\aj] {10.3847/1538-3881/153/1/34},
  \href {https://ui.adsabs.harvard.edu/abs/2017AJ....153...34C} {153, 34}

\bibitem[\protect\citeauthoryear{{Chakrabarty} \& {Sengupta}}{{Chakrabarty} \&
  {Sengupta}}{2019}]{2019AJ....158...39C}
{Chakrabarty} A.,  {Sengupta} S.,  2019, \mn@doi [\aj]
  {10.3847/1538-3881/ab24dd}, \href
  {https://ui.adsabs.harvard.edu/abs/2019AJ....158...39C} {158, 39}

\bibitem[\protect\citeauthoryear{{Chazelas} et~al.,}{{Chazelas}
  et~al.}{2012}]{Chazelas2012}
{Chazelas} B.,  et~al., 2012, in Ground-based and Airborne Telescopes IV. p.
  84440E, \mn@doi{10.1117/12.925755}

\bibitem[\protect\citeauthoryear{{Chen} et~al.,}{{Chen}
  et~al.}{2014}]{2014A&A...563A..40C}
{Chen} G.,  et~al., 2014, \mn@doi [\aap] {10.1051/0004-6361/201322740}, \href
  {https://ui.adsabs.harvard.edu/abs/2014A&A...563A..40C} {563, A40}

\bibitem[\protect\citeauthoryear{{Collier Cameron} \& {Jardine}}{{Collier
  Cameron} \& {Jardine}}{2018}]{acc2018}
{Collier Cameron} A.,  {Jardine} M.,  2018, \mn@doi [\mnras]
  {10.1093/mnras/sty292}, \href
  {http://adsabs.harvard.edu/abs/2018MNRAS.476.2542C} {476, 2542}

\bibitem[\protect\citeauthoryear{Coppejans et~al.,}{Coppejans
  et~al.}{2013}]{Coppejans:2013gx}
Coppejans R.,  et~al., 2013, Publications of the Astronomical Society of the
  Pacific, 125, 976

\bibitem[\protect\citeauthoryear{{Craig} et~al.,}{{Craig}
  et~al.}{2015}]{2015ascl.soft10007C}
{Craig} M.~W.,  et~al., 2015, {ccdproc: CCD data reduction software},
  Astrophysics Source Code Library (\mn@eprint {ascl} {1510.007})

\bibitem[\protect\citeauthoryear{{Dobbs-Dixon}, {Lin}  \&
  {Mardling}}{{Dobbs-Dixon} et~al.}{2004}]{dobbsdixon2004}
{Dobbs-Dixon} I.,  {Lin} D.~N.~C.,   {Mardling} R.~A.,  2004, \mn@doi [\apj]
  {10.1086/421510}, \href {http://adsabs.harvard.edu/abs/2004ApJ...610..464D}
  {610, 464}

\bibitem[\protect\citeauthoryear{{Doyle} et~al.,}{{Doyle}
  et~al.}{2013}]{2013MNRAS.428.3164D}
{Doyle} A.~P.,  et~al., 2013, \mn@doi [\mnras] {10.1093/mnras/sts267}, \href
  {http://adsabs.harvard.edu/abs/2013MNRAS.428.3164D} {428, 3164}

\bibitem[\protect\citeauthoryear{{Doyle}, {Davies}, {Smalley}, {Chaplin}  \&
  {Elsworth}}{{Doyle} et~al.}{2014}]{2014MNRAS.444.3592D}
{Doyle} A.~P.,  {Davies} G.~R.,  {Smalley} B.,  {Chaplin} W.~J.,   {Elsworth}
  Y.,  2014, \mn@doi [\mnras] {10.1093/mnras/stu1692}, \href
  {https://ui.adsabs.harvard.edu/abs/2014MNRAS.444.3592D} {444, 3592}

\bibitem[\protect\citeauthoryear{{Eggleton}, {Kiseleva}  \& {Hut}}{{Eggleton}
  et~al.}{1998}]{eggleton1998}
{Eggleton} P.~P.,  {Kiseleva} L.~G.,   {Hut} P.,  1998, \mn@doi [\apj]
  {10.1086/305670}, \href {http://adsabs.harvard.edu/abs/1998ApJ...499..853E}
  {499, 853}

\bibitem[\protect\citeauthoryear{{Eigm{\"u}ller} et~al.,}{{Eigm{\"u}ller}
  et~al.}{2019}]{2019A&A...625A.142E}
{Eigm{\"u}ller} P.,  et~al., 2019, \mn@doi [\aap]
  {10.1051/0004-6361/201935206}, \href
  {https://ui.adsabs.harvard.edu/abs/2019A&A...625A.142E} {625, A142}

\bibitem[\protect\citeauthoryear{{Espinoza} et~al.,}{{Espinoza}
  et~al.}{2019}]{2019MNRAS.482.2065E}
{Espinoza} N.,  et~al., 2019, \mn@doi [\mnras] {10.1093/mnras/sty2691}, \href
  {https://ui.adsabs.harvard.edu/abs/2019MNRAS.482.2065E} {482, 2065}

\bibitem[\protect\citeauthoryear{{Fitzpatrick}}{{Fitzpatrick}}{1999}]{Fitzpatrick99}
{Fitzpatrick} E.~L.,  1999, \mn@doi [\pasp] {10.1086/316293}, \href
  {http://adsabs.harvard.edu/abs/1999PASP..111...63F} {111, 63}

\bibitem[\protect\citeauthoryear{{Foreman-Mackey}, {Hogg}, {Lang}  \&
  {Goodman}}{{Foreman-Mackey} et~al.}{2013}]{foremanmackey13}
{Foreman-Mackey} D.,  {Hogg} D.~W.,  {Lang} D.,   {Goodman} J.,  2013, \mn@doi
  [\pasp] {10.1086/670067}, \href
  {http://adsabs.harvard.edu/abs/2013PASP..125..306F} {125, 306}

\bibitem[\protect\citeauthoryear{{Gaia Collaboration} et~al.,}{{Gaia
  Collaboration} et~al.}{2016a}]{GAIA}
{Gaia Collaboration} et~al., 2016a, \mn@doi [\aap]
  {10.1051/0004-6361/201629272}, \href
  {http://adsabs.harvard.edu/abs/2016A%26A...595A...1G} {595, A1}

\bibitem[\protect\citeauthoryear{{Gaia Collaboration} et~al.,}{{Gaia
  Collaboration} et~al.}{2016b}]{gaiadr1}
{Gaia Collaboration} et~al., 2016b, \mn@doi [\aap]
  {10.1051/0004-6361/201629512}, \href
  {http://adsabs.harvard.edu/abs/2016A%26A...595A...2G} {595, A2}

\bibitem[\protect\citeauthoryear{{Gaia Collaboration}, {Brown}, {Vallenari},
  {Prusti}, {de Bruijne}, {Babusiaux}  \& {Bailer-Jones}}{{Gaia Collaboration}
  et~al.}{2018}]{gaiadr2}
{Gaia Collaboration} {Brown} A.~G.~A.,  {Vallenari} A.,  {Prusti} T.,  {de
  Bruijne} J.~H.~J.,  {Babusiaux} C.,   {Bailer-Jones} C.~A.~L.,  2018,
  preprint, \href {http://adsabs.harvard.edu/abs/2018arXiv180409365G} {}
  (\mn@eprint {arXiv} {1804.09365})

\bibitem[\protect\citeauthoryear{{Gandhi} \& {Madhusudhan}}{{Gandhi} \&
  {Madhusudhan}}{2018}]{2018MNRAS.474..271G}
{Gandhi} S.,  {Madhusudhan} N.,  2018, \mn@doi [\mnras]
  {10.1093/mnras/stx2748}, \href
  {https://ui.adsabs.harvard.edu/abs/2018MNRAS.474..271G} {474, 271}

\bibitem[\protect\citeauthoryear{{Gillen}, {Hillenbrand}, {David}, {Aigrain},
  {Rebull}, {Stauffer}, {Cody}  \& {Queloz}}{{Gillen} et~al.}{2017}]{Gillen17}
{Gillen} E.,  {Hillenbrand} L.~A.,  {David} T.~J.,  {Aigrain} S.,  {Rebull} L.,
   {Stauffer} J.,  {Cody} A.~M.,   {Queloz} D.,  2017, \mn@doi [\apj]
  {10.3847/1538-4357/849/1/11}, \href
  {http://adsabs.harvard.edu/abs/2017arXiv170603084G} {849, 11}

\bibitem[\protect\citeauthoryear{{Gillen} et~al.,}{{Gillen}
  et~al.}{2019}]{2019arXiv191109705G}
{Gillen} E.,  et~al., 2019, arXiv e-prints, \href
  {https://ui.adsabs.harvard.edu/abs/2019arXiv191109705G} {p. arXiv:1911.09705}

\bibitem[\protect\citeauthoryear{{Gillon} et~al.,}{{Gillon}
  et~al.}{2014}]{wasp103}
{Gillon} M.,  et~al., 2014, \mn@doi [\aap] {10.1051/0004-6361/201323014}, \href
  {http://adsabs.harvard.edu/abs/2014A%26A...562L...3G} {562, L3}

\bibitem[\protect\citeauthoryear{{Ginsburg} et~al.,}{{Ginsburg}
  et~al.}{2019}]{2019AJ....157...98G}
{Ginsburg} A.,  et~al., 2019, \mn@doi [\aj] {10.3847/1538-3881/aafc33}, \href
  {https://ui.adsabs.harvard.edu/abs/2019AJ....157...98G} {157, 98}

\bibitem[\protect\citeauthoryear{{Gray}}{{Gray}}{2008}]{2008oasp.book.....G}
{Gray} D.~F.,  2008, {The Observation and Analysis of Stellar Photospheres}

\bibitem[\protect\citeauthoryear{{G{\"u}nther} et~al.,}{{G{\"u}nther}
  et~al.}{2017}]{Guenther17b}
{G{\"u}nther} M.~N.,  et~al., 2017, \mn@doi [\mnras] {10.1093/mnras/stx1920},
  472, 295

\bibitem[\protect\citeauthoryear{{G{\"u}nther} et~al.,}{{G{\"u}nther}
  et~al.}{2018}]{2018MNRAS.478.4720G}
{G{\"u}nther} M.~N.,  et~al., 2018, \mn@doi [\mnras] {10.1093/mnras/sty1193},
  \href {https://ui.adsabs.harvard.edu/abs/2018MNRAS.478.4720G} {478, 4720}

\bibitem[\protect\citeauthoryear{{Hebb} et~al.,}{{Hebb} et~al.}{2009}]{wasp12}
{Hebb} L.,  et~al., 2009, \mn@doi [\apj] {10.1088/0004-637X/693/2/1920}, \href
  {http://adsabs.harvard.edu/abs/2009ApJ...693.1920H} {693, 1920}

\bibitem[\protect\citeauthoryear{{Hebb} et~al.,}{{Hebb} et~al.}{2010}]{wasp19}
{Hebb} L.,  et~al., 2010, \mn@doi [\apj] {10.1088/0004-637X/708/1/224}, \href
  {http://adsabs.harvard.edu/abs/2010ApJ...708..224H} {708, 224}

\bibitem[\protect\citeauthoryear{{Heller}}{{Heller}}{2019}]{Heller19}
{Heller} R.,  2019, \mn@doi [\aap] {10.1051/0004-6361/201833486}, \href
  {https://ui.adsabs.harvard.edu/abs/2019A&A...628A..42H} {628, A42}

\bibitem[\protect\citeauthoryear{{Hellier} et~al.,}{{Hellier}
  et~al.}{2009}]{wasp18}
{Hellier} C.,  et~al., 2009, \mn@doi [\nat] {10.1038/nature08245}, \href
  {http://adsabs.harvard.edu/abs/2009Natur.460.1098H} {460, 1098}

\bibitem[\protect\citeauthoryear{{Hellier} et~al.,}{{Hellier}
  et~al.}{2011}]{wasp43}
{Hellier} C.,  et~al., 2011, \mn@doi [\aap] {10.1051/0004-6361/201117081},
  \href {http://adsabs.harvard.edu/abs/2011A%26A...535L...7H} {535, L7}

\bibitem[\protect\citeauthoryear{{Helling}, {Gourbin}, {Woitke}  \&
  {Parmentier}}{{Helling} et~al.}{2019}]{2019A&A...626A.133H}
{Helling} C.,  {Gourbin} P.,  {Woitke} P.,   {Parmentier} V.,  2019, \mn@doi
  [\aap] {10.1051/0004-6361/201834085}, \href
  {https://ui.adsabs.harvard.edu/abs/2019A&A...626A.133H} {626, A133}

\bibitem[\protect\citeauthoryear{{Henden} \& {Munari}}{{Henden} \&
  {Munari}}{2014}]{APASS}
{Henden} A.,  {Munari} U.,  2014, Contributions of the Astronomical Observatory
  Skalnate Pleso, \href {http://adsabs.harvard.edu/abs/2014CoSka..43..518H}
  {43, 518}

\bibitem[\protect\citeauthoryear{{H{\o}g} et~al.,}{{H{\o}g}
  et~al.}{2000}]{tycho2}
{H{\o}g} E.,  et~al., 2000, \aap, \href
  {http://adsabs.harvard.edu/abs/2000A%26A...355L..27H} {355, L27}

\bibitem[\protect\citeauthoryear{{Hoyer}, {Pall{\'e}}, {Dragomir}  \&
  {Murgas}}{{Hoyer} et~al.}{2016}]{2016AJ....151..137H}
{Hoyer} S.,  {Pall{\'e}} E.,  {Dragomir} D.,   {Murgas} F.,  2016, \mn@doi
  [\aj] {10.3847/0004-6256/151/6/137}, \href
  {https://ui.adsabs.harvard.edu/abs/2016AJ....151..137H} {151, 137}

\bibitem[\protect\citeauthoryear{Hunter}{Hunter}{2007}]{hunter2007matplotlib}
Hunter J.~D.,  2007, Computing in science \& engineering, 9, 90

\bibitem[\protect\citeauthoryear{{Husser}, {Wende-von Berg}, {Dreizler},
  {Homeier}, {Reiners}, {Barman}  \& {Hauschildt}}{{Husser}
  et~al.}{2013}]{Husser13}
{Husser} T.~O.,  {Wende-von Berg} S.,  {Dreizler} S.,  {Homeier} D.,  {Reiners}
  A.,  {Barman} T.,   {Hauschildt} P.~H.,  2013, \mn@doi [\aap]
  {10.1051/0004-6361/201219058}, \href
  {https://ui.adsabs.harvard.edu/#abs/2013A&A...553A...6H} {553, A6}

\bibitem[\protect\citeauthoryear{{Indebetouw} et~al.,}{{Indebetouw}
  et~al.}{2005}]{Indebetouw05}
{Indebetouw} R.,  et~al., 2005, \mn@doi [\apj] {10.1086/426679}, \href
  {http://adsabs.harvard.edu/abs/2005ApJ...619..931I} {619, 931}

\bibitem[\protect\citeauthoryear{{Jackman} et~al.,}{{Jackman}
  et~al.}{2019}]{2019arXiv190608219J}
{Jackman} J. A.~G.,  et~al., 2019, arXiv e-prints, \href
  {https://ui.adsabs.harvard.edu/abs/2019arXiv190608219J} {p. arXiv:1906.08219}

\bibitem[\protect\citeauthoryear{{Jiang}, {Lai}, {Savushkin}, {Mkrtichian},
  {Antonyuk}, {Griv}, {Hsieh}  \& {Yeh}}{{Jiang}
  et~al.}{2016}]{2016AJ....151...17J}
{Jiang} I.-G.,  {Lai} C.-Y.,  {Savushkin} A.,  {Mkrtichian} D.,  {Antonyuk} K.,
   {Griv} E.,  {Hsieh} H.-F.,   {Yeh} L.-C.,  2016, \mn@doi [\aj]
  {10.3847/0004-6256/151/1/17}, \href
  {https://ui.adsabs.harvard.edu/abs/2016AJ....151...17J} {151, 17}

\bibitem[\protect\citeauthoryear{Jones, Oliphant, Peterson  et~al.}{Jones
  et~al.}{2001}]{scipy2001}
Jones E.,  Oliphant T.,  Peterson P.,   et~al., 2001, {SciPy}: Open source
  scientific tools for {Python}, \url {http://www.scipy.org/}

\bibitem[\protect\citeauthoryear{{Keating} \& {Cowan}}{{Keating} \&
  {Cowan}}{2017}]{2017ApJ...849L...5K}
{Keating} D.,  {Cowan} N.~B.,  2017, \mn@doi [\apjl]
  {10.3847/2041-8213/aa8b6b}, \href
  {https://ui.adsabs.harvard.edu/abs/2017ApJ...849L...5K} {849, L5}

\bibitem[\protect\citeauthoryear{Kluyver et~al.,}{Kluyver
  et~al.}{2016}]{Kluyver:2016aa}
Kluyver T.,  et~al., 2016, in Loizides F.,  Schmidt B.,  eds, Positioning and
  Power in Academic Publishing: Players, Agents and Agendas. pp 87 -- 90

\bibitem[\protect\citeauthoryear{{Komacek} \& {Showman}}{{Komacek} \&
  {Showman}}{2016}]{2016ApJ...821...16K}
{Komacek} T.~D.,  {Showman} A.~P.,  2016, \mn@doi [\apj]
  {10.3847/0004-637X/821/1/16}, \href
  {https://ui.adsabs.harvard.edu/abs/2016ApJ...821...16K} {821, 16}

\bibitem[\protect\citeauthoryear{{Kov{\'a}cs}, {Zucker}  \&
  {Mazeh}}{{Kov{\'a}cs} et~al.}{2002}]{Kovacs2002}
{Kov{\'a}cs} G.,  {Zucker} S.,   {Mazeh} T.,  2002, \mn@doi [\aap]
  {10.1051/0004-6361:20020802}, \href
  {http://adsabs.harvard.edu/abs/2002A%26A...391..369K} {391, 369}

\bibitem[\protect\citeauthoryear{{Lendl} et~al.,}{{Lendl}
  et~al.}{2012}]{Lendl2012}
{Lendl} M.,  et~al., 2012, \mn@doi [\aap] {10.1051/0004-6361/201219585}, \href
  {http://adsabs.harvard.edu/abs/2012A%26A...544A..72L} {544, A72}

\bibitem[\protect\citeauthoryear{{Lendl}, {Cubillos}, {Hagelberg},
  {M{\"u}ller}, {Juvan}  \& {Fossati}}{{Lendl}
  et~al.}{2017}]{2017A&A...606A..18L}
{Lendl} M.,  {Cubillos} P.~E.,  {Hagelberg} J.,  {M{\"u}ller} A.,  {Juvan} I.,
   {Fossati} L.,  2017, \mn@doi [\aap] {10.1051/0004-6361/201731242}, \href
  {https://ui.adsabs.harvard.edu/abs/2017A&A...606A..18L} {606, A18}

\bibitem[\protect\citeauthoryear{{Maciejewski} et~al.,}{{Maciejewski}
  et~al.}{2018}]{2018AcA....68..371M}
{Maciejewski} G.,  et~al., 2018, \mn@doi [\actaa] {10.32023/0001-5237/68.4.4},
  \href {https://ui.adsabs.harvard.edu/abs/2018AcA....68..371M} {68, 371}

\bibitem[\protect\citeauthoryear{{Mandel} \& {Agol}}{{Mandel} \&
  {Agol}}{2002}]{Mandel02}
{Mandel} K.,  {Agol} E.,  2002, \mn@doi [\apjl] {10.1086/345520}, \href
  {http://adsabs.harvard.edu/abs/2002ApJ...580L.171M} {580, L171}

\bibitem[\protect\citeauthoryear{{Maxted}, {Serenelli}  \&
  {Southworth}}{{Maxted} et~al.}{2015}]{2015AA...575A..36M}
{Maxted} P.~F.~L.,  {Serenelli} A.~M.,   {Southworth} J.,  2015, \mn@doi [\aap]
  {10.1051/0004-6361/201425331}, \href
  {https://ui.adsabs.harvard.edu/abs/2015A&A...575A..36M} {575, A36}

\bibitem[\protect\citeauthoryear{{Mayor} et~al.,}{{Mayor}
  et~al.}{2003}]{2003Msngr.114...20M}
{Mayor} M.,  et~al., 2003, The Messenger, \href
  {http://adsabs.harvard.edu/abs/2003Msngr.114...20M} {114, 20}

\bibitem[\protect\citeauthoryear{{McCormac}, {Pollacco}, {Skillen}, {Faedi},
  {Todd}  \& {Watson}}{{McCormac} et~al.}{2013}]{donuts}
{McCormac} J.,  {Pollacco} D.,  {Skillen} I.,  {Faedi} F.,  {Todd} I.,
  {Watson} C.~A.,  2013, \mn@doi [\pasp] {10.1086/670940}, \href
  {https://ui.adsabs.harvard.edu/abs/2013PASP..125..548M} {125, 548}

\bibitem[\protect\citeauthoryear{{McCormac} et~al.,}{{McCormac}
  et~al.}{2017}]{McCormac2017}
{McCormac} J.,  et~al., 2017, \mn@doi [\pasp]
  {10.1088/1538-3873/129/972/025002}, \href
  {http://adsabs.harvard.edu/abs/2017PASP..129b5002M} {129, 025002}

\bibitem[\protect\citeauthoryear{{McQuillan}, {Mazeh}  \&
  {Aigrain}}{{McQuillan} et~al.}{2014}]{mcquillan2014}
{McQuillan} A.,  {Mazeh} T.,   {Aigrain} S.,  2014, \mn@doi [The Astrophysical
  Journal Supplement Series] {10.1088/0067-0049/211/2/24}, \href
  {https://ui.adsabs.harvard.edu/#abs/2014ApJS..211...24M} {211, 24}

\bibitem[\protect\citeauthoryear{{Mendon{\c{c}}a}, {Malik}, {Demory}  \&
  {Heng}}{{Mendon{\c{c}}a} et~al.}{2018a}]{2018AJ....155..150M}
{Mendon{\c{c}}a} J.~M.,  {Malik} M.,  {Demory} B.-O.,   {Heng} K.,  2018a,
  \mn@doi [\aj] {10.3847/1538-3881/aaaebc}, \href
  {https://ui.adsabs.harvard.edu/abs/2018AJ....155..150M} {155, 150}

\bibitem[\protect\citeauthoryear{{Mendon{\c{c}}a}, {Tsai}, {Malik}, {Grimm}  \&
  {Heng}}{{Mendon{\c{c}}a} et~al.}{2018b}]{2018ApJ...869..107M}
{Mendon{\c{c}}a} J.~M.,  {Tsai} S.-m.,  {Malik} M.,  {Grimm} S.~L.,   {Heng}
  K.,  2018b, \mn@doi [\apj] {10.3847/1538-4357/aaed23}, \href
  {https://ui.adsabs.harvard.edu/abs/2018ApJ...869..107M} {869, 107}

\bibitem[\protect\citeauthoryear{{Murgas}, {Pall{\'e}}, {Zapatero Osorio},
  {Nortmann}, {Hoyer}  \& {Cabrera-Lavers}}{{Murgas}
  et~al.}{2014}]{2014A&A...563A..41M}
{Murgas} F.,  {Pall{\'e}} E.,  {Zapatero Osorio} M.~R.,  {Nortmann} L.,
  {Hoyer} S.,   {Cabrera-Lavers} A.,  2014, \mn@doi [\aap]
  {10.1051/0004-6361/201322374}, \href
  {https://ui.adsabs.harvard.edu/abs/2014A&A...563A..41M} {563, A41}

\bibitem[\protect\citeauthoryear{{Ni}}{{Ni}}{2018}]{2018A&A...613A..32N}
{Ni} D.,  2018, \mn@doi [\aap] {10.1051/0004-6361/201732183}, \href
  {https://ui.adsabs.harvard.edu/abs/2018A&A...613A..32N} {613, A32}

\bibitem[\protect\citeauthoryear{Oberst et~al.,}{Oberst et~al.}{2017}]{kelt16}
Oberst T.~E.,  et~al., 2017, The Astronomical Journal, 153, 97

\bibitem[\protect\citeauthoryear{{Ogilvie} \& {Lin}}{{Ogilvie} \&
  {Lin}}{2007}]{OgilvieLin07}
{Ogilvie} G.~I.,  {Lin} D.~N.~C.,  2007, \mn@doi [\apj] {10.1086/515435}, \href
  {https://ui.adsabs.harvard.edu/abs/2007ApJ...661.1180O} {661, 1180}

\bibitem[\protect\citeauthoryear{Oliphant}{Oliphant}{2006}]{oliphant2006guide}
Oliphant T.~E.,  2006, A guide to NumPy.
 Vol. 1, Trelgol Publishing USA

\bibitem[\protect\citeauthoryear{{Parviainen} \& {Aigrain}}{{Parviainen} \&
  {Aigrain}}{2015}]{Parviainen15}
{Parviainen} H.,  {Aigrain} S.,  2015, \mn@doi [\mnras]
  {10.1093/mnras/stv1857}, \href
  {http://adsabs.harvard.edu/abs/2015MNRAS.453.3821P} {453, 3821}

\bibitem[\protect\citeauthoryear{{Penev} et~al.,}{{Penev}
  et~al.}{2016a}]{hats18}
{Penev} K.,  et~al., 2016a, \mn@doi [\aj] {10.3847/0004-6256/152/5/127}, \href
  {http://adsabs.harvard.edu/abs/2016AJ....152..127P} {152, 127}

\bibitem[\protect\citeauthoryear{{Penev} et~al.,}{{Penev}
  et~al.}{2016b}]{2016AJ....152..127P}
{Penev} K.,  et~al., 2016b, \mn@doi [\aj] {10.3847/0004-6256/152/5/127}, \href
  {https://ui.adsabs.harvard.edu/abs/2016AJ....152..127P} {152, 127}

\bibitem[\protect\citeauthoryear{{Penev}, {Bouma}, {Winn}  \&
  {Hartman}}{{Penev} et~al.}{2018}]{penev2018}
{Penev} K.,  {Bouma} L.~G.,  {Winn} J.~N.,   {Hartman} J.~D.,  2018, \mn@doi
  [\aj] {10.3847/1538-3881/aaaf71}, \href
  {http://adsabs.harvard.edu/abs/2018AJ....155..165P} {155, 165}

\bibitem[\protect\citeauthoryear{{Pepper} et~al.,}{{Pepper}
  et~al.}{2007}]{keltproject1}
{Pepper} J.,  et~al., 2007, \mn@doi [\pasp] {10.1086/521836}, \href
  {http://adsabs.harvard.edu/abs/2007PASP..119..923P} {119, 923}

\bibitem[\protect\citeauthoryear{{Pepper}, {Kuhn}, {Siverd}, {James}  \&
  {Stassun}}{{Pepper} et~al.}{2012}]{keltproject2}
{Pepper} J.,  {Kuhn} R.~B.,  {Siverd} R.,  {James} D.,   {Stassun} K.,  2012,
  \mn@doi [\pasp] {10.1086/665044}, \href
  {http://adsabs.harvard.edu/abs/2012PASP..124..230P} {124, 230}

\bibitem[\protect\citeauthoryear{{Petrucci}, {Jofr{\'e}}, {G{\'o}mez Maqueo
  Chew}, {Hinse}, {Ma{\v{s}}ek}, {Tan}  \& {G{\'o}mez}}{{Petrucci}
  et~al.}{2019}]{2019arXiv191011930P}
{Petrucci} R.,  {Jofr{\'e}} E.,  {G{\'o}mez Maqueo Chew} Y.,  {Hinse} T.~C.,
  {Ma{\v{s}}ek} M.,  {Tan} T.~G.,   {G{\'o}mez} M.,  2019, arXiv e-prints,
  \href {https://ui.adsabs.harvard.edu/abs/2019arXiv191011930P} {p.
  arXiv:1910.11930}

\bibitem[\protect\citeauthoryear{{Pinhas}, {Madhusudhan}, {Gandhi}  \&
  {MacDonald}}{{Pinhas} et~al.}{2019}]{2019MNRAS.482.1485P}
{Pinhas} A.,  {Madhusudhan} N.,  {Gandhi} S.,   {MacDonald} R.,  2019, \mn@doi
  [\mnras] {10.1093/mnras/sty2544}, \href
  {https://ui.adsabs.harvard.edu/abs/2019MNRAS.482.1485P} {482, 1485}

\bibitem[\protect\citeauthoryear{{Pollacco} et~al.,}{{Pollacco}
  et~al.}{2006}]{waspproject}
{Pollacco} D.~L.,  et~al., 2006, \mn@doi [\pasp] {10.1086/508556}, \href
  {http://adsabs.harvard.edu/abs/2006PASP..118.1407P} {118, 1407}

\bibitem[\protect\citeauthoryear{{Press}, {Teukolsky}, {Vetterling}  \&
  {Flannery}}{{Press} et~al.}{1992}]{press1992}
{Press} W.~H.,  {Teukolsky} S.~A.,  {Vetterling} W.~T.,   {Flannery} B.~P.,
  1992, {Numerical recipes in C. The art of scientific computing}

\bibitem[\protect\citeauthoryear{{Price-Whelan} et~al.,}{{Price-Whelan}
  et~al.}{2018}]{astropy:2018}
{Price-Whelan} A.~M.,  et~al., 2018, \mn@doi [\aj] {10.3847/1538-3881/aabc4f},
  \href {https://ui.adsabs.harvard.edu/#abs/2018AJ....156..123T} {156, 123}

\bibitem[\protect\citeauthoryear{{Queloz} et~al.,}{{Queloz}
  et~al.}{2001}]{queloz2001}
{Queloz} D.,  et~al., 2001, \mn@doi [\aap] {10.1051/0004-6361:20011308}, \href
  {http://adsabs.harvard.edu/abs/2001A%26A...379..279Q} {379, 279}

\bibitem[\protect\citeauthoryear{{Raynard} et~al.,}{{Raynard}
  et~al.}{2018}]{2018MNRAS.481.4960R}
{Raynard} L.,  et~al., 2018, \mn@doi [\mnras] {10.1093/mnras/sty2581}, \href
  {https://ui.adsabs.harvard.edu/abs/2018MNRAS.481.4960R} {481, 4960}

\bibitem[\protect\citeauthoryear{{Ricci} et~al.,}{{Ricci}
  et~al.}{2015}]{2015PASP..127..143R}
{Ricci} D.,  et~al., 2015, \mn@doi [\pasp] {10.1086/680233}, \href
  {https://ui.adsabs.harvard.edu/abs/2015PASP..127..143R} {127, 143}

\bibitem[\protect\citeauthoryear{{Sedaghati} et~al.,}{{Sedaghati}
  et~al.}{2017}]{2017Natur.549..238S}
{Sedaghati} E.,  et~al., 2017, \mn@doi [\nat] {10.1038/nature23651}, \href
  {https://ui.adsabs.harvard.edu/abs/2017Natur.549..238S} {549, 238}

\bibitem[\protect\citeauthoryear{{Serenelli}, {Bergemann}, {Ruchti}  \&
  {Casagrande}}{{Serenelli} et~al.}{2013}]{2013MNRAS.429.3645S}
{Serenelli} A.~M.,  {Bergemann} M.,  {Ruchti} G.,   {Casagrande} L.,  2013,
  \mn@doi [\mnras] {10.1093/mnras/sts648}, \href
  {https://ui.adsabs.harvard.edu/abs/2013MNRAS.429.3645S} {429, 3645}

\bibitem[\protect\citeauthoryear{{Sheppard}, {Mandell}, {Tamburo}, {Gand hi},
  {Pinhas}, {Madhusudhan}  \& {Deming}}{{Sheppard}
  et~al.}{2017}]{2017ApJ...850L..32S}
{Sheppard} K.~B.,  {Mandell} A.~M.,  {Tamburo} P.,  {Gand hi} S.,  {Pinhas} A.,
   {Madhusudhan} N.,   {Deming} D.,  2017, \mn@doi [\apjl]
  {10.3847/2041-8213/aa9ae9}, \href
  {https://ui.adsabs.harvard.edu/abs/2017ApJ...850L..32S} {850, L32}

\bibitem[\protect\citeauthoryear{{Shporer} et~al.,}{{Shporer}
  et~al.}{2019}]{2019AJ....157..178S}
{Shporer} A.,  et~al., 2019, \mn@doi [\aj] {10.3847/1538-3881/ab0f96}, \href
  {https://ui.adsabs.harvard.edu/abs/2019AJ....157..178S} {p.~178}

\bibitem[\protect\citeauthoryear{{Sing} et~al.,}{{Sing}
  et~al.}{2016}]{2016Natur.529...59S}
{Sing} D.~K.,  et~al., 2016, \mn@doi [\nat] {10.1038/nature16068}, \href
  {https://ui.adsabs.harvard.edu/abs/2016Natur.529...59S} {529, 59}

\bibitem[\protect\citeauthoryear{{Skrutskie} et~al.,}{{Skrutskie}
  et~al.}{2006}]{2MASS}
{Skrutskie} M.~F.,  et~al., 2006, \mn@doi [\aj] {10.1086/498708}, \href
  {http://adsabs.harvard.edu/abs/2006AJ....131.1163S} {131, 1163}

\bibitem[\protect\citeauthoryear{{Stassun}, {Collins}  \& {Gaudi}}{{Stassun}
  et~al.}{2017}]{stassun2017}
{Stassun} K.~G.,  {Collins} K.~A.,   {Gaudi} B.~S.,  2017, \mn@doi [\aj]
  {10.3847/1538-3881/aa5df3}, \href
  {https://ui.adsabs.harvard.edu/abs/2017AJ....153..136S} {153, 136}

\bibitem[\protect\citeauthoryear{{Su{\'a}rez Mascare{\~n}o}, {Rebolo},
  {Gonz{\'a}lez Hern{\'a}ndez}  \& {Esposito}}{{Su{\'a}rez Mascare{\~n}o}
  et~al.}{2017}]{2017MNRAS.468.4772S}
{Su{\'a}rez Mascare{\~n}o} A.,  {Rebolo} R.,  {Gonz{\'a}lez Hern{\'a}ndez}
  J.~I.,   {Esposito} M.,  2017, \mn@doi [\mnras] {10.1093/mnras/stx771}, \href
  {https://ui.adsabs.harvard.edu/abs/2017MNRAS.468.4772S} {468, 4772}

\bibitem[\protect\citeauthoryear{{Tamuz}, {Mazeh}  \& {Zucker}}{{Tamuz}
  et~al.}{2005}]{Tamuz2005}
{Tamuz} O.,  {Mazeh} T.,   {Zucker} S.,  2005, \mn@doi [\mnras]
  {10.1111/j.1365-2966.2004.08585.x}, \href
  {http://adsabs.harvard.edu/abs/2005MNRAS.356.1466T} {356, 1466}

\bibitem[\protect\citeauthoryear{{Vines} et~al.,}{{Vines}
  et~al.}{2019}]{2019arXiv190407997V}
{Vines} J.~I.,  et~al., 2019, arXiv e-prints, \href
  {https://ui.adsabs.harvard.edu/abs/2019arXiv190407997V} {p. arXiv:1904.07997}

\bibitem[\protect\citeauthoryear{{Weiss} \& {Schlattl}}{{Weiss} \&
  {Schlattl}}{2008}]{2008ApSS.316...99W}
{Weiss} A.,  {Schlattl} H.,  2008, \mn@doi [\apss] {10.1007/s10509-007-9606-5},
  \href {https://ui.adsabs.harvard.edu/abs/2008Ap&SS.316...99W} {316, 99}

\bibitem[\protect\citeauthoryear{{West} et~al.,}{{West}
  et~al.}{2019}]{2019MNRAS.486.5094W}
{West} R.~G.,  et~al., 2019, \mn@doi [\mnras] {10.1093/mnras/stz1084}, \href
  {https://ui.adsabs.harvard.edu/abs/2019MNRAS.486.5094W} {486, 5094}

\bibitem[\protect\citeauthoryear{{Wheatley} et~al.,}{{Wheatley}
  et~al.}{2013}]{2013EPJWC..4713002W}
{Wheatley} P.~J.,  et~al., 2013, in European Physical Journal Web of
  Conferences. p. 13002 (\mn@eprint {arXiv} {1302.6592}),
  \mn@doi{10.1051/epjconf/20134713002}

\bibitem[\protect\citeauthoryear{{Wheatley} et~al.,}{{Wheatley}
  et~al.}{2017}]{project2018}
{Wheatley} P.,  et~al., 2017, \mnras, \href
  {http://adsabs.harvard.edu/abs/2017MNRAS....999.999W} {accepted}

\bibitem[\protect\citeauthoryear{{Wilkins}, {Delrez}, {Barker}, {Deming},
  {Hamilton}, {Gillon}  \& {Jehin}}{{Wilkins}
  et~al.}{2017}]{2017ApJ...836L..24W}
{Wilkins} A.~N.,  {Delrez} L.,  {Barker} A.~J.,  {Deming} D.,  {Hamilton} D.,
  {Gillon} M.,   {Jehin} E.,  2017, \mn@doi [\apjl] {10.3847/2041-8213/aa5d9f},
  \href {https://ui.adsabs.harvard.edu/abs/2017ApJ...836L..24W} {836, L24}

\bibitem[\protect\citeauthoryear{{Wong} et~al.,}{{Wong}
  et~al.}{2016}]{wong2016}
{Wong} I.,  et~al., 2016, \mn@doi [\apj] {10.3847/0004-637X/823/2/122}, \href
  {https://ui.adsabs.harvard.edu/abs/2016ApJ...823..122W} {823, 122}

\bibitem[\protect\citeauthoryear{{Wright} et~al.,}{{Wright}
  et~al.}{2010}]{WISE}
{Wright} E.~L.,  et~al., 2010, \mn@doi [\aj] {10.1088/0004-6256/140/6/1868},
  \href {http://adsabs.harvard.edu/abs/2010AJ....140.1868W} {140, 1868}

\bibitem[\protect\citeauthoryear{{del Peloso}, {da Silva}, {Porto de Mello}  \&
  {Arany-Prado}}{{del Peloso} et~al.}{2005}]{2005A&A...440.1153D}
{del Peloso} E.~F.,  {da Silva} L.,  {Porto de Mello} G.~F.,   {Arany-Prado}
  L.~I.,  2005, \mn@doi [\aap] {10.1051/0004-6361:20053307}, \href
  {https://ui.adsabs.harvard.edu/abs/2005A&A...440.1153D} {440, 1153}

\makeatother
\end{thebibliography}

\appendix

\section{Undetrended photometry}
\label{apendix:undetrended_phot}

In this section we show the undetrended follow-up transit photometry of \NPlanet{} for completeness. 


\begin{figure*}
	\includegraphics[width=5.8cm]{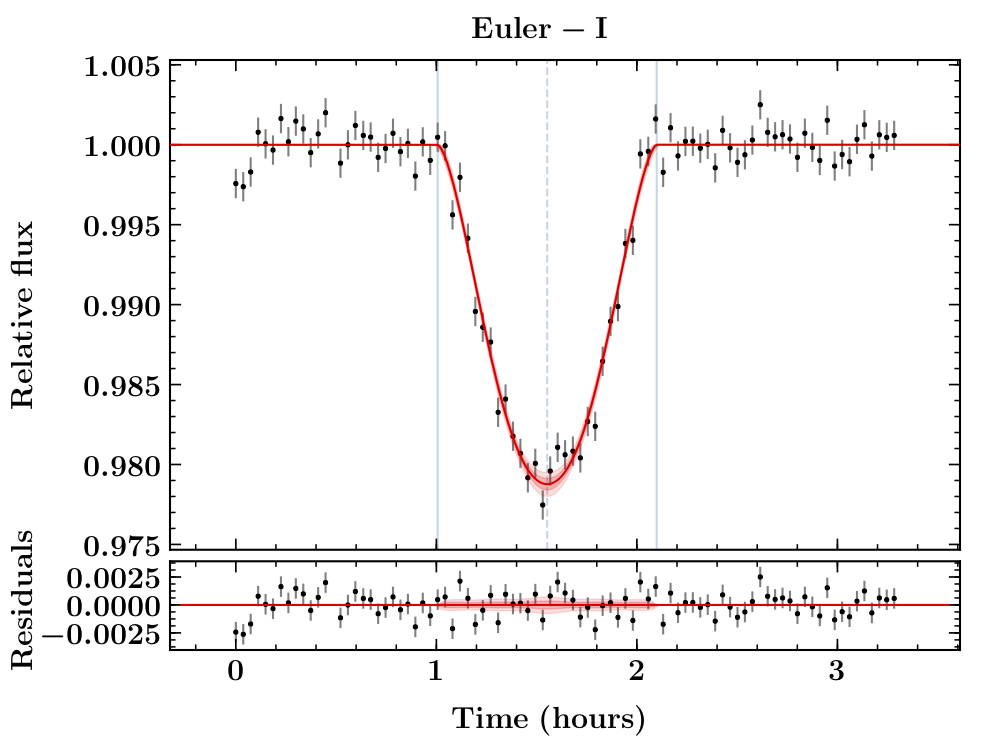}
	\includegraphics[width=5.8cm]{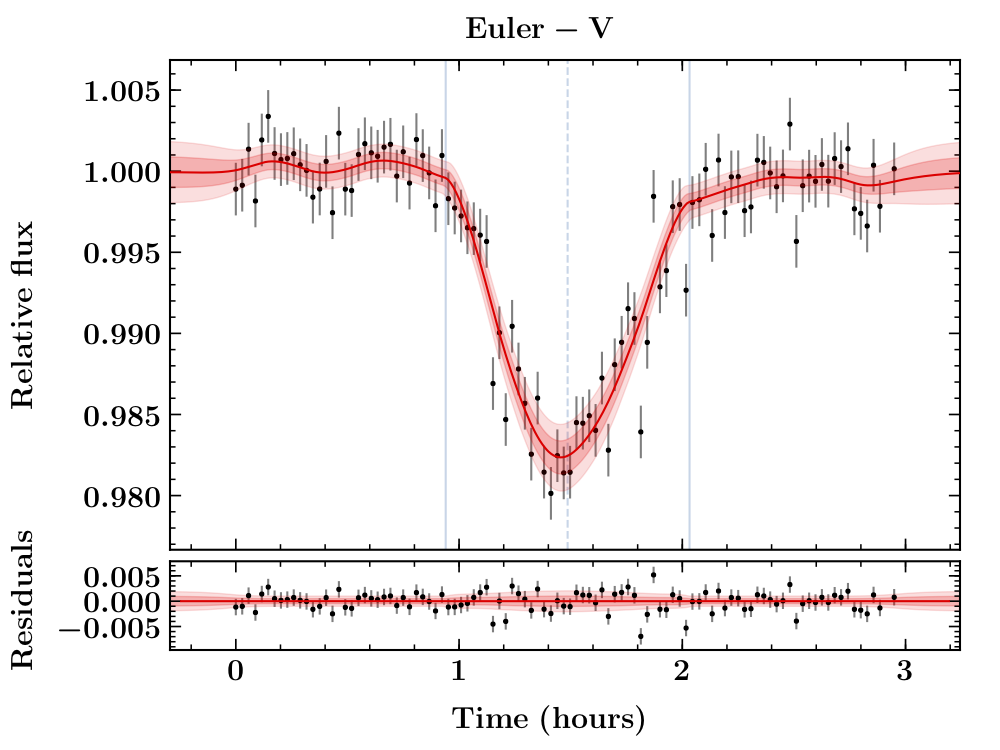}
	\includegraphics[width=5.8cm]{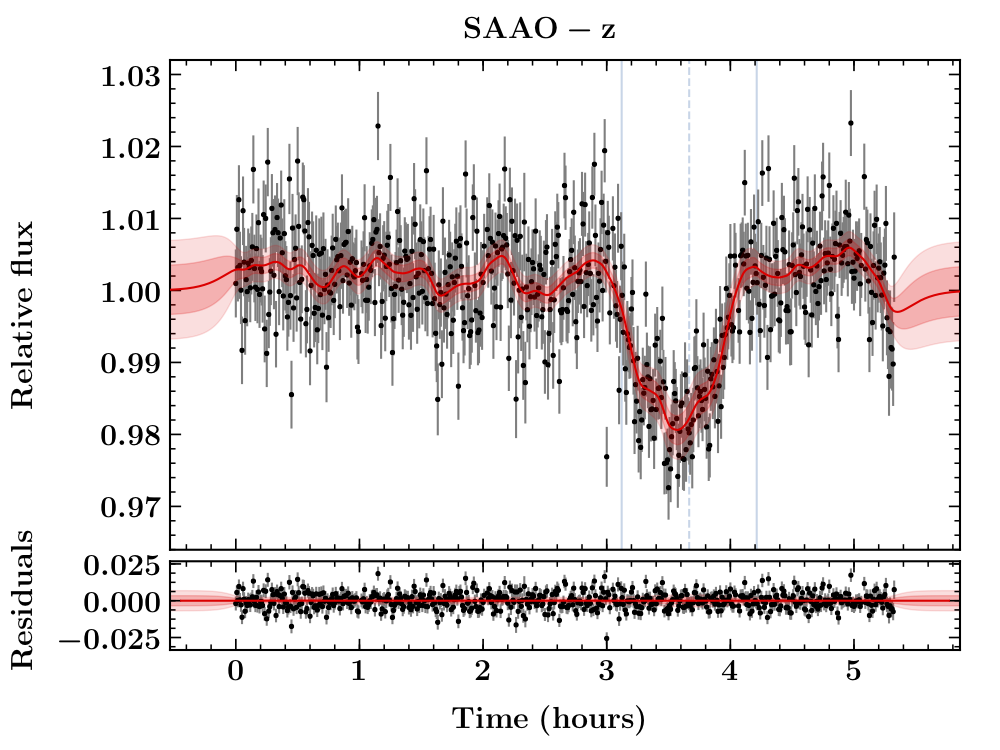}
	\includegraphics[width=5.8cm]{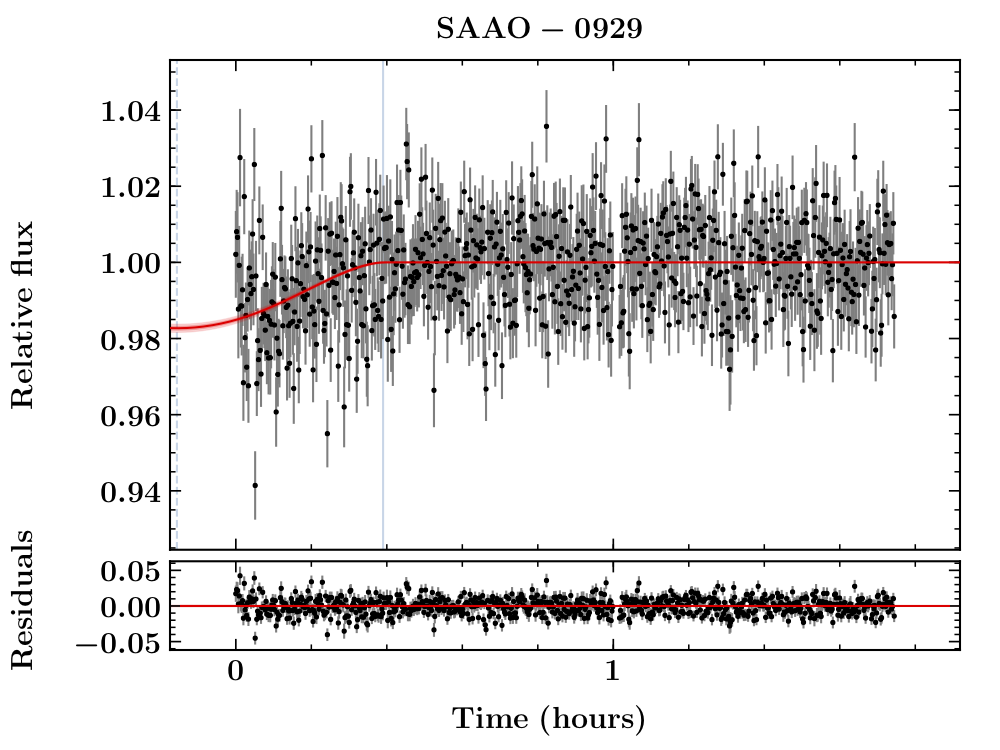}
    \includegraphics[width=5.8cm]{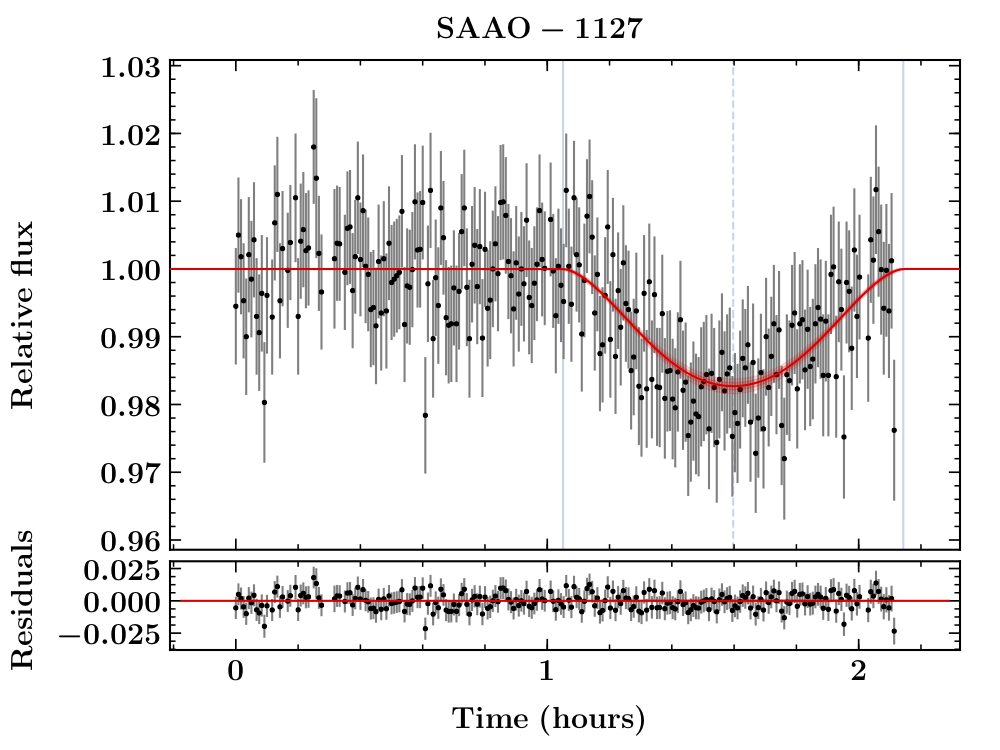}
    \includegraphics[width=5.8cm]{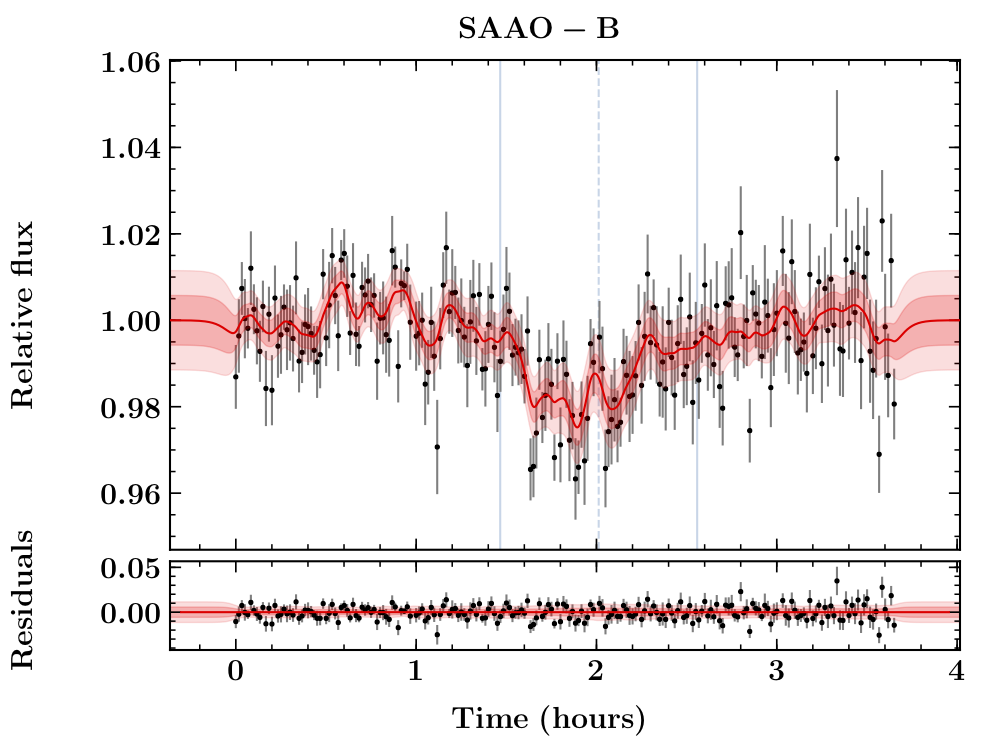}
    \caption{Same as Figure \ref{fig:followup_phot_all_detrended} but for the undetrended photometry}
    \label{fig:followup_phot_all_undetrended}
\end{figure*}

\bsp	
\label{lastpage}
\end{document}